\documentclass[%
 reprint,
superscriptaddress,
 amsmath,amssymb,
prb,
]{revtex4-2}
\usepackage[usenames,dvipsnames]{xcolor}
\usepackage[version=4]{mhchem}
\usepackage[none]{hyphenat}
\usepackage{graphicx}
\usepackage{dcolumn}
\usepackage{bm}
\usepackage{hyperref}

\hypersetup{
  colorlinks = true,
  urlcolor = blue,
  linkcolor = blue,
  citecolor = blue
}
\usepackage{chemformula}
\let\ce\ch
\usepackage{amsmath}
\usepackage{tikz}
\usetikzlibrary{shapes.geometric, arrows, calc, backgrounds}

\usepackage{natbib}

\usepackage{tikz}
\usetikzlibrary{arrows.meta,
                chains,
                decorations.pathreplacing,calligraphy,
                positioning,
                quotes,
                shapes.geometric
                }

\tikzstyle{startstop} = [rectangle, rounded corners, 
minimum width=3cm, 
minimum height=1cm,
text centered, 
draw=black, 
fill=red!30]

\tikzstyle{io} = [trapezium, 
trapezium stretches=true, 
trapezium left angle=70, 
trapezium right angle=110, 
minimum width=3cm, 
minimum height=1cm, text centered, 
draw=black, fill=blue!30]

\tikzstyle{process} = [rectangle, 
minimum width=3cm, 
minimum height=1cm, 
text centered, 
text width=3cm, 
draw=black, 
fill=orange!30]

\tikzstyle{decision} = [diamond, 
minimum width=3cm, 
minimum height=1cm, 
text centered, 
draw=black, 
fill=green!30]
\tikzstyle{arrow} = [thick,->,>=stealth]

\usepackage{booktabs}
\usepackage{multirow}
\usepackage{longtable}
\usepackage{threeparttable}
\usepackage{textcomp}
\usepackage{array}
\usepackage{adjustbox}
\usepackage{makecell}

\usepackage{enumitem}
\usepackage[caption=false, font = bf]{subfig}

\usepackage{relsize}

\begin{document}
\setcitestyle{super}
\makeatletter
\renewcommand*{\fnum@figure}{{\normalfont\bfseries \figurename~\thefigure}}
\renewcommand*{\@caption@fignum@sep}{\textbf{. }}

\preprint{APS/123-QED}

\title{
High-$T_c$ superconductor candidates
proposed by machine learning 
}

\author{Siwoo Lee}
\affiliation{Department of Chemistry, University of Toronto, St. George campus, Toronto, ON, Canada}
\affiliation{Acceleration Consortium, University of Toronto, St. George campus, Toronto, ON, Canada}

\author{Jason Hattrick-Simpers}
\email{jason.hattrick.simpers@utoronto.ca}
\affiliation{Acceleration Consortium, University of Toronto, St. George campus, Toronto, ON, Canada}
\affiliation{Department of Materials Science \& Engineering, University of Toronto, St. George campus, Toronto, ON, Canada}

\author{Young-June Kim}
\email{youngjune.kim@utoronto.ca}
\affiliation{Department of Physics, University of Toronto, St. George campus, Toronto, ON, Canada}

\author{O. Anatole von Lilienfeld}
\email{anatole.vonlilienfeld@utoronto.ca}
\affiliation{Department of Chemistry, University of Toronto, St. George campus, Toronto, ON, Canada}
\affiliation{Acceleration Consortium, University of Toronto,  St. George campus, Toronto, ON, Canada}
\affiliation{Department of Materials Science \& Engineering, University of Toronto, St. George campus, Toronto, ON, Canada}
\affiliation{Department of Physics, University of Toronto, St. George campus, Toronto, ON, Canada}
\affiliation{Vector Institute for Artificial Intelligence, Toronto, ON, Canada}

\begin{abstract}
We cast the relation between the chemical composition of a solid-state material and its superconducting critical temperature ($T_c$) as a statistical learning problem with reduced complexity. 
Training of query-aware similarity-based ridge regression models on experimental SuperCon data achieve average $T_c$ prediction errors of $\pm$5 K for unseen out-of-sample materials. 
Two models were trained with one excluding high pressure data in training (``ambient'' model) and a second also including high pressure data (``implicit'' model).
Subsequent utilization of the approach to scan $\sim$153k materials in the Materials Project enables the ranking of candidates by $T_c$ while accounting for thermodynamic stability and small band gap. 
The ambient model is used to predict stable top three high-$T_c$ candidate materials that include those with large band gaps of \ce{LiCuF4} (316 K), \ce{Ag2H12S(NO)4} (316 K), and \ce{Na2H6PtO6} (315 K). 
Filtering these candidates for those with small band gaps correspondingly yields \ce{LiCuF4} (316 K), \ce{Cu2P2O7} (311 K), and \ce{Cu3P2H2O9} (307 K). 
\end{abstract}

\maketitle

\section{\label{sec: Introduction}Introduction}

\begin{figure*}
\centering
\subfloat{\includegraphics[width=1\textwidth]{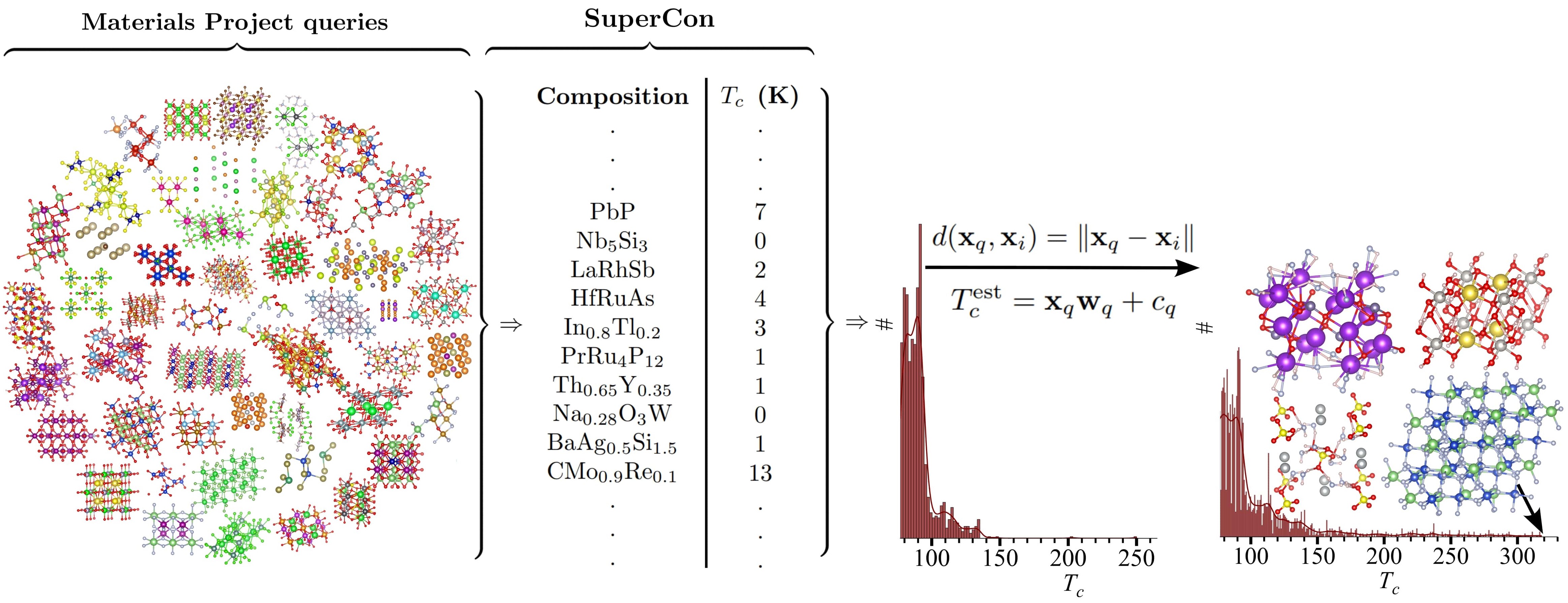}}
\hfill
\caption{\label{fig: Graphical abstract}
From left to right: query candidates (structural depictions drawn from the Materials Project~\cite{Materials_Project_1}) are used to select closest instances (in terms of $d$) from the SuperCon data set~\cite{SuperCon} with known experimentally-measured $T_c$'s for training ridge regression models.
Materials representations do not require any structural information and are based only on features derived from chemical compositions.
The application of this similarity-based machine learning to $\sim$153k materials in the Materials Project enriches the existing $T_c$ distribution and includes multiple promising high-$T_c$ superconductors not present in SuperCon.
}
\end{figure*}

A major unsolved question in the physical sciences is whether there exists at ambient pressure a superconductor with its superconducting critical temperature, $T_c$, at or above room temperature\cite{holy_grail}.
Under these conditions, such a material must exhibit the hallmark properties of superconductivity \cite{superconductor_properties_1,superconductor_properties_2}, \textit{i.e.}\ the persistent absence of electrical resistance and the expulsion of magnetic fields (the Meissner effect) below a critical magnetic field strength ($H_c$).
Its discovery would be of immense significance to various scientific and industrial applications, encompassing energy storage and electrical power transmission\cite{superconductor_energy_applications_1, superconductor_energy_applications_2, superconductor_energy_applications_3, superconductor_energy_applications_4}, nuclear fusion energy\cite{superconductor_fusion_application_1, superconductor_fusion_application_2, superconductor_fusion_application_3}, computing\cite{superconductor_computer_applications_1, superconductor_computer_applications_2, superconductor_computer_applications_3, superconductor_computer_applications_4}, medicine\cite{superconductor_medicine_applications_1, superconductor_medicine_applications_2}, public transport\cite{superconductor_transportation_applications_1, superconductor_transportation_applications_2, superconductor_transportation_applications_3}, to particle physics\cite{superconductor_particle_physics_applications_1, superconductor_particle_physics_applications_2, superconductor_particle_physics_applications_3, superconductor_particle_physics_applications_4}.
Unfortunately, the identification of candidates is difficult because the overwhelming majority of known superconductors have $T_c$'s near 0 K\cite{SuperCon},
making it challenging to learn from the properties of high-$T_c$ samples and thereby propose novel candidates with higher $T_c$'s. 
While the Bardeen-Cooper-Schrieffer (BCS) theory\cite{BCS_theory_1, BCS_theory_2} can successfully explain superconductivity at low temperatures as a phonon-mediated process to form a Cooper pair, there currently exists no complete microscopic theory of high-$T_c$ superconductivity\cite{superconductivity_theory_unknown}.
Therefore, approaches to designing new higher-$T_c$ materials have historically been largely empirical without the guidance of theory\cite{there_is_hope_2}.  
While \textit{ab initio} electronic structures methods could be more efficient than experiments, the lack of predictive $T_c$ modeling also impedes the computational design and discovery. 
This poses a severe bottleneck on the search for new superconductors within the vast materials compound space estimated by some to be populated by $\sim$$10^{100}$ plausible materials\cite{size_of_materials_space}.\\

In this work, we describe a similarity-based machine learning (ML) approach that is capable of extrapolating beyond the distribution of experimentally-measured $T_c$ values found in the SuperCon data set\cite{SuperCon} to potentially identify materials with $T_c$'s greater than the currently known highest value, which is 250 K for \ce{LaH10}\cite{lanthanum_hydride_superconductor}. 
Recently, there have been many efforts to predict $T_c$'s directly from the crystal structures and chemical compositions of materials using ML\cite{ML_method_background_1, ML_method_background_2, ML_method_background_3, ML_method_background_4, ML_method_background_5, ML_method_background_6, ML_method_background_7, ML_method_background_8, rediscovery_1, ML_method_background_9}.
However, as far as we are aware, none of these works has successfully dealt with the out-of-domain (OOD) problem\cite{OOD_1, OOD_2} that is inherent to attempts of making accurate predictions for samples with label values found beyond the range that the ML models were trained on.
This is, in fact, a well-known issue in the ML field\cite{OOD_ML_problem_1, OOD_ML_problem_2} that must be tackled if ML models are to be used to propose novel high-$T_c$ candidates.\\

We have addressed this problem using the procedure outlined in \autoref{fig: Graphical abstract}.
First, for a given material, its $n$-nearest neighbors in the SuperCon~\cite{SuperCon,SuperCon2} training data set are queried.
These $n$-nearest neighbors, defined as those with feature vectors that yield the smallest Euclidean distances relative to the test sample's feature vector, are then used to train a ridge regression model, from which the test sample's $T_c$ is predicted.
Leave-one-out prediction tests of the resulting models trained on sets drawn either from all of the $\sim$13.6k samples in SuperCon~\cite{SuperCon} (implicit model), or from its subset exclusively containing entries at ambient pressures (ambient model), suggest good predictive performance throughout the full range of measured $T_c$'s. 
Subsequently, we have applied this approach to predict $T_c$'s of $\sim$153k materials for which calculated density functional theory (DFT) property results are recorded in the Materials Project\cite{Materials_Project_1} database.
After filtering for thermodynamic stability, we have ranked all remaining materials by our ML $T_c$ estimates. 
The ambient model identifies sixty-four candidate materials with predicted $T_c$'s above 250 K, of which thirty-four have DFT-computed band gaps smaller than 1 eV.

\section{\label{sec: Methodology}Methodology}

\subsection{\label{sec: Superconductors Data Set}Superconductors Data Set}

\begin{figure}[h]
\resizebox{\columnwidth}{!}{\includegraphics{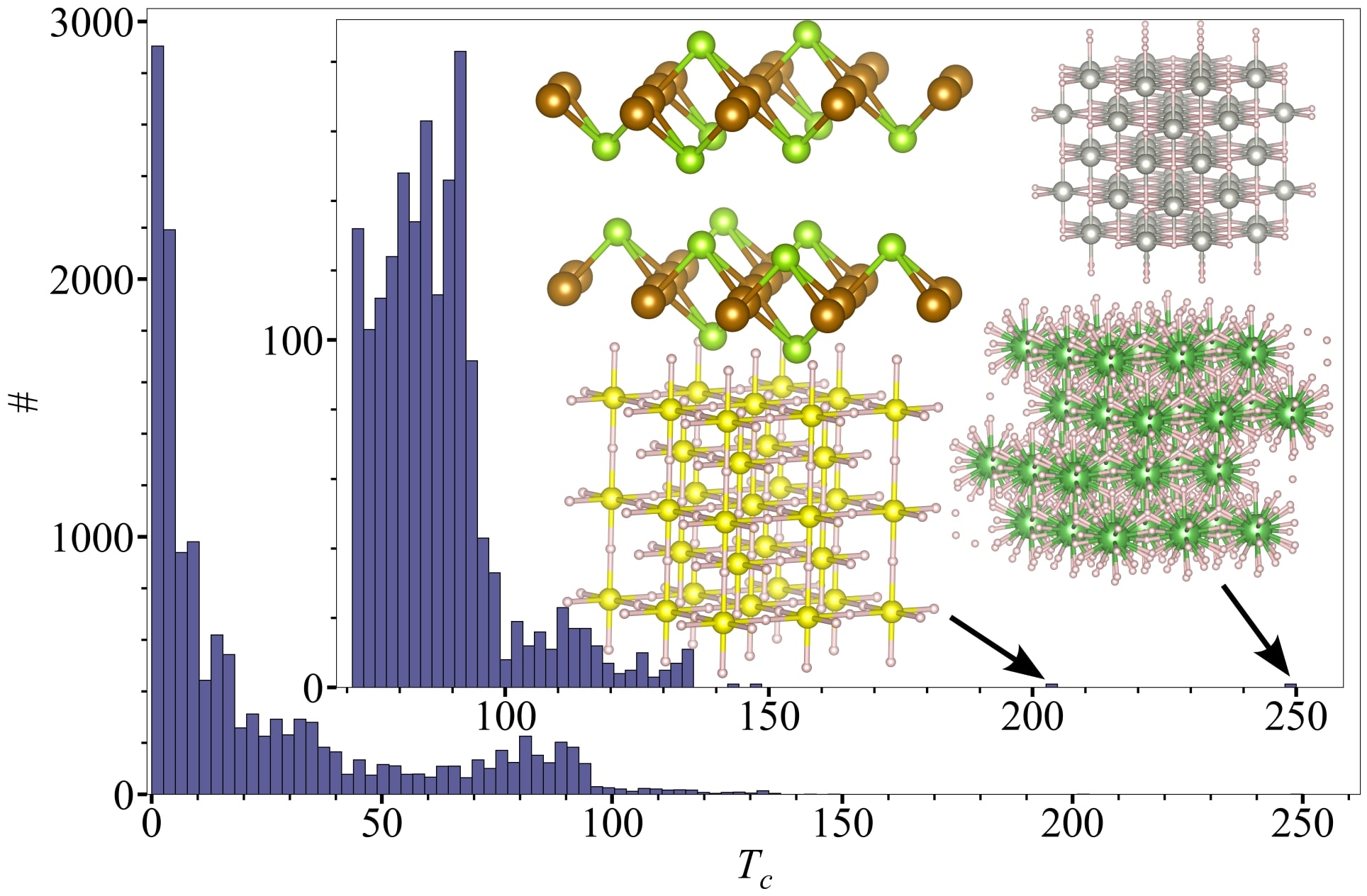}}
\caption{\label{fig: SuperCon distribution}
Distribution of 13,661 $T_c$ measurements in the curated SuperCon data set, including samples with $T_c$'s measured under ambient and applied pressures.
Inset shows the distribution of samples with $T_c$'s measured to be greater than 70 K and four exemplary crystal structures.  
}
\end{figure}

In this work, the SuperCon data set\cite{SuperCon} was obtained from the Materials Data Repository, which is maintained by the Japanese National Institute for Materials Science (NIMS).
During the time which this work was conducted, SuperCon listed 26,321 materials, each with their associated chemical compositions, experimentally-measured $T_c$'s (Kelvin), and article references.
It does not include atomic coordinates required to construct unit cells and does not mention whether the $T_c$ measurements were performed while external pressures were applied to the samples. 
We cleaned the data set by assigning to stoichiometries with multiple $T_c$ measurements their mean values.
Some contentious high-$T_c$ measurements \cite{bad_article_1, bad_article_2, bad_article_3} were also removed.
As it was deemed to be prohibitively time-consuming to verify via human-effort the repute and accuracy of each measurement in SuperCon, it was assumed that the issue of the inclusion of measurements from untrustworthy experiments could be largely neglected in the low-$T_c$ regime where the overwhelming majority of samples lie. 
Further, samples consisting of one or ten different element types were removed, as were those specified by arbitrary doping concentrations (e.g. \ce{Hg2Ba2YCu2O_{8-x}}).\\

The final, cleaned data set includes 13,661 unique stoichiometries with $T_c$'s ranging 0--250 K, with a median value of 10 K (\autoref{fig: SuperCon distribution}). 
Using the definition of high-temperature superconductors as those with $T_c$'s greater than the boiling point of liquid nitrogen, 77 K\cite{nitrogen_boiling_point}, there exists only 1,372 such samples in the data set, constituting 10 $\%$ of its total size -- clearly, the distribution is highly right-skewed.
To illustrate, the four samples in the data set with the highest $T_c$'s are \ce{LaH10}, \ce{H2S}, \ce{H3S}, \ce{Hg_{0.66}Pb_{0.34}Ba2Ca_{1.98}Cu_{2.9}O_{8.4}}, with corresponding $T_c$'s of 250\cite{lanthanum_hydride_superconductor}, 203\cite{203K_superconductor}, 147\cite{147K_superconductor}, and 143\cite{143K_superconductor} K, respectively. 
It may be of interest to note that experimental studies suggest the superconducting phase of \ce{H2S} which occurs under high pressures (gigapascals) actually exists as \ce{H3S} due to decomposition\cite{203K_superconductor}.
While this mechanism may make redundant the inclusion in the data set of both \ce{H2S} and \ce{H3S}, they were retained for chemical diversity in this low-data regime.\\

The SuperCon$^2$ data set\cite{SuperCon2} -- which was compiled by automatically scraping the literature to extract data related to superconductors -- was then used to determine the samples in SuperCon with $T_c$ measurements performed under applied pressure. 
This analysis suggests that thirty-seven samples were measured under applied pressures.
These were removed from SuperCon to create a separate data set containing only samples with $T_c$'s measured under ambient conditions (0--135 K).

\subsection{\label{sec: Similarity-Based Machine Learning}Similarity-Based Machine Learning}

To estimate the $T_c$ of a given test sample via similarity-based ML\cite{Dominik_similarity}, the training data set is first queried to find its $n$-nearest neighbors. 
The distance metric we employ is the Euclidean distance, calculated as
\begin{equation}
    d(\textbf{x}_q, \textbf{x}_t) = \lVert \textbf{x}_q - \textbf{x}_t \rVert
\end{equation}
where $\textbf{x}_q$ is the feature vector of the query test sample and $\textbf{x}_t$ is the feature vector of $t$'th training set sample.\\

The $n$ training samples with the smallest norm values are then chosen to train a ridge regression model with the optimal $\alpha$ hyperparameter\cite{ridge_regression_notes} selected as the one yielding the best performance on the training set (smallest mean squared error), evaluated by $n$-fold cross-validation see supplementary information on the justification of our selection of the learning algorithm).
A prediction for a given test sample is made as 
\begin{equation}
    T_c^{\rm est} = \textbf{x}_q \textbf{w}_q + c_q
\end{equation}
where $c_q$ is the intercept and $\textbf{w}_q$ is the vector of weight coefficients obtained from the training set as the closed-form solution of
\begin{equation}
    \textbf{w}_q = \left( \textbf{X}\textbf{X}^\top + \alpha \textbf{I} \right)^{-1} \textbf{X}^\top\textbf{y}
\end{equation}
for $\textbf{X}$ the $n$-sample feature matrix and $\textbf{y}$ the corresponding $n$-dimensional labels vector ($T_c$).
We perform all our computations with the \texttt{scikit-learn}\cite{sklearn, sklearn_ridge, sklearn_ridgecv} library in \texttt{Python} and perform the matrix inversion using the Cholesky decomposition\cite{ridge_regression_notes}. 
The absolute values of predictions are used for the final estimates to avoid non-physical negative $T_c$ values.\\

\subsection{\label{sec: Machine Learning Representations}Machine Learning Representations}

Since only chemical compositions are provided in SuperCon, they form the basis of the representations used to train our ML models.
Despite their apparent simplicity, chemical compositions have been shown to be sufficient in accurately learning various properties of materials\cite{formation_energies_from_composition, crystal_structure_type_from_composition, ElemNet, ROOST, CrabNet}.
They are also less restrictive in suggesting interesting stoichiometries to experimentalists, as the predictions derived from them are not specific to particular crystal structures contained in their convex hulls\cite{materials_ML_representations}.\\

147 features were generated for each sample from its composition using the materials informatics \texttt{Python} library Matminer\cite{Matminer}. 
The minimum, maximum, range, mean, average deviation, and mode were calculated from the following atomic properties, with weights given by stoichiometric coefficients:
atomic number; Mendeleev number; atomic weight; melting temperature of elemental solids; periodic table group number; periodic table period number; covalent radius; electronegativity; number of filled \textit{s, p, d, f} orbitals; number of valence electrons; number of unfilled \textit{s, p, d, f} orbitals; number of unfilled valence orbitals; volume of elemental solid; band gap of elemental solid; magnetic moment of elemental solids; and space group number of elemental solids.
Matminer was used to additionally calculate transition metal fractions; stoichiometric 0-, 2-, 3-, 5-, 7-, and 10-norms; average number of valence electrons in each of \textit{s, p, d, f} orbitals; and the fractions of valence electrons in each of \textit{s, p, d, f} orbitals.\\

\subsection{\label{sec: Materials Project}Materials Project}

We apply our similarity-based ML method to $\sim$153k samples listed in the Materials Project database to screen for potential novel superconductors with high-$T_c$'s. 
For each material represented as described in \autoref{sec: Machine Learning Representations}, predictions are made at implicit/ambient pressure by selecting $n$ training samples from the cleaned implicit/ambient pressure SuperCon data set and training a ridge regression model, as described in \autoref{sec: Similarity-Based Machine Learning}.

\section{\label{sec: Results and Discussion}Results and Discussion}

\subsection{\label{sec: SuperCon Analysis}SuperCon based validation of our approach}

\begin{figure}[h]
\resizebox{\columnwidth}{!}{\includegraphics{SuperCon_learning_curves.jpg}}
\caption{\label{fig: Tc learning curves}
Implicit pressure (left) and ambient pressure (right) model learning curves displaying the mean prediction errors (mean absolute errors) on five different sets of unseen out-of-sample test materials in SuperCon. 
Each test sample's prediction was made by either training ridge regression on $n$ random training samples, or on its $n$-nearest neighbors, as indicated in the legend with ``Random" and ``Similarity", respectively. 
Two different test sets were evaluated, in which the ``full $T_c$" test set was composed of randomly-selected test samples with $T_c$'s in the range of 0--250 K and the ``high-$T_c$" test set was composed of samples with $T_c$'s greater than 110 K. 
The results of the ridge regression models are compared to those obtained from baseline $k$-nearest neighbors regression models. 
}
\end{figure}

The predictive performances of ridge regression models in mapping chemical compositions to their superconducting $T_c$'s in SuperCon are first assessed with learning curves\cite{learning_curves}.
Due to their characteristic shape, learning curves are useful for evaluating not only the data-efficiency of ML models but also the learnability of the problem. 
In \autoref{fig: Tc learning curves}, the learning curves show the prediction error (mean absolute error, MAE) on the test set after training on $n$ training samples drawn either randomly or by a similarity measure (labelled in the legend as ``Random" and ``Similarity", respectively). 
The shadings indicate the standard deviations in the predictions on five different test sets which were held constant for all $n$.\\

From the ``full-$T_c$" curves in which the test samples' $T_c$'s corresponded to the range of 0--250 K, it is clear that similarity-based ML can achieve low average prediction errors of $\sim$5 K even with training on as low as 10 samples.
This error is less than the $\sim$9 K average $T_c$ prediction errors achieved with boosted trees and neural networks as reported by others\cite{ML_method_background_6, rediscovery_1}, while using an altogether simpler learning algorithm that is not incapable of extrapolation.
The advantage of the similarity measure is rather profound as the ``Random" curves lie significantly above the ``Similarity" curves, with the two sets of curves only converging to $\sim$10 K error in the full data set limit.\\

We also considered a second set of test sets composed of $T_c$'s greater than 110 K to better assess the models' capabilities of making accurate predictions in the high-$T_c$ regime.
Again, the advantage of similarity-based ML is clear as it achieves errors of $\sim$15 K in this low-data regime with training only on 10 nearest neighbors, whereas random training samples result in errors of $\sim$80 K that converge only to $\sim$50 K in the full data set limit.
In the context of the large range of experimentally-measured $T_c$ values spanned by different materials in SuperCon, we assume in the following that the result of the learning curves suggest that our method is sufficiently accurate and transferable for obtaining robust rankings among high-$T_c$ candidate materials.\\

\begin{figure*}
\centering
\subfloat[\label{fig: SuperCon predictions}]{\includegraphics[width=0.78\textwidth]{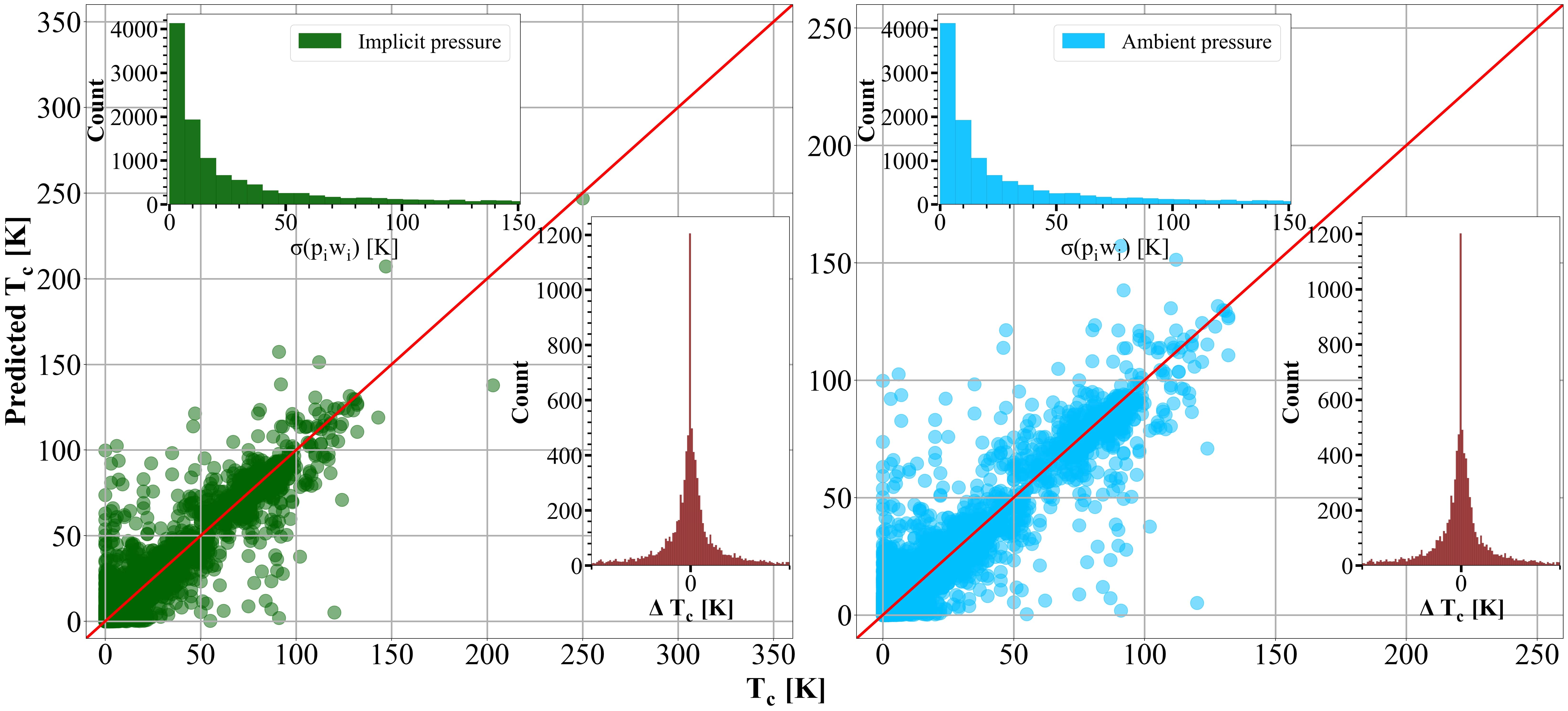}}\\
\vspace{0.3in}
\centering 
\subfloat[\label{fig: SuperCon number of elements errors}]{\includegraphics[width=0.26\textwidth]{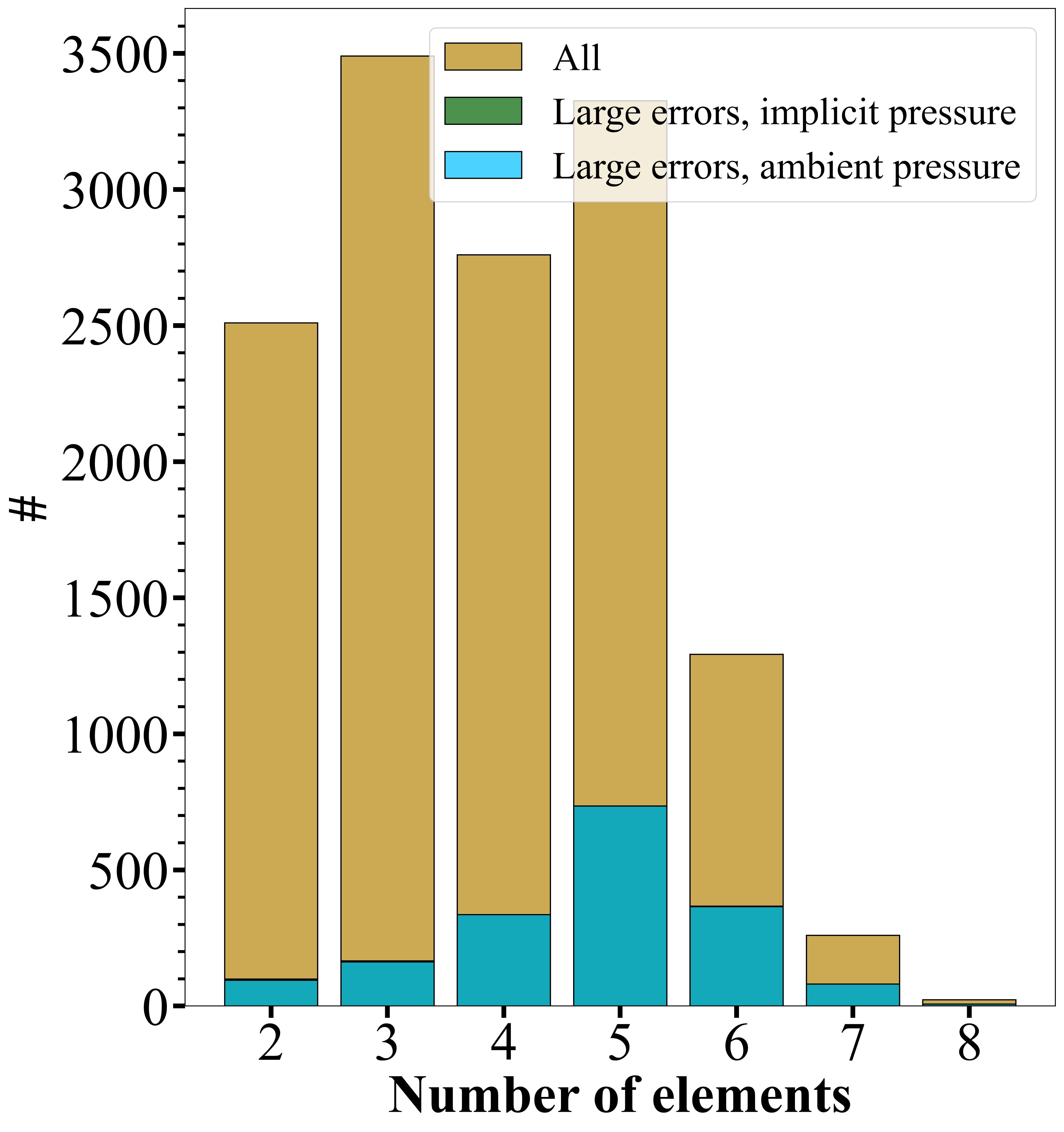}}
\hspace{0.3cm} 
\centering 
\subfloat[\label{fig: SuperCon atomic numbers errors}]{\includegraphics[width=0.70\textwidth]{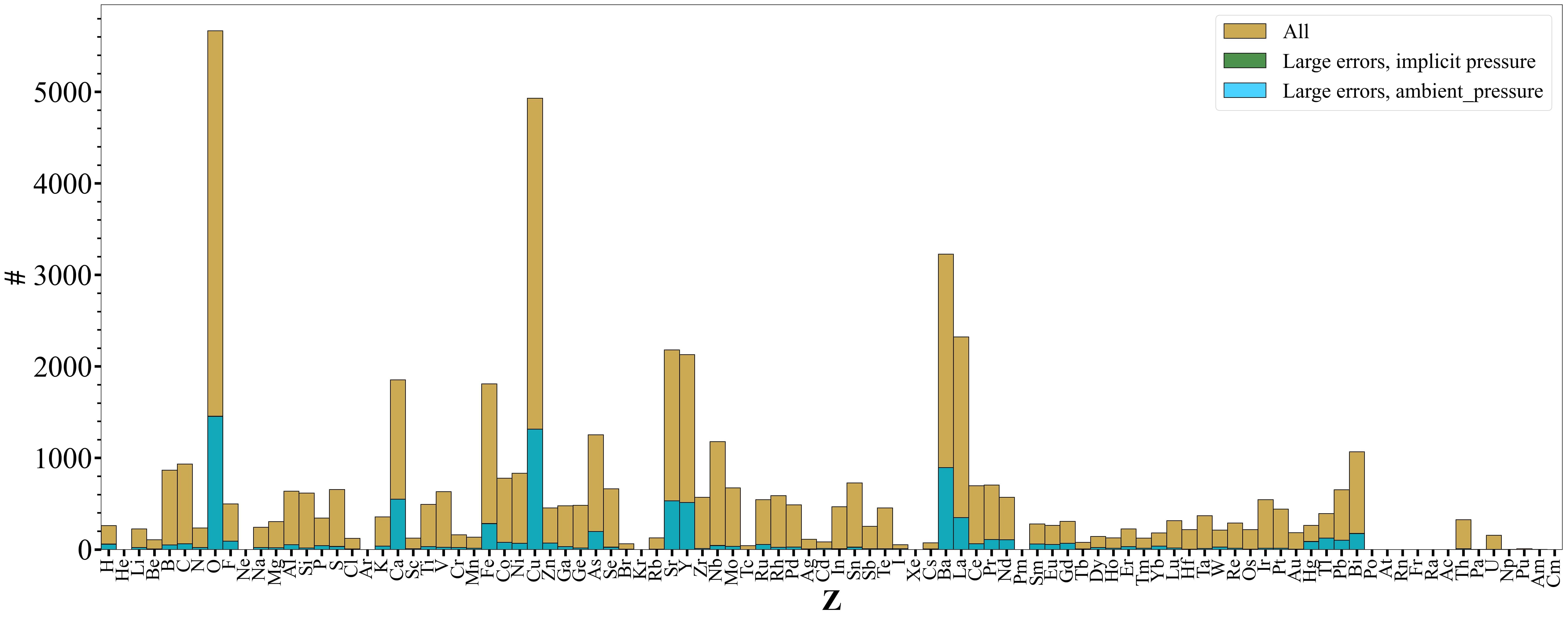}}
\caption{\label{fig: toy problem solutions}
\textbf{(a)} Leave-one-out $T_c$ predictions under implicit/ambient pressure for 9,247/9,211 samples in the implicit/ambient pressure SuperCon data set with spreads in their index-wise feature-weight products of less than 50 K, achieving MAE $\sim$3/3 K and $R^2$ $\sim$0.89/0.89.
Insets display these samples' distributions of index-wise feature-weight products and signed-errors.
\textbf{(b)} Distributions of the number of different elements in SuperCon. 
\textbf{(c)} and element types in the compositions of the SuperCon samples with absolute prediction errors greater than 10 K (without filtering for spreads in feature-weight products), compared against all SuperCon samples.
}
\end{figure*}

With the similarity-based ML approach, leave-one-out $T_c$ predictions under unknown/ambient pressure are made for each of the 13,661/13,624 materials in the implicit/ambient pressure data set.
For both pressure cases, each individual model was trained on the test sample's 10 nearest neighbors in the respective data sets as $n = 10$ was found from \autoref{fig: Tc learning curves} to be sufficient in reaching an MAE of $\sim$5 K while avoiding the increased computational cost of the matrix inversion involved in ridge regression that would be incurred from training on a larger number of samples. 
The resulting execution of this method (in \texttt{Python}) is extremely efficient as the total wall time required to search for $n$-nearest neighbors, tune for the $\alpha$ hyperparameter, train on the $n$ training samples, and predict for the test sample is several milliseconds on a laptop (12th gen. Intel i7-1260P, 12 cores, 2.10 GHz clock rate).\\ 

Upon examination of the distribution of the index-wise products in the dot product between a query sample's feature vector, $\textbf{x}_q$, and the vector of weight coefficients, $\textbf{w}_q$
\begin{equation}
    \textbf{x}_q \textbf{w}_q = \sum_{i = 1}^{147}x_{q, i} w_{q, i}
\end{equation}
it emerges that some samples have large standard deviations in their product values (\autoref{fig: SuperCon predictions}). 
This suggests that the $\alpha$ parameter, obtained from the training set, is not generalizable to the test sample as it fails to penalize the large coefficients that produce large $\lvert x_{q,i} w_{q,i} \rvert$ values. 
The removal of such implicit/ambient pressure samples with large spreads, defined as those less than 50 K, improves predictive performances in terms of leave-one-out predictions on SuperCon (\autoref{fig: SuperCon predictions}), as the MAE can be reduced to $\sim$3K. 
Although there are few samples with large prediction errors, similarity-based ML appears to achieve relatively good performances across the full range of $T_c$ values.
For instance, in the implicit pressure data set, \ce{LaH10} with $T_c$ of 250 K is accurately predicted, which corroborates the robustness of the models. 
The signed-errors distribution is symmetrical but not normally distributed. 
The former suggests that the models' predictions do not carry a systematic error. 
The non-normal distribution with large tails is due to the unphysical statistical nature of the outliers (particularly prominent in the scarce data regime the similarity-based ML model is operating in), and has previously already been observed and discussed~\cite{pernot2020impact}.\\

An examination of the samples with absolute prediction errors greater than 10 K reveals they mostly comprise of compositions consisting of five different elements and that the error-rate proportion increases with greater number of differing elements in the system (\autoref{fig: SuperCon number of elements errors}). 
This may not be unexpected since the different number of environments an element can experience in its crystal system increases with increasing number of differing elements. 
However, in SuperCon, the number of samples as a function of differing elements does not appear to increase sufficiently for the different environments to be adequately accounted for in our ML models.
We also note that large errors are associated with materials consisting of oxygen, copper, or barium (\autoref{fig: SuperCon atomic numbers errors}).
This may be the effect of a combination of a bias incurred from intense research by the superconductor community into yttrium barium copper oxide and related compounds, and that oxygen as a strong oxidant complicates the chemistry involved in these systems.\\

The weights obtained from the training of ridge regression models were also inspected to better understand which features made the most significant contributions to the predictions of $T_c$ labels. 
On average, for the samples in both the implicit and ambient pressure data sets, the five most significant features, in decreasing order of importance as calculated by the magnitude of the absolute value of the product of the weight with its corresponding feature value $\lvert x_{q,i} w_{q,i} \rvert$, are the mean melting temperature, average deviation of melting temperature, mean atomic weight, mean covalent radius, and the mean space group number. 
Clearly, the $T_c$ predictions are greatly influenced by the crystal structure and the mass of the elemental species of a given material's stoichiometry. 
These results are logical considering the sensitive relations between crystal structures and their bulk properties, and the observations of the isotope effect in both conventional and unconventional superconductors\cite{isotope_effect_1, isotope_effect_2, unconventional_isotope_effects_1, unconventional_isotope_effects_2}, in which their $T_c$'s are inversely proportional to the square-root of the atomic masses of the isotopes in their compositions as lighter ones produce higher phonon frequencies.

\subsection{\label{sec: Comparison to Other Models}Comparison to Other Models}

The predictive power of our models are further assessed by predicting $T_c$'s for some of the candidates with the highest $T_c$'s predicted from different works in the literature that were proposed using ML either for direct regression or in a screen prior to electronic structures-based calculations.
Such candidates include \ce{RbH12} (133 K)\cite{hutcheon_2020}, \ce{AlBaCaF7} (46 K)\cite{roter_2020}, and \ce{MoN} (33 K)\cite{ML_method_background_8}, for which our models predict 64 K, 25 K, and 15 K, respectively. 
Although our predictions likely do not perfectly correspond to those reported by the other models due to the different learning algorithms employed, both sets of predicted $T_c$'s interestingly produce the same relative ranking of the materials.
This is an additional encouraging benchmark of our models' generalization capabilities that we believe will allow for the ranking of candidate materials with sufficient accuracy in the high-$T_c$ regime.

\subsection{\label{sec: Materials Projecct Analysis}Application to the Materials Project}

\begin{figure}[h]
\resizebox{\columnwidth}{!}{\includegraphics{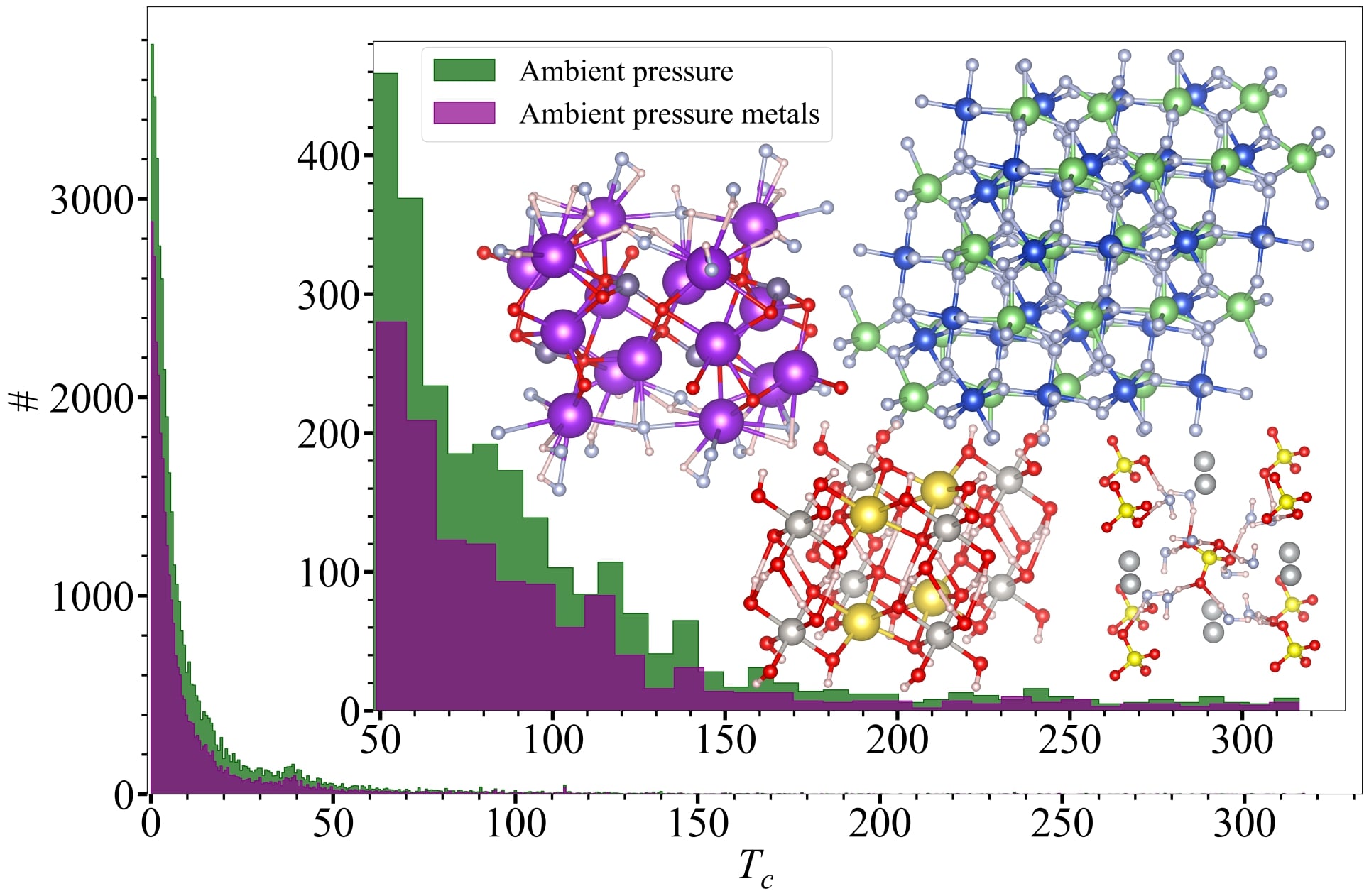}}
\caption{\label{fig: Materials Project distribution}
Distribution of ambient pressure $T_c$ predictions for Materials Project samples with energies above convex hulls of less than 0.030 eV/atom (41,559 samples), with subsets filtered for band gaps of less than 1.000 eV (30,725 samples). 
Inset shows the distributions of samples with predicted $T_c$'s greater than 50 K and representative crystal structures.
}
\end{figure}

Our similarity-based ML approach has been used to estimate $T_c$'s of $\sim$153k materials listed in the Materials Project. 
The Materials Project was queried because it provides DFT-computed estimates of various materials properties, such as stability and band gap. 
Due to the computational efficiency of our approach, the entire scan consumed only $\sim$2 node hours. 
We note that due to the generality of our approach,
{\em any} other materials library (e.g.~with experimental stability and band gap values) could have been used just as well.\\

Each sample's predictions under implicit pressure ($n = 10$) and ambient pressure ($n = 10$) were made by training on its nearest neighbors found in the implicit and ambient pressure SuperCon data sets, respectively. 
Samples with implicit and ambient pressure predictions associated with standard deviations larger than 50 K, respectively, are disregarded.
Those with computed energies above their convex hulls of greater than 0.030 eV/atom are also disregarded as being thermodynamically unstable.\\

\begin{figure*}
\centering
\subfloat[\label{fig: mp_765341}]{\includegraphics[width=0.36\textwidth]{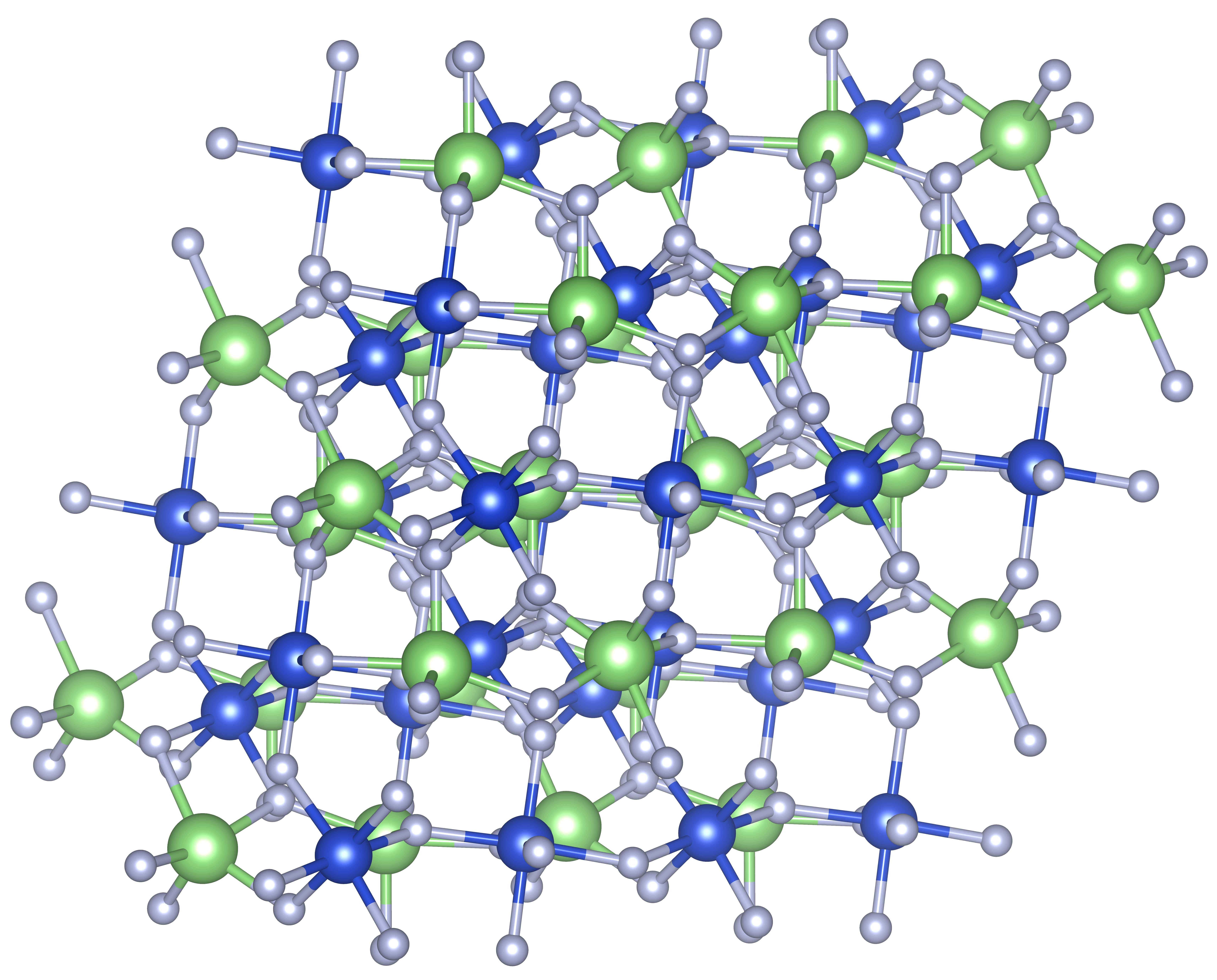}}
\hfill
\centering
\subfloat[\label{fig: mp_765341_band_structure_and_dos}]{\includegraphics[width=0.63\textwidth]{mp_765341_band_structure_and_dos.jpg}}
\hfill
\caption{
Crystal structure and band structure of material in Table~\ref{table: ambient pressure top ten} with highest predicted $T_c$. 
{\bf (a)} Crystal structure of \ce{LiCuF4}, and \textbf{(b)} its electronic band structures and density of states. 
The Fermi level is set to 0 eV, indicated by the dashed line.
All data downloaded from Materials Project~\cite{MP_765341_1, MP_765341_2}. 
}
\end{figure*}

\begin{table}[h]
\begin{center}
\caption{\label{table: ambient pressure top ten}
Compositions of materials in the Materials Project with the highest predicted $T_c$'s under ambient pressure, that lie $<0.030$ eV/atom above their convex hulls.
}
\begin{ruledtabular}
\begin{tabular}{p{3cm}ll}
\textbf{ID} & \textbf{Composition} & \textbf{$T_c^{\rm est}$ (K)}\\
\hline 
mp-765341, mp-753273, mp-753257 & \ce{LiCuF4} & 316\\ 
mp-723002 & \ce{Ag2H12S(NO)4} & 316\\ 
mp-632760 & \ce{Na2H6PtO6} & 315\\ 
mp-720197 & \ce{K4GeH4N2O3} & 315\\
mp-8090, mp-1225821, mp-3638 & \ce{Cu2P2O7} & 311\\ 
mp-567820 & \ce{K6NbTlAs4} & 308\\ 
mp-1204687 & \ce{Cu3P2H2O9} & 307\\ 
mp-26304 & \ce{Cu9(PO4)8} & 306\\ 
mp-1190512 & \ce{Na2UTeO7} & 302\\ 
mp-721293 & \ce{Na4GeH28(Se2O7)2} & 302\\ 
\end{tabular}
\end{ruledtabular}
\end{center}
\end{table}

The resulting distributions in $T_c$ predictions is shown in \autoref{fig: Materials Project distribution}.
We note that since SuperCon contains experimentally-measured $T_c$ values from laboratory-synthesized materials of varying degrees of purity, our $T_c$ predictions for the samples in the Materials Project likely contains this inherent noise - \textit{i.e.}\ our predictions would not correspond exactly to those derived from pure compounds as assumed by the Materials Project in their DFT calculations.
Our analysis suggests that at ambient pressures, sixty-four materials may have $T_c$'s greater than 250 K.
Of these materials, forty-four may transition to their superconducting phase at temperatures greater than 273 K, with the highest $T_c$ predicted to be 316 K for \ce{LiCuF4}\cite{MP_765341_1, MP_765341_2, MP_765341_3} (\autoref{fig: mp_765341}).
When filtering for metal-like materials with small band gaps (less than 1.000 eV), thirty-four samples are predicted to have $T_c$'s greater than 250 K, of which twenty-three are greater than 273 K.\\

It is believed that one of the identifying characteristics of high-$T_c$ superconductors are strongly correlated bands that allow for unorthodox Cooper pair formations\cite{flat_bands_1, flat_bands_2}. 
For instance, from \autoref{fig: mp_765341_band_structure_and_dos}, it appears that \ce{LiCuF4} (monoclinic P$2_1$/c space group) has a pair of bands near its Fermi level that are relatively flat and narrow with a maximum bandwidth of $\sim$600 meV and even extends beyond the Fermi level.
By the assumptions about band flatness correlating with superconductivity along with the Materials Project reporting its band gap to be 0.000 eV, this result may suggest that this material can transition to its superconducting phase at higher temperatures. 
However, band structure calculations for any given material often involve severe approximations and produce results that can deviate significantly (usually as underestimates) from experimental results\cite{band_structure_problem_1, band_structure_problem_2}. 
Therefore, it would be of great interest to verify the superconducting properties and electronic band structures of \ce{LiCuF4} via experimental synthesis and measurements. 
Moreover, it would be interesting to conduct experiments of the thermodynamic stability of \ce{LiCuF4} because the Materials Project states that two of its three computed crystal structure for this stoichiometry are unstable as they lie $\sim$0.027 and 0.020 ev/atom above the convex hull\cite{MP_765341_2, MP_765341_3}. 
After all, it would be of very limited practical usefulness if its superconducting phase occurs only under extremely high pressures.
It appears that there currently has been no efforts to conduct such experiments, as this material is not reported in SuperCon and, to the best of our knowledge, it has not been studied in the literature.\\

\section{\label{sec: Conclusion}Conclusion}

We have introduced a data-efficient similarity-based ML approach to estimating the superconducting critical temperatures of materials at both implicit and ambient pressures.
Predictions for novel materials require training a query aware ridge regression model `on the fly' using only the $n$ nearest neighbours in the training data, with $n$ as low as 10. 
This is feasible thanks to the extremely low computational cost required by our model (several milliseconds/material on a twelve-core CPU).
When this simple method was evaluated via leave-one-out predictions on the full SuperCon data set, it was found to be relatively robust in making accurate predictions across the full range of $T_c$ values, 0--250 K. 
Moreover, it was observed that one may be able to identify predictions with large uncertainties as samples with large standard deviations in their index-wise products contained in their feature-weights vectors dot products. 
The analysis of the weight coefficients revealed that material properties related to the crystal structures and atomic weights and radii of the elements in the stoichiometries were the most significant contributors to the predictions of $T_c$'s, suggesting that the ML models were able to at least capture the basic physical principles involved in superconductivity.\\

We have used the model to rank the entire Materials Project data base of $\sim$153k materials by the estimated $T_c$.
Several materials were identified as potentially being able to exist as superconductors near room-temperature, with \ce{LiCuF4} possibly having a metallic character with flat bands near and above its Fermi level, as computed with DFT.
As superconductivity experiments appear to have not yet been performed for \ce{LiCuF4}, nor for the other high-$T_c$ materials we have identified, they might be valuable targets for future research.\\

Our results also exemplify the usefulness of  similarity-based ML in accelerating the virtual design and discovery of novel materials and molecules with interesting properties.
We note this may be facilitated by the specific choice of a representation which is more or less agnostic (depending on use case). 
For example, as done in this study, features only derived from chemical composition may be advantageous in allowing for the screening of a much greater number of materials as it would be trivial  to create new stoichiometries by combinatorics of element types and stoichiometric coefficients. 
Further important criteria, such as measures of synthesizability or stability, could be added as constraints later on.\\

\section{\label{sec: Data and Code Availability}Data and Code Availability}

The supplementary information contains a discussion, using toy problems, on why we chose to develop our ML models with ridge regression, rather than other learning algorithms.
It displays a table listing the SuperCon samples that we have identified as having $T_c$'s measured under applied pressure, and a table ranking the relative importance of each of the feature descriptors used in our ML models. 
It also displays two tables ranking the one-hundred highest $T_c$ materials in the Materials Project identified by our implicit/ambient pressure models, as well as two additional tables obtained after subsequent filtering for band gaps lower than 1 eV.\\

Refer to \url{https://zenodo.org/records/14052692} for: \texttt{Python} code to generate ML features and to implement our similarity-based ML models; chemical compositions, $T_c$'s, pressures, ML features, and implicit/ambient pressure $T_c$ predictions for our SuperCon samples; and chemical compositions, identifiers, energies above convex hulls, band gaps, ML features, and ambient pressure $T_c$ predictions for materials in the Materials Project.\\

\section{\label{sec: Acknowledgments}Acknowledgments}

We acknowledge the support of the Natural Sciences and Engineering Research Council of Canada (NSERC), [funding reference number RGPIN-2023-04853]. Cette recherche a été financée par le Conseil de recherches en sciences naturelles et en génie du Canada (CRSNG), [numéro de référence RGPIN-2023-04853].
This research was undertaken thanks in part to funding provided to the University of Toronto's Acceleration Consortium from the Canada First Research Excellence Fund,
grant number: CFREF-2022-00042.
O.A.v.L. has received support as the Ed Clark Chair of Advanced Materials and as a Canada CIFAR AI Chair.

\section{\label{sec: Author Contributions}Author Contributions}

S.L. and O.A.v.L. conceived the idea. 
S.L. developed and implemented the methodology and performed all experiments, with guidance from O.A.v.L. 
All authors discussed the results, and contributed suggestions and edits to the manuscript.

\bibliography{References}

\begin{thebibliography}{88}%
\makeatletter
\providecommand \@ifxundefined [1]{%
 \@ifx{#1\undefined}
}%
\providecommand \@ifnum [1]{%
 \ifnum #1\expandafter \@firstoftwo
 \else \expandafter \@secondoftwo
 \fi
}%
\providecommand \@ifx [1]{%
 \ifx #1\expandafter \@firstoftwo
 \else \expandafter \@secondoftwo
 \fi
}%
\providecommand \natexlab [1]{#1}%
\providecommand \enquote  [1]{``#1''}%
\providecommand \bibnamefont  [1]{#1}%
\providecommand \bibfnamefont [1]{#1}%
\providecommand \citenamefont [1]{#1}%
\providecommand \href@noop [0]{\@secondoftwo}%
\providecommand \href [0]{\begingroup \@sanitize@url \@href}%
\providecommand \@href[1]{\@@startlink{#1}\@@href}%
\providecommand \@@href[1]{\endgroup#1\@@endlink}%
\providecommand \@sanitize@url [0]{\catcode `\\12\catcode `\$12\catcode `\&12\catcode `\#12\catcode `\^12\catcode `\_12\catcode `\%12\relax}%
\providecommand \@@startlink[1]{}%
\providecommand \@@endlink[0]{}%
\providecommand \url  [0]{\begingroup\@sanitize@url \@url }%
\providecommand \@url [1]{\endgroup\@href {#1}{\urlprefix }}%
\providecommand \urlprefix  [0]{URL }%
\providecommand \Eprint [0]{\href }%
\providecommand \doibase [0]{http://dx.doi.org/}%
\providecommand \selectlanguage [0]{\@gobble}%
\providecommand \bibinfo  [0]{\@secondoftwo}%
\providecommand \bibfield  [0]{\@secondoftwo}%
\providecommand \translation [1]{[#1]}%
\providecommand \BibitemOpen [0]{}%
\providecommand \bibitemStop [0]{}%
\providecommand \bibitemNoStop [0]{.\EOS\space}%
\providecommand \EOS [0]{\spacefactor3000\relax}%
\providecommand \BibitemShut  [1]{\csname bibitem#1\endcsname}%
\let\auto@bib@innerbib\@empty
\bibitem [{\citenamefont {Jain}\ \emph {et~al.}(2013)\citenamefont {Jain}, \citenamefont {Ong}, \citenamefont {Hautier}, \citenamefont {Chen}, \citenamefont {Richards}, \citenamefont {Dacek}, \citenamefont {Cholia}, \citenamefont {Gunter}, \citenamefont {Skinner}, \citenamefont {Ceder} \emph {et~al.}}]{Materials_Project_1}%
  \BibitemOpen
  \bibfield  {author} {\bibinfo {author} {\bibfnamefont {A.}~\bibnamefont {Jain}}, \bibinfo {author} {\bibfnamefont {S.~P.}\ \bibnamefont {Ong}}, \bibinfo {author} {\bibfnamefont {G.}~\bibnamefont {Hautier}}, \bibinfo {author} {\bibfnamefont {W.}~\bibnamefont {Chen}}, \bibinfo {author} {\bibfnamefont {W.~D.}\ \bibnamefont {Richards}}, \bibinfo {author} {\bibfnamefont {S.}~\bibnamefont {Dacek}}, \bibinfo {author} {\bibfnamefont {S.}~\bibnamefont {Cholia}}, \bibinfo {author} {\bibfnamefont {D.}~\bibnamefont {Gunter}}, \bibinfo {author} {\bibfnamefont {D.}~\bibnamefont {Skinner}}, \bibinfo {author} {\bibfnamefont {G.}~\bibnamefont {Ceder}},  \emph {et~al.},\ }\href@noop {} {\bibfield  {journal} {\bibinfo  {journal} {APL materials}\ }\textbf {\bibinfo {volume} {1}} (\bibinfo {year} {2013})}\BibitemShut {NoStop}%
\bibitem [{Sup(2024)}]{SuperCon}%
  \BibitemOpen
  \href@noop {} {\enquote {\bibinfo {title} {{M}{D}{R} {S}uper{C}on {D}atasheet // {M}{D}{R} --- mdr.nims.go.jp},}\ }\bibinfo {howpublished} {\url{https://mdr.nims.go.jp/collections/5712mb227}} (\bibinfo {year} {2024}),\ \bibinfo {note} {[Accessed 17-01-2024]}\BibitemShut {NoStop}%
\bibitem [{\citenamefont {Sleight}(1995)}]{holy_grail}%
  \BibitemOpen
  \bibfield  {author} {\bibinfo {author} {\bibfnamefont {A.~W.}\ \bibnamefont {Sleight}},\ }\href@noop {} {\bibfield  {journal} {\bibinfo  {journal} {Accounts of chemical research}\ }\textbf {\bibinfo {volume} {28}},\ \bibinfo {pages} {103} (\bibinfo {year} {1995})}\BibitemShut {NoStop}%
\bibitem [{\citenamefont {Tinkham}(1974)}]{superconductor_properties_1}%
  \BibitemOpen
  \bibfield  {author} {\bibinfo {author} {\bibfnamefont {M.}~\bibnamefont {Tinkham}},\ }\href@noop {} {\bibfield  {journal} {\bibinfo  {journal} {Reviews of Modern Physics}\ }\textbf {\bibinfo {volume} {46}},\ \bibinfo {pages} {587} (\bibinfo {year} {1974})}\BibitemShut {NoStop}%
\bibitem [{\citenamefont {Berger}\ and\ \citenamefont {Roberts}(1997)}]{superconductor_properties_2}%
  \BibitemOpen
  \bibfield  {author} {\bibinfo {author} {\bibfnamefont {L.}~\bibnamefont {Berger}}\ and\ \bibinfo {author} {\bibfnamefont {B.}~\bibnamefont {Roberts}},\ }\href@noop {} {\bibfield  {journal} {\bibinfo  {journal} {CRC Handbook of Chemistry and Physics}\ ,\ \bibinfo {pages} {12}} (\bibinfo {year} {1997})}\BibitemShut {NoStop}%
\bibitem [{\citenamefont {Mukherjee}\ and\ \citenamefont {Rao}(2019)}]{superconductor_energy_applications_1}%
  \BibitemOpen
  \bibfield  {author} {\bibinfo {author} {\bibfnamefont {P.}~\bibnamefont {Mukherjee}}\ and\ \bibinfo {author} {\bibfnamefont {V.}~\bibnamefont {Rao}},\ }\href@noop {} {\bibfield  {journal} {\bibinfo  {journal} {Physica C: Superconductivity and its applications}\ }\textbf {\bibinfo {volume} {563}},\ \bibinfo {pages} {67} (\bibinfo {year} {2019})}\BibitemShut {NoStop}%
\bibitem [{\citenamefont {Hull}(2003)}]{superconductor_energy_applications_2}%
  \BibitemOpen
  \bibfield  {author} {\bibinfo {author} {\bibfnamefont {J.~R.}\ \bibnamefont {Hull}},\ }\href@noop {} {\bibfield  {journal} {\bibinfo  {journal} {Reports on Progress in Physics}\ }\textbf {\bibinfo {volume} {66}},\ \bibinfo {pages} {1865} (\bibinfo {year} {2003})}\BibitemShut {NoStop}%
\bibitem [{\citenamefont {Scanlan}\ \emph {et~al.}(2004)\citenamefont {Scanlan}, \citenamefont {Malozemoff},\ and\ \citenamefont {Larbalestier}}]{superconductor_energy_applications_3}%
  \BibitemOpen
  \bibfield  {author} {\bibinfo {author} {\bibfnamefont {R.~M.}\ \bibnamefont {Scanlan}}, \bibinfo {author} {\bibfnamefont {A.~P.}\ \bibnamefont {Malozemoff}}, \ and\ \bibinfo {author} {\bibfnamefont {D.~C.}\ \bibnamefont {Larbalestier}},\ }\href@noop {} {\bibfield  {journal} {\bibinfo  {journal} {Proceedings of the IEEE}\ }\textbf {\bibinfo {volume} {92}},\ \bibinfo {pages} {1639} (\bibinfo {year} {2004})}\BibitemShut {NoStop}%
\bibitem [{\citenamefont {Hassenzahl}\ \emph {et~al.}(2004)\citenamefont {Hassenzahl}, \citenamefont {Hazelton}, \citenamefont {Johnson}, \citenamefont {Komarek}, \citenamefont {Noe},\ and\ \citenamefont {Reis}}]{superconductor_energy_applications_4}%
  \BibitemOpen
  \bibfield  {author} {\bibinfo {author} {\bibfnamefont {W.~V.}\ \bibnamefont {Hassenzahl}}, \bibinfo {author} {\bibfnamefont {D.~W.}\ \bibnamefont {Hazelton}}, \bibinfo {author} {\bibfnamefont {B.~K.}\ \bibnamefont {Johnson}}, \bibinfo {author} {\bibfnamefont {P.}~\bibnamefont {Komarek}}, \bibinfo {author} {\bibfnamefont {M.}~\bibnamefont {Noe}}, \ and\ \bibinfo {author} {\bibfnamefont {C.~T.}\ \bibnamefont {Reis}},\ }\href@noop {} {\bibfield  {journal} {\bibinfo  {journal} {Proceedings of the IEEE}\ }\textbf {\bibinfo {volume} {92}},\ \bibinfo {pages} {1655} (\bibinfo {year} {2004})}\BibitemShut {NoStop}%
\bibitem [{\citenamefont {Bruzzone}(2010)}]{superconductor_fusion_application_1}%
  \BibitemOpen
  \bibfield  {author} {\bibinfo {author} {\bibfnamefont {P.}~\bibnamefont {Bruzzone}},\ }\href@noop {} {\bibfield  {journal} {\bibinfo  {journal} {Physica C: Superconductivity and its applications}\ }\textbf {\bibinfo {volume} {470}},\ \bibinfo {pages} {1734} (\bibinfo {year} {2010})}\BibitemShut {NoStop}%
\bibitem [{\citenamefont {Bruzzone}(2014)}]{superconductor_fusion_application_2}%
  \BibitemOpen
  \bibfield  {author} {\bibinfo {author} {\bibfnamefont {P.}~\bibnamefont {Bruzzone}},\ }\href@noop {} {\bibfield  {journal} {\bibinfo  {journal} {Superconductor Science and Technology}\ }\textbf {\bibinfo {volume} {28}},\ \bibinfo {pages} {024001} (\bibinfo {year} {2014})}\BibitemShut {NoStop}%
\bibitem [{\citenamefont {Whyte}\ \emph {et~al.}(2016)\citenamefont {Whyte}, \citenamefont {Minervini}, \citenamefont {LaBombard}, \citenamefont {Marmar}, \citenamefont {Bromberg},\ and\ \citenamefont {Greenwald}}]{superconductor_fusion_application_3}%
  \BibitemOpen
  \bibfield  {author} {\bibinfo {author} {\bibfnamefont {D.}~\bibnamefont {Whyte}}, \bibinfo {author} {\bibfnamefont {J.}~\bibnamefont {Minervini}}, \bibinfo {author} {\bibfnamefont {B.}~\bibnamefont {LaBombard}}, \bibinfo {author} {\bibfnamefont {E.}~\bibnamefont {Marmar}}, \bibinfo {author} {\bibfnamefont {L.}~\bibnamefont {Bromberg}}, \ and\ \bibinfo {author} {\bibfnamefont {M.}~\bibnamefont {Greenwald}},\ }\href@noop {} {\bibfield  {journal} {\bibinfo  {journal} {Journal of Fusion Energy}\ }\textbf {\bibinfo {volume} {35}},\ \bibinfo {pages} {41} (\bibinfo {year} {2016})}\BibitemShut {NoStop}%
\bibitem [{\citenamefont {Malozemoff}(1988)}]{superconductor_computer_applications_1}%
  \BibitemOpen
  \bibfield  {author} {\bibinfo {author} {\bibfnamefont {A.}~\bibnamefont {Malozemoff}},\ }\href@noop {} {\bibfield  {journal} {\bibinfo  {journal} {Physica C: Superconductivity}\ }\textbf {\bibinfo {volume} {153}},\ \bibinfo {pages} {1049} (\bibinfo {year} {1988})}\BibitemShut {NoStop}%
\bibitem [{\citenamefont {Braginski}(2019)}]{superconductor_computer_applications_2}%
  \BibitemOpen
  \bibfield  {author} {\bibinfo {author} {\bibfnamefont {A.~I.}\ \bibnamefont {Braginski}},\ }\href@noop {} {\bibfield  {journal} {\bibinfo  {journal} {Journal of superconductivity and novel magnetism}\ }\textbf {\bibinfo {volume} {32}},\ \bibinfo {pages} {23} (\bibinfo {year} {2019})}\BibitemShut {NoStop}%
\bibitem [{\citenamefont {Berggren}(2004)}]{superconductor_computer_applications_3}%
  \BibitemOpen
  \bibfield  {author} {\bibinfo {author} {\bibfnamefont {K.~K.}\ \bibnamefont {Berggren}},\ }\href@noop {} {\bibfield  {journal} {\bibinfo  {journal} {Proceedings of the IEEE}\ }\textbf {\bibinfo {volume} {92}},\ \bibinfo {pages} {1630} (\bibinfo {year} {2004})}\BibitemShut {NoStop}%
\bibitem [{\citenamefont {Bravyi}\ \emph {et~al.}(2022)\citenamefont {Bravyi}, \citenamefont {Dial}, \citenamefont {Gambetta}, \citenamefont {Gil},\ and\ \citenamefont {Nazario}}]{superconductor_computer_applications_4}%
  \BibitemOpen
  \bibfield  {author} {\bibinfo {author} {\bibfnamefont {S.}~\bibnamefont {Bravyi}}, \bibinfo {author} {\bibfnamefont {O.}~\bibnamefont {Dial}}, \bibinfo {author} {\bibfnamefont {J.~M.}\ \bibnamefont {Gambetta}}, \bibinfo {author} {\bibfnamefont {D.}~\bibnamefont {Gil}}, \ and\ \bibinfo {author} {\bibfnamefont {Z.}~\bibnamefont {Nazario}},\ }\href@noop {} {\bibfield  {journal} {\bibinfo  {journal} {Journal of Applied Physics}\ }\textbf {\bibinfo {volume} {132}} (\bibinfo {year} {2022})}\BibitemShut {NoStop}%
\bibitem [{\citenamefont {Alonso}\ and\ \citenamefont {Antaya}(2012)}]{superconductor_medicine_applications_1}%
  \BibitemOpen
  \bibfield  {author} {\bibinfo {author} {\bibfnamefont {J.~R.}\ \bibnamefont {Alonso}}\ and\ \bibinfo {author} {\bibfnamefont {T.~A.}\ \bibnamefont {Antaya}},\ }\href@noop {} {\bibfield  {journal} {\bibinfo  {journal} {Reviews of Accelerator Science and Technology}\ }\textbf {\bibinfo {volume} {5}},\ \bibinfo {pages} {227} (\bibinfo {year} {2012})}\BibitemShut {NoStop}%
\bibitem [{\citenamefont {Ali}\ and\ \citenamefont {Zulqarnain}(2022)}]{superconductor_medicine_applications_2}%
  \BibitemOpen
  \bibfield  {author} {\bibinfo {author} {\bibfnamefont {S.}~\bibnamefont {Ali}}\ and\ \bibinfo {author} {\bibfnamefont {M.}~\bibnamefont {Zulqarnain}},\ }\href@noop {} {\bibfield  {journal} {\bibinfo  {journal} {Superconductors: Materials and Applications}\ }\textbf {\bibinfo {volume} {132}},\ \bibinfo {pages} {211} (\bibinfo {year} {2022})}\BibitemShut {NoStop}%
\bibitem [{\citenamefont {Werfel}\ \emph {et~al.}(2011)\citenamefont {Werfel}, \citenamefont {Floegel-Delor}, \citenamefont {Rothfeld}, \citenamefont {Riedel}, \citenamefont {Goebel}, \citenamefont {Wippich},\ and\ \citenamefont {Schirrmeister}}]{superconductor_transportation_applications_1}%
  \BibitemOpen
  \bibfield  {author} {\bibinfo {author} {\bibfnamefont {F.}~\bibnamefont {Werfel}}, \bibinfo {author} {\bibfnamefont {U.}~\bibnamefont {Floegel-Delor}}, \bibinfo {author} {\bibfnamefont {R.}~\bibnamefont {Rothfeld}}, \bibinfo {author} {\bibfnamefont {T.}~\bibnamefont {Riedel}}, \bibinfo {author} {\bibfnamefont {B.}~\bibnamefont {Goebel}}, \bibinfo {author} {\bibfnamefont {D.}~\bibnamefont {Wippich}}, \ and\ \bibinfo {author} {\bibfnamefont {P.}~\bibnamefont {Schirrmeister}},\ }\href@noop {} {\bibfield  {journal} {\bibinfo  {journal} {Superconductor Science and Technology}\ }\textbf {\bibinfo {volume} {25}},\ \bibinfo {pages} {014007} (\bibinfo {year} {2011})}\BibitemShut {NoStop}%
\bibitem [{\citenamefont {Werfel}\ \emph {et~al.}(2012)\citenamefont {Werfel}, \citenamefont {Floegel-Delor}, \citenamefont {Rothfeld}, \citenamefont {Riedel}, \citenamefont {Wippich}, \citenamefont {Goebel},\ and\ \citenamefont {Schirrmeister}}]{superconductor_transportation_applications_2}%
  \BibitemOpen
  \bibfield  {author} {\bibinfo {author} {\bibfnamefont {F.}~\bibnamefont {Werfel}}, \bibinfo {author} {\bibfnamefont {U.}~\bibnamefont {Floegel-Delor}}, \bibinfo {author} {\bibfnamefont {R.}~\bibnamefont {Rothfeld}}, \bibinfo {author} {\bibfnamefont {T.}~\bibnamefont {Riedel}}, \bibinfo {author} {\bibfnamefont {D.}~\bibnamefont {Wippich}}, \bibinfo {author} {\bibfnamefont {B.}~\bibnamefont {Goebel}}, \ and\ \bibinfo {author} {\bibfnamefont {P.}~\bibnamefont {Schirrmeister}},\ }\href@noop {} {\bibfield  {journal} {\bibinfo  {journal} {Physics Procedia}\ }\textbf {\bibinfo {volume} {36}},\ \bibinfo {pages} {948} (\bibinfo {year} {2012})}\BibitemShut {NoStop}%
\bibitem [{\citenamefont {Schultz}\ \emph {et~al.}(2005)\citenamefont {Schultz}, \citenamefont {de~Haas}, \citenamefont {Verges}, \citenamefont {Beyer}, \citenamefont {Rohlig}, \citenamefont {Olsen}, \citenamefont {Kuhn}, \citenamefont {Berger}, \citenamefont {Noteboom},\ and\ \citenamefont {Funk}}]{superconductor_transportation_applications_3}%
  \BibitemOpen
  \bibfield  {author} {\bibinfo {author} {\bibfnamefont {L.}~\bibnamefont {Schultz}}, \bibinfo {author} {\bibfnamefont {O.}~\bibnamefont {de~Haas}}, \bibinfo {author} {\bibfnamefont {P.}~\bibnamefont {Verges}}, \bibinfo {author} {\bibfnamefont {C.}~\bibnamefont {Beyer}}, \bibinfo {author} {\bibfnamefont {S.}~\bibnamefont {Rohlig}}, \bibinfo {author} {\bibfnamefont {H.}~\bibnamefont {Olsen}}, \bibinfo {author} {\bibfnamefont {L.}~\bibnamefont {Kuhn}}, \bibinfo {author} {\bibfnamefont {D.}~\bibnamefont {Berger}}, \bibinfo {author} {\bibfnamefont {U.}~\bibnamefont {Noteboom}}, \ and\ \bibinfo {author} {\bibfnamefont {U.}~\bibnamefont {Funk}},\ }\href@noop {} {\bibfield  {journal} {\bibinfo  {journal} {IEEE Transactions on Applied Superconductivity}\ }\textbf {\bibinfo {volume} {15}},\ \bibinfo {pages} {2301} (\bibinfo {year} {2005})}\BibitemShut {NoStop}%
\bibitem [{\citenamefont {Dam}\ \emph {et~al.}(2020)\citenamefont {Dam}, \citenamefont {Battiston}, \citenamefont {Burger}, \citenamefont {Carpentiero}, \citenamefont {Chesta}, \citenamefont {Iuppa}, \citenamefont {de~Rijk},\ and\ \citenamefont {Rossi}}]{superconductor_particle_physics_applications_1}%
  \BibitemOpen
  \bibfield  {author} {\bibinfo {author} {\bibfnamefont {M.}~\bibnamefont {Dam}}, \bibinfo {author} {\bibfnamefont {R.}~\bibnamefont {Battiston}}, \bibinfo {author} {\bibfnamefont {W.~J.}\ \bibnamefont {Burger}}, \bibinfo {author} {\bibfnamefont {R.}~\bibnamefont {Carpentiero}}, \bibinfo {author} {\bibfnamefont {E.}~\bibnamefont {Chesta}}, \bibinfo {author} {\bibfnamefont {R.}~\bibnamefont {Iuppa}}, \bibinfo {author} {\bibfnamefont {G.}~\bibnamefont {de~Rijk}}, \ and\ \bibinfo {author} {\bibfnamefont {L.}~\bibnamefont {Rossi}},\ }\href@noop {} {\bibfield  {journal} {\bibinfo  {journal} {Superconductor Science and Technology}\ }\textbf {\bibinfo {volume} {33}},\ \bibinfo {pages} {044012} (\bibinfo {year} {2020})}\BibitemShut {NoStop}%
\bibitem [{\citenamefont {Rossi}\ and\ \citenamefont {Bottura}(2012)}]{superconductor_particle_physics_applications_2}%
  \BibitemOpen
  \bibfield  {author} {\bibinfo {author} {\bibfnamefont {L.}~\bibnamefont {Rossi}}\ and\ \bibinfo {author} {\bibfnamefont {L.}~\bibnamefont {Bottura}},\ }\href@noop {} {\bibfield  {journal} {\bibinfo  {journal} {Reviews of accelerator science and technology}\ }\textbf {\bibinfo {volume} {5}},\ \bibinfo {pages} {51} (\bibinfo {year} {2012})}\BibitemShut {NoStop}%
\bibitem [{\citenamefont {Tollestrup}\ and\ \citenamefont {Todesco}(2008)}]{superconductor_particle_physics_applications_3}%
  \BibitemOpen
  \bibfield  {author} {\bibinfo {author} {\bibfnamefont {A.}~\bibnamefont {Tollestrup}}\ and\ \bibinfo {author} {\bibfnamefont {E.}~\bibnamefont {Todesco}},\ }\href@noop {} {\bibfield  {journal} {\bibinfo  {journal} {Reviews of Accelerator Science and Technology}\ }\textbf {\bibinfo {volume} {1}},\ \bibinfo {pages} {185} (\bibinfo {year} {2008})}\BibitemShut {NoStop}%
\bibitem [{\citenamefont {Bottura}\ \emph {et~al.}(2015)\citenamefont {Bottura}, \citenamefont {Gourlay}, \citenamefont {Yamamoto},\ and\ \citenamefont {Zlobin}}]{superconductor_particle_physics_applications_4}%
  \BibitemOpen
  \bibfield  {author} {\bibinfo {author} {\bibfnamefont {L.}~\bibnamefont {Bottura}}, \bibinfo {author} {\bibfnamefont {S.~A.}\ \bibnamefont {Gourlay}}, \bibinfo {author} {\bibfnamefont {A.}~\bibnamefont {Yamamoto}}, \ and\ \bibinfo {author} {\bibfnamefont {A.~V.}\ \bibnamefont {Zlobin}},\ }\href@noop {} {\bibfield  {journal} {\bibinfo  {journal} {IEEE Transactions on Nuclear Science}\ }\textbf {\bibinfo {volume} {63}},\ \bibinfo {pages} {751} (\bibinfo {year} {2015})}\BibitemShut {NoStop}%
\bibitem [{\citenamefont {Bardeen}\ \emph {et~al.}(1957{\natexlab{a}})\citenamefont {Bardeen}, \citenamefont {Cooper},\ and\ \citenamefont {Schrieffer}}]{BCS_theory_1}%
  \BibitemOpen
  \bibfield  {author} {\bibinfo {author} {\bibfnamefont {J.}~\bibnamefont {Bardeen}}, \bibinfo {author} {\bibfnamefont {L.~N.}\ \bibnamefont {Cooper}}, \ and\ \bibinfo {author} {\bibfnamefont {J.~R.}\ \bibnamefont {Schrieffer}},\ }\href@noop {} {\bibfield  {journal} {\bibinfo  {journal} {Physical review}\ }\textbf {\bibinfo {volume} {108}},\ \bibinfo {pages} {1175} (\bibinfo {year} {1957}{\natexlab{a}})}\BibitemShut {NoStop}%
\bibitem [{\citenamefont {Bardeen}\ \emph {et~al.}(1957{\natexlab{b}})\citenamefont {Bardeen}, \citenamefont {Cooper},\ and\ \citenamefont {Schrieffer}}]{BCS_theory_2}%
  \BibitemOpen
  \bibfield  {author} {\bibinfo {author} {\bibfnamefont {J.}~\bibnamefont {Bardeen}}, \bibinfo {author} {\bibfnamefont {L.~N.}\ \bibnamefont {Cooper}}, \ and\ \bibinfo {author} {\bibfnamefont {J.~R.}\ \bibnamefont {Schrieffer}},\ }\href@noop {} {\bibfield  {journal} {\bibinfo  {journal} {Physical Review}\ }\textbf {\bibinfo {volume} {106}},\ \bibinfo {pages} {162} (\bibinfo {year} {1957}{\natexlab{b}})}\BibitemShut {NoStop}%
\bibitem [{\citenamefont {Mann}(2011)}]{superconductivity_theory_unknown}%
  \BibitemOpen
  \bibfield  {author} {\bibinfo {author} {\bibfnamefont {A.}~\bibnamefont {Mann}},\ }\href@noop {} {\bibfield  {journal} {\bibinfo  {journal} {Nature}\ }\textbf {\bibinfo {volume} {475}},\ \bibinfo {pages} {280} (\bibinfo {year} {2011})}\BibitemShut {NoStop}%
\bibitem [{\citenamefont {Dew-Hughes}(2001)}]{there_is_hope_2}%
  \BibitemOpen
  \bibfield  {author} {\bibinfo {author} {\bibfnamefont {D.}~\bibnamefont {Dew-Hughes}},\ }\href@noop {} {\bibfield  {journal} {\bibinfo  {journal} {Low temperature physics}\ }\textbf {\bibinfo {volume} {27}},\ \bibinfo {pages} {713} (\bibinfo {year} {2001})}\BibitemShut {NoStop}%
\bibitem [{\citenamefont {Walsh}(2015)}]{size_of_materials_space}%
  \BibitemOpen
  \bibfield  {author} {\bibinfo {author} {\bibfnamefont {A.}~\bibnamefont {Walsh}},\ }\href@noop {} {\bibfield  {journal} {\bibinfo  {journal} {Nature chemistry}\ }\textbf {\bibinfo {volume} {7}},\ \bibinfo {pages} {274} (\bibinfo {year} {2015})}\BibitemShut {NoStop}%
\bibitem [{\citenamefont {Drozdov}\ \emph {et~al.}(2019)\citenamefont {Drozdov}, \citenamefont {Kong}, \citenamefont {Minkov}, \citenamefont {Besedin}, \citenamefont {Kuzovnikov}, \citenamefont {Mozaffari}, \citenamefont {Balicas}, \citenamefont {Balakirev}, \citenamefont {Graf}, \citenamefont {Prakapenka} \emph {et~al.}}]{lanthanum_hydride_superconductor}%
  \BibitemOpen
  \bibfield  {author} {\bibinfo {author} {\bibfnamefont {A.}~\bibnamefont {Drozdov}}, \bibinfo {author} {\bibfnamefont {P.}~\bibnamefont {Kong}}, \bibinfo {author} {\bibfnamefont {V.}~\bibnamefont {Minkov}}, \bibinfo {author} {\bibfnamefont {S.}~\bibnamefont {Besedin}}, \bibinfo {author} {\bibfnamefont {M.}~\bibnamefont {Kuzovnikov}}, \bibinfo {author} {\bibfnamefont {S.}~\bibnamefont {Mozaffari}}, \bibinfo {author} {\bibfnamefont {L.}~\bibnamefont {Balicas}}, \bibinfo {author} {\bibfnamefont {F.}~\bibnamefont {Balakirev}}, \bibinfo {author} {\bibfnamefont {D.}~\bibnamefont {Graf}}, \bibinfo {author} {\bibfnamefont {V.}~\bibnamefont {Prakapenka}},  \emph {et~al.},\ }\href@noop {} {\bibfield  {journal} {\bibinfo  {journal} {Nature}\ }\textbf {\bibinfo {volume} {569}},\ \bibinfo {pages} {528} (\bibinfo {year} {2019})}\BibitemShut {NoStop}%
\bibitem [{\citenamefont {Stanev}\ \emph {et~al.}(2018)\citenamefont {Stanev}, \citenamefont {Oses}, \citenamefont {Kusne}, \citenamefont {Rodriguez}, \citenamefont {Paglione}, \citenamefont {Curtarolo},\ and\ \citenamefont {Takeuchi}}]{ML_method_background_1}%
  \BibitemOpen
  \bibfield  {author} {\bibinfo {author} {\bibfnamefont {V.}~\bibnamefont {Stanev}}, \bibinfo {author} {\bibfnamefont {C.}~\bibnamefont {Oses}}, \bibinfo {author} {\bibfnamefont {A.~G.}\ \bibnamefont {Kusne}}, \bibinfo {author} {\bibfnamefont {E.}~\bibnamefont {Rodriguez}}, \bibinfo {author} {\bibfnamefont {J.}~\bibnamefont {Paglione}}, \bibinfo {author} {\bibfnamefont {S.}~\bibnamefont {Curtarolo}}, \ and\ \bibinfo {author} {\bibfnamefont {I.}~\bibnamefont {Takeuchi}},\ }\href@noop {} {\bibfield  {journal} {\bibinfo  {journal} {npj Computational Materials}\ }\textbf {\bibinfo {volume} {4}},\ \bibinfo {pages} {29} (\bibinfo {year} {2018})}\BibitemShut {NoStop}%
\bibitem [{\citenamefont {Sommer}\ \emph {et~al.}(2023)\citenamefont {Sommer}, \citenamefont {Willa}, \citenamefont {Schmalian},\ and\ \citenamefont {Friederich}}]{ML_method_background_2}%
  \BibitemOpen
  \bibfield  {author} {\bibinfo {author} {\bibfnamefont {T.}~\bibnamefont {Sommer}}, \bibinfo {author} {\bibfnamefont {R.}~\bibnamefont {Willa}}, \bibinfo {author} {\bibfnamefont {J.}~\bibnamefont {Schmalian}}, \ and\ \bibinfo {author} {\bibfnamefont {P.}~\bibnamefont {Friederich}},\ }\href@noop {} {\bibfield  {journal} {\bibinfo  {journal} {Scientific Data}\ }\textbf {\bibinfo {volume} {10}},\ \bibinfo {pages} {816} (\bibinfo {year} {2023})}\BibitemShut {NoStop}%
\bibitem [{\citenamefont {Tran}\ and\ \citenamefont {Vu}(2023)}]{ML_method_background_3}%
  \BibitemOpen
  \bibfield  {author} {\bibinfo {author} {\bibfnamefont {H.}~\bibnamefont {Tran}}\ and\ \bibinfo {author} {\bibfnamefont {T.~N.}\ \bibnamefont {Vu}},\ }\href@noop {} {\bibfield  {journal} {\bibinfo  {journal} {Physical Review Materials}\ }\textbf {\bibinfo {volume} {7}},\ \bibinfo {pages} {054805} (\bibinfo {year} {2023})}\BibitemShut {NoStop}%
\bibitem [{\citenamefont {Novakovic}\ \emph {et~al.}(2023)\citenamefont {Novakovic}, \citenamefont {Salamat},\ and\ \citenamefont {Lawler}}]{ML_method_background_4}%
  \BibitemOpen
  \bibfield  {author} {\bibinfo {author} {\bibfnamefont {L.}~\bibnamefont {Novakovic}}, \bibinfo {author} {\bibfnamefont {A.}~\bibnamefont {Salamat}}, \ and\ \bibinfo {author} {\bibfnamefont {K.~V.}\ \bibnamefont {Lawler}},\ }\href@noop {} {\bibfield  {journal} {\bibinfo  {journal} {arXiv preprint arXiv:2301.10474}\ } (\bibinfo {year} {2023})}\BibitemShut {NoStop}%
\bibitem [{\citenamefont {Konno}\ \emph {et~al.}(2021)\citenamefont {Konno}, \citenamefont {Kurokawa}, \citenamefont {Nabeshima}, \citenamefont {Sakishita}, \citenamefont {Ogawa}, \citenamefont {Hosako},\ and\ \citenamefont {Maeda}}]{ML_method_background_5}%
  \BibitemOpen
  \bibfield  {author} {\bibinfo {author} {\bibfnamefont {T.}~\bibnamefont {Konno}}, \bibinfo {author} {\bibfnamefont {H.}~\bibnamefont {Kurokawa}}, \bibinfo {author} {\bibfnamefont {F.}~\bibnamefont {Nabeshima}}, \bibinfo {author} {\bibfnamefont {Y.}~\bibnamefont {Sakishita}}, \bibinfo {author} {\bibfnamefont {R.}~\bibnamefont {Ogawa}}, \bibinfo {author} {\bibfnamefont {I.}~\bibnamefont {Hosako}}, \ and\ \bibinfo {author} {\bibfnamefont {A.}~\bibnamefont {Maeda}},\ }\href@noop {} {\bibfield  {journal} {\bibinfo  {journal} {Physical Review B}\ }\textbf {\bibinfo {volume} {103}},\ \bibinfo {pages} {014509} (\bibinfo {year} {2021})}\BibitemShut {NoStop}%
\bibitem [{\citenamefont {Moscato}\ \emph {et~al.}(2023)\citenamefont {Moscato}, \citenamefont {Haque}, \citenamefont {Huang}, \citenamefont {Sloan},\ and\ \citenamefont {Corrales~de Oliveira}}]{ML_method_background_6}%
  \BibitemOpen
  \bibfield  {author} {\bibinfo {author} {\bibfnamefont {P.}~\bibnamefont {Moscato}}, \bibinfo {author} {\bibfnamefont {M.~N.}\ \bibnamefont {Haque}}, \bibinfo {author} {\bibfnamefont {K.}~\bibnamefont {Huang}}, \bibinfo {author} {\bibfnamefont {J.}~\bibnamefont {Sloan}}, \ and\ \bibinfo {author} {\bibfnamefont {J.}~\bibnamefont {Corrales~de Oliveira}},\ }\href@noop {} {\bibfield  {journal} {\bibinfo  {journal} {Algorithms}\ }\textbf {\bibinfo {volume} {16}},\ \bibinfo {pages} {382} (\bibinfo {year} {2023})}\BibitemShut {NoStop}%
\bibitem [{\citenamefont {Meredig}\ \emph {et~al.}(2018)\citenamefont {Meredig}, \citenamefont {Antono}, \citenamefont {Church}, \citenamefont {Hutchinson}, \citenamefont {Ling}, \citenamefont {Paradiso}, \citenamefont {Blaiszik}, \citenamefont {Foster}, \citenamefont {Gibbons}, \citenamefont {Hattrick-Simpers} \emph {et~al.}}]{ML_method_background_7}%
  \BibitemOpen
  \bibfield  {author} {\bibinfo {author} {\bibfnamefont {B.}~\bibnamefont {Meredig}}, \bibinfo {author} {\bibfnamefont {E.}~\bibnamefont {Antono}}, \bibinfo {author} {\bibfnamefont {C.}~\bibnamefont {Church}}, \bibinfo {author} {\bibfnamefont {M.}~\bibnamefont {Hutchinson}}, \bibinfo {author} {\bibfnamefont {J.}~\bibnamefont {Ling}}, \bibinfo {author} {\bibfnamefont {S.}~\bibnamefont {Paradiso}}, \bibinfo {author} {\bibfnamefont {B.}~\bibnamefont {Blaiszik}}, \bibinfo {author} {\bibfnamefont {I.}~\bibnamefont {Foster}}, \bibinfo {author} {\bibfnamefont {B.}~\bibnamefont {Gibbons}}, \bibinfo {author} {\bibfnamefont {J.}~\bibnamefont {Hattrick-Simpers}},  \emph {et~al.},\ }\href@noop {} {\bibfield  {journal} {\bibinfo  {journal} {Molecular Systems Design \& Engineering}\ }\textbf {\bibinfo {volume} {3}},\ \bibinfo {pages} {819} (\bibinfo {year} {2018})}\BibitemShut {NoStop}%
\bibitem [{\citenamefont {Choudhary}\ and\ \citenamefont {Garrity}(2022)}]{ML_method_background_8}%
  \BibitemOpen
  \bibfield  {author} {\bibinfo {author} {\bibfnamefont {K.}~\bibnamefont {Choudhary}}\ and\ \bibinfo {author} {\bibfnamefont {K.}~\bibnamefont {Garrity}},\ }\href@noop {} {\bibfield  {journal} {\bibinfo  {journal} {npj Computational Materials}\ }\textbf {\bibinfo {volume} {8}},\ \bibinfo {pages} {244} (\bibinfo {year} {2022})}\BibitemShut {NoStop}%
\bibitem [{\citenamefont {Pereti}\ \emph {et~al.}(2023)\citenamefont {Pereti}, \citenamefont {Bernot}, \citenamefont {Guizouarn}, \citenamefont {Laufek}, \citenamefont {Vymazalov{\'a}}, \citenamefont {Bindi}, \citenamefont {Sessoli},\ and\ \citenamefont {Fanelli}}]{rediscovery_1}%
  \BibitemOpen
  \bibfield  {author} {\bibinfo {author} {\bibfnamefont {C.}~\bibnamefont {Pereti}}, \bibinfo {author} {\bibfnamefont {K.}~\bibnamefont {Bernot}}, \bibinfo {author} {\bibfnamefont {T.}~\bibnamefont {Guizouarn}}, \bibinfo {author} {\bibfnamefont {F.}~\bibnamefont {Laufek}}, \bibinfo {author} {\bibfnamefont {A.}~\bibnamefont {Vymazalov{\'a}}}, \bibinfo {author} {\bibfnamefont {L.}~\bibnamefont {Bindi}}, \bibinfo {author} {\bibfnamefont {R.}~\bibnamefont {Sessoli}}, \ and\ \bibinfo {author} {\bibfnamefont {D.}~\bibnamefont {Fanelli}},\ }\href@noop {} {\bibfield  {journal} {\bibinfo  {journal} {Npj Computational Materials}\ }\textbf {\bibinfo {volume} {9}},\ \bibinfo {pages} {71} (\bibinfo {year} {2023})}\BibitemShut {NoStop}%
\bibitem [{\citenamefont {Gashmard}\ \emph {et~al.}(2024)\citenamefont {Gashmard}, \citenamefont {Shakeripour},\ and\ \citenamefont {Alaei}}]{ML_method_background_9}%
  \BibitemOpen
  \bibfield  {author} {\bibinfo {author} {\bibfnamefont {H.}~\bibnamefont {Gashmard}}, \bibinfo {author} {\bibfnamefont {H.}~\bibnamefont {Shakeripour}}, \ and\ \bibinfo {author} {\bibfnamefont {M.}~\bibnamefont {Alaei}},\ }\href@noop {} {\bibfield  {journal} {\bibinfo  {journal} {Scientific Reports}\ }\textbf {\bibinfo {volume} {14}},\ \bibinfo {pages} {3965} (\bibinfo {year} {2024})}\BibitemShut {NoStop}%
\bibitem [{\citenamefont {Koh}\ \emph {et~al.}(2021)\citenamefont {Koh}, \citenamefont {Sagawa}, \citenamefont {Marklund}, \citenamefont {Xie}, \citenamefont {Zhang}, \citenamefont {Balsubramani}, \citenamefont {Hu}, \citenamefont {Yasunaga}, \citenamefont {Phillips}, \citenamefont {Gao} \emph {et~al.}}]{OOD_1}%
  \BibitemOpen
  \bibfield  {author} {\bibinfo {author} {\bibfnamefont {P.~W.}\ \bibnamefont {Koh}}, \bibinfo {author} {\bibfnamefont {S.}~\bibnamefont {Sagawa}}, \bibinfo {author} {\bibfnamefont {H.}~\bibnamefont {Marklund}}, \bibinfo {author} {\bibfnamefont {S.~M.}\ \bibnamefont {Xie}}, \bibinfo {author} {\bibfnamefont {M.}~\bibnamefont {Zhang}}, \bibinfo {author} {\bibfnamefont {A.}~\bibnamefont {Balsubramani}}, \bibinfo {author} {\bibfnamefont {W.}~\bibnamefont {Hu}}, \bibinfo {author} {\bibfnamefont {M.}~\bibnamefont {Yasunaga}}, \bibinfo {author} {\bibfnamefont {R.~L.}\ \bibnamefont {Phillips}}, \bibinfo {author} {\bibfnamefont {I.}~\bibnamefont {Gao}},  \emph {et~al.},\ }in\ \href@noop {} {\emph {\bibinfo {booktitle} {International Conference on Machine Learning}}}\ (\bibinfo {organization} {PMLR},\ \bibinfo {year} {2021})\ pp.\ \bibinfo {pages} {5637--5664}\BibitemShut {NoStop}%
\bibitem [{\citenamefont {Wald}\ \emph {et~al.}(2021)\citenamefont {Wald}, \citenamefont {Feder}, \citenamefont {Greenfeld},\ and\ \citenamefont {Shalit}}]{OOD_2}%
  \BibitemOpen
  \bibfield  {author} {\bibinfo {author} {\bibfnamefont {Y.}~\bibnamefont {Wald}}, \bibinfo {author} {\bibfnamefont {A.}~\bibnamefont {Feder}}, \bibinfo {author} {\bibfnamefont {D.}~\bibnamefont {Greenfeld}}, \ and\ \bibinfo {author} {\bibfnamefont {U.}~\bibnamefont {Shalit}},\ }\href@noop {} {\bibfield  {journal} {\bibinfo  {journal} {Advances in neural information processing systems}\ }\textbf {\bibinfo {volume} {34}},\ \bibinfo {pages} {2215} (\bibinfo {year} {2021})}\BibitemShut {NoStop}%
\bibitem [{\citenamefont {Gulrajani}\ and\ \citenamefont {Lopez-Paz}(2020)}]{OOD_ML_problem_1}%
  \BibitemOpen
  \bibfield  {author} {\bibinfo {author} {\bibfnamefont {I.}~\bibnamefont {Gulrajani}}\ and\ \bibinfo {author} {\bibfnamefont {D.}~\bibnamefont {Lopez-Paz}},\ }\href@noop {} {\bibfield  {journal} {\bibinfo  {journal} {arXiv preprint arXiv:2007.01434}\ } (\bibinfo {year} {2020})}\BibitemShut {NoStop}%
\bibitem [{\citenamefont {Akrout}\ \emph {et~al.}(2023)\citenamefont {Akrout}, \citenamefont {Feriani}, \citenamefont {Bellili}, \citenamefont {Mezghani},\ and\ \citenamefont {Hossain}}]{OOD_ML_problem_2}%
  \BibitemOpen
  \bibfield  {author} {\bibinfo {author} {\bibfnamefont {M.}~\bibnamefont {Akrout}}, \bibinfo {author} {\bibfnamefont {A.}~\bibnamefont {Feriani}}, \bibinfo {author} {\bibfnamefont {F.}~\bibnamefont {Bellili}}, \bibinfo {author} {\bibfnamefont {A.}~\bibnamefont {Mezghani}}, \ and\ \bibinfo {author} {\bibfnamefont {E.}~\bibnamefont {Hossain}},\ }\href@noop {} {\bibfield  {journal} {\bibinfo  {journal} {IEEE Communications Surveys \& Tutorials}\ } (\bibinfo {year} {2023})}\BibitemShut {NoStop}%
\bibitem [{\citenamefont {Foppiano}\ \emph {et~al.}(2023)\citenamefont {Foppiano}, \citenamefont {Castro}, \citenamefont {Ortiz~Suarez}, \citenamefont {Terashima}, \citenamefont {Takano},\ and\ \citenamefont {Ishii}}]{SuperCon2}%
  \BibitemOpen
  \bibfield  {author} {\bibinfo {author} {\bibfnamefont {L.}~\bibnamefont {Foppiano}}, \bibinfo {author} {\bibfnamefont {P.~B.}\ \bibnamefont {Castro}}, \bibinfo {author} {\bibfnamefont {P.}~\bibnamefont {Ortiz~Suarez}}, \bibinfo {author} {\bibfnamefont {K.}~\bibnamefont {Terashima}}, \bibinfo {author} {\bibfnamefont {Y.}~\bibnamefont {Takano}}, \ and\ \bibinfo {author} {\bibfnamefont {M.}~\bibnamefont {Ishii}},\ }\href@noop {} {\bibfield  {journal} {\bibinfo  {journal} {Science and Technology of Advanced Materials: Methods}\ }\textbf {\bibinfo {volume} {3}},\ \bibinfo {pages} {2153633} (\bibinfo {year} {2023})}\BibitemShut {NoStop}%
\bibitem [{\citenamefont {Djurek}\ \emph {et~al.}(2001)\citenamefont {Djurek}, \citenamefont {Meduni{\'c}}, \citenamefont {Tonejc},\ and\ \citenamefont {Paljevi{\'c}}}]{bad_article_1}%
  \BibitemOpen
  \bibfield  {author} {\bibinfo {author} {\bibfnamefont {D.}~\bibnamefont {Djurek}}, \bibinfo {author} {\bibfnamefont {Z.}~\bibnamefont {Meduni{\'c}}}, \bibinfo {author} {\bibfnamefont {A.}~\bibnamefont {Tonejc}}, \ and\ \bibinfo {author} {\bibfnamefont {M.}~\bibnamefont {Paljevi{\'c}}},\ }\href@noop {} {\bibfield  {journal} {\bibinfo  {journal} {Physica C: Superconductivity}\ }\textbf {\bibinfo {volume} {351}},\ \bibinfo {pages} {78} (\bibinfo {year} {2001})}\BibitemShut {NoStop}%
\bibitem [{\citenamefont {Snider}\ \emph {et~al.}(2020)\citenamefont {Snider}, \citenamefont {Dasenbrock-Gammon}, \citenamefont {McBride}, \citenamefont {Debessai}, \citenamefont {Vindana}, \citenamefont {Vencatasamy}, \citenamefont {Lawler}, \citenamefont {Salamat},\ and\ \citenamefont {Dias}}]{bad_article_2}%
  \BibitemOpen
  \bibfield  {author} {\bibinfo {author} {\bibfnamefont {E.}~\bibnamefont {Snider}}, \bibinfo {author} {\bibfnamefont {N.}~\bibnamefont {Dasenbrock-Gammon}}, \bibinfo {author} {\bibfnamefont {R.}~\bibnamefont {McBride}}, \bibinfo {author} {\bibfnamefont {M.}~\bibnamefont {Debessai}}, \bibinfo {author} {\bibfnamefont {H.}~\bibnamefont {Vindana}}, \bibinfo {author} {\bibfnamefont {K.}~\bibnamefont {Vencatasamy}}, \bibinfo {author} {\bibfnamefont {K.~V.}\ \bibnamefont {Lawler}}, \bibinfo {author} {\bibfnamefont {A.}~\bibnamefont {Salamat}}, \ and\ \bibinfo {author} {\bibfnamefont {R.~P.}\ \bibnamefont {Dias}},\ }\href@noop {} {\bibfield  {journal} {\bibinfo  {journal} {Nature}\ }\textbf {\bibinfo {volume} {588}},\ \bibinfo {pages} {E18} (\bibinfo {year} {2020})}\BibitemShut {NoStop}%
\bibitem [{\citenamefont {Ayyub}\ \emph {et~al.}(1987)\citenamefont {Ayyub}, \citenamefont {Guptasarma}, \citenamefont {Rajarajan}, \citenamefont {Gupta}, \citenamefont {Vijayaraghavan},\ and\ \citenamefont {Multani}}]{bad_article_3}%
  \BibitemOpen
  \bibfield  {author} {\bibinfo {author} {\bibfnamefont {P.}~\bibnamefont {Ayyub}}, \bibinfo {author} {\bibfnamefont {P.}~\bibnamefont {Guptasarma}}, \bibinfo {author} {\bibfnamefont {A.}~\bibnamefont {Rajarajan}}, \bibinfo {author} {\bibfnamefont {L.}~\bibnamefont {Gupta}}, \bibinfo {author} {\bibfnamefont {R.}~\bibnamefont {Vijayaraghavan}}, \ and\ \bibinfo {author} {\bibfnamefont {M.}~\bibnamefont {Multani}},\ }\href@noop {} {\bibfield  {journal} {\bibinfo  {journal} {Journal of Physics C: Solid State Physics}\ }\textbf {\bibinfo {volume} {20}},\ \bibinfo {pages} {L673} (\bibinfo {year} {1987})}\BibitemShut {NoStop}%
\bibitem [{\citenamefont {Henshaw}\ \emph {et~al.}(1953)\citenamefont {Henshaw}, \citenamefont {Hurst},\ and\ \citenamefont {Pope}}]{nitrogen_boiling_point}%
  \BibitemOpen
  \bibfield  {author} {\bibinfo {author} {\bibfnamefont {D.}~\bibnamefont {Henshaw}}, \bibinfo {author} {\bibfnamefont {D.}~\bibnamefont {Hurst}}, \ and\ \bibinfo {author} {\bibfnamefont {N.}~\bibnamefont {Pope}},\ }\href@noop {} {\bibfield  {journal} {\bibinfo  {journal} {Physical Review}\ }\textbf {\bibinfo {volume} {92}},\ \bibinfo {pages} {1229} (\bibinfo {year} {1953})}\BibitemShut {NoStop}%
\bibitem [{\citenamefont {Drozdov}\ \emph {et~al.}(2015)\citenamefont {Drozdov}, \citenamefont {Eremets}, \citenamefont {Troyan}, \citenamefont {Ksenofontov},\ and\ \citenamefont {Shylin}}]{203K_superconductor}%
  \BibitemOpen
  \bibfield  {author} {\bibinfo {author} {\bibfnamefont {A.}~\bibnamefont {Drozdov}}, \bibinfo {author} {\bibfnamefont {M.}~\bibnamefont {Eremets}}, \bibinfo {author} {\bibfnamefont {I.}~\bibnamefont {Troyan}}, \bibinfo {author} {\bibfnamefont {V.}~\bibnamefont {Ksenofontov}}, \ and\ \bibinfo {author} {\bibfnamefont {S.~I.}\ \bibnamefont {Shylin}},\ }\href@noop {} {\bibfield  {journal} {\bibinfo  {journal} {Nature}\ }\textbf {\bibinfo {volume} {525}},\ \bibinfo {pages} {73} (\bibinfo {year} {2015})}\BibitemShut {NoStop}%
\bibitem [{\citenamefont {Shimizu}\ \emph {et~al.}(2018)\citenamefont {Shimizu}, \citenamefont {Einaga}, \citenamefont {Sakata}, \citenamefont {Masuda}, \citenamefont {Nakao}, \citenamefont {Eremets}, \citenamefont {Drozdov}, \citenamefont {Troyan}, \citenamefont {Hirao}, \citenamefont {Kawaguchi} \emph {et~al.}}]{147K_superconductor}%
  \BibitemOpen
  \bibfield  {author} {\bibinfo {author} {\bibfnamefont {K.}~\bibnamefont {Shimizu}}, \bibinfo {author} {\bibfnamefont {M.}~\bibnamefont {Einaga}}, \bibinfo {author} {\bibfnamefont {M.}~\bibnamefont {Sakata}}, \bibinfo {author} {\bibfnamefont {A.}~\bibnamefont {Masuda}}, \bibinfo {author} {\bibfnamefont {H.}~\bibnamefont {Nakao}}, \bibinfo {author} {\bibfnamefont {M.}~\bibnamefont {Eremets}}, \bibinfo {author} {\bibfnamefont {A.}~\bibnamefont {Drozdov}}, \bibinfo {author} {\bibfnamefont {I.}~\bibnamefont {Troyan}}, \bibinfo {author} {\bibfnamefont {N.}~\bibnamefont {Hirao}}, \bibinfo {author} {\bibfnamefont {S.}~\bibnamefont {Kawaguchi}},  \emph {et~al.},\ }\href@noop {} {\bibfield  {journal} {\bibinfo  {journal} {Physica C: Superconductivity and its applications}\ }\textbf {\bibinfo {volume} {552}},\ \bibinfo {pages} {27} (\bibinfo {year} {2018})}\BibitemShut {NoStop}%
\bibitem [{\citenamefont {Shao}\ \emph {et~al.}(1995)\citenamefont {Shao}, \citenamefont {Lam}, \citenamefont {Fung}, \citenamefont {Wu}, \citenamefont {Du}, \citenamefont {Shen}, \citenamefont {Chow}, \citenamefont {Ho}, \citenamefont {Hung},\ and\ \citenamefont {Yao}}]{143K_superconductor}%
  \BibitemOpen
  \bibfield  {author} {\bibinfo {author} {\bibfnamefont {H.}~\bibnamefont {Shao}}, \bibinfo {author} {\bibfnamefont {C.}~\bibnamefont {Lam}}, \bibinfo {author} {\bibfnamefont {P.}~\bibnamefont {Fung}}, \bibinfo {author} {\bibfnamefont {X.}~\bibnamefont {Wu}}, \bibinfo {author} {\bibfnamefont {J.}~\bibnamefont {Du}}, \bibinfo {author} {\bibfnamefont {G.}~\bibnamefont {Shen}}, \bibinfo {author} {\bibfnamefont {J.}~\bibnamefont {Chow}}, \bibinfo {author} {\bibfnamefont {S.}~\bibnamefont {Ho}}, \bibinfo {author} {\bibfnamefont {K.}~\bibnamefont {Hung}}, \ and\ \bibinfo {author} {\bibfnamefont {X.}~\bibnamefont {Yao}},\ }\href@noop {} {\bibfield  {journal} {\bibinfo  {journal} {Physica C: Superconductivity}\ }\textbf {\bibinfo {volume} {246}},\ \bibinfo {pages} {207} (\bibinfo {year} {1995})}\BibitemShut {NoStop}%
\bibitem [{\citenamefont {Lemm}\ \emph {et~al.}(2023)\citenamefont {Lemm}, \citenamefont {von Rudorff},\ and\ \citenamefont {von Lilienfeld}}]{Dominik_similarity}%
  \BibitemOpen
  \bibfield  {author} {\bibinfo {author} {\bibfnamefont {D.}~\bibnamefont {Lemm}}, \bibinfo {author} {\bibfnamefont {G.~F.}\ \bibnamefont {von Rudorff}}, \ and\ \bibinfo {author} {\bibfnamefont {O.~A.}\ \bibnamefont {von Lilienfeld}},\ }\href@noop {} {\bibfield  {journal} {\bibinfo  {journal} {Machine Learning: Science and Technology}\ }\textbf {\bibinfo {volume} {4}},\ \bibinfo {pages} {045043} (\bibinfo {year} {2023})}\BibitemShut {NoStop}%
\bibitem [{\citenamefont {van Wieringen}(2015)}]{ridge_regression_notes}%
  \BibitemOpen
  \bibfield  {author} {\bibinfo {author} {\bibfnamefont {W.~N.}\ \bibnamefont {van Wieringen}},\ }\href@noop {} {\bibfield  {journal} {\bibinfo  {journal} {arXiv preprint arXiv:1509.09169}\ } (\bibinfo {year} {2015})}\BibitemShut {NoStop}%
\bibitem [{\citenamefont {Pedregosa}\ \emph {et~al.}(2011)\citenamefont {Pedregosa}, \citenamefont {Varoquaux}, \citenamefont {Gramfort}, \citenamefont {Michel}, \citenamefont {Thirion}, \citenamefont {Grisel}, \citenamefont {Blondel}, \citenamefont {Prettenhofer}, \citenamefont {Weiss}, \citenamefont {Dubourg}, \citenamefont {Vanderplas}, \citenamefont {Passos}, \citenamefont {Cournapeau}, \citenamefont {Brucher}, \citenamefont {Perrot},\ and\ \citenamefont {Duchesnay}}]{sklearn}%
  \BibitemOpen
  \bibfield  {author} {\bibinfo {author} {\bibfnamefont {F.}~\bibnamefont {Pedregosa}}, \bibinfo {author} {\bibfnamefont {G.}~\bibnamefont {Varoquaux}}, \bibinfo {author} {\bibfnamefont {A.}~\bibnamefont {Gramfort}}, \bibinfo {author} {\bibfnamefont {V.}~\bibnamefont {Michel}}, \bibinfo {author} {\bibfnamefont {B.}~\bibnamefont {Thirion}}, \bibinfo {author} {\bibfnamefont {O.}~\bibnamefont {Grisel}}, \bibinfo {author} {\bibfnamefont {M.}~\bibnamefont {Blondel}}, \bibinfo {author} {\bibfnamefont {P.}~\bibnamefont {Prettenhofer}}, \bibinfo {author} {\bibfnamefont {R.}~\bibnamefont {Weiss}}, \bibinfo {author} {\bibfnamefont {V.}~\bibnamefont {Dubourg}}, \bibinfo {author} {\bibfnamefont {J.}~\bibnamefont {Vanderplas}}, \bibinfo {author} {\bibfnamefont {A.}~\bibnamefont {Passos}}, \bibinfo {author} {\bibfnamefont {D.}~\bibnamefont {Cournapeau}}, \bibinfo {author} {\bibfnamefont {M.}~\bibnamefont {Brucher}}, \bibinfo {author} {\bibfnamefont {M.}~\bibnamefont {Perrot}}, \ and\ \bibinfo {author} {\bibfnamefont
  {E.}~\bibnamefont {Duchesnay}},\ }\href@noop {} {\bibfield  {journal} {\bibinfo  {journal} {Journal of Machine Learning Research}\ }\textbf {\bibinfo {volume} {12}},\ \bibinfo {pages} {2825} (\bibinfo {year} {2011})}\BibitemShut {NoStop}%
\bibitem [{skl(2024{\natexlab{a}})}]{sklearn_ridge}%
  \BibitemOpen
  \href@noop {} {\enquote {\bibinfo {title} {sklearn.linear\_model.{R}idge --- scikit-learn.org},}\ }\bibinfo {howpublished} {\url{https://scikit-learn.org/stable/modules/generated/sklearn.linear_model.Ridge.html}} (\bibinfo {year} {2024}{\natexlab{a}})\BibitemShut {NoStop}%
\bibitem [{skl(2024{\natexlab{b}})}]{sklearn_ridgecv}%
  \BibitemOpen
  \href@noop {} {\enquote {\bibinfo {title} {sklearn.linear\_model.{R}idge{C}{V} --- scikit-learn.org},}\ }\bibinfo {howpublished} {\url{https://scikit-learn.org/stable/modules/generated/sklearn.linear_model.RidgeCV.html}} (\bibinfo {year} {2024}{\natexlab{b}})\BibitemShut {NoStop}%
\bibitem [{\citenamefont {Meredig}\ \emph {et~al.}(2014)\citenamefont {Meredig}, \citenamefont {Agrawal}, \citenamefont {Kirklin}, \citenamefont {Saal}, \citenamefont {Doak}, \citenamefont {Thompson}, \citenamefont {Zhang}, \citenamefont {Choudhary},\ and\ \citenamefont {Wolverton}}]{formation_energies_from_composition}%
  \BibitemOpen
  \bibfield  {author} {\bibinfo {author} {\bibfnamefont {B.}~\bibnamefont {Meredig}}, \bibinfo {author} {\bibfnamefont {A.}~\bibnamefont {Agrawal}}, \bibinfo {author} {\bibfnamefont {S.}~\bibnamefont {Kirklin}}, \bibinfo {author} {\bibfnamefont {J.~E.}\ \bibnamefont {Saal}}, \bibinfo {author} {\bibfnamefont {J.~W.}\ \bibnamefont {Doak}}, \bibinfo {author} {\bibfnamefont {A.}~\bibnamefont {Thompson}}, \bibinfo {author} {\bibfnamefont {K.}~\bibnamefont {Zhang}}, \bibinfo {author} {\bibfnamefont {A.}~\bibnamefont {Choudhary}}, \ and\ \bibinfo {author} {\bibfnamefont {C.}~\bibnamefont {Wolverton}},\ }\href@noop {} {\bibfield  {journal} {\bibinfo  {journal} {Physical Review B}\ }\textbf {\bibinfo {volume} {89}},\ \bibinfo {pages} {094104} (\bibinfo {year} {2014})}\BibitemShut {NoStop}%
\bibitem [{\citenamefont {Ghiringhelli}\ \emph {et~al.}(2015)\citenamefont {Ghiringhelli}, \citenamefont {Vybiral}, \citenamefont {Levchenko}, \citenamefont {Draxl},\ and\ \citenamefont {Scheffler}}]{crystal_structure_type_from_composition}%
  \BibitemOpen
  \bibfield  {author} {\bibinfo {author} {\bibfnamefont {L.~M.}\ \bibnamefont {Ghiringhelli}}, \bibinfo {author} {\bibfnamefont {J.}~\bibnamefont {Vybiral}}, \bibinfo {author} {\bibfnamefont {S.~V.}\ \bibnamefont {Levchenko}}, \bibinfo {author} {\bibfnamefont {C.}~\bibnamefont {Draxl}}, \ and\ \bibinfo {author} {\bibfnamefont {M.}~\bibnamefont {Scheffler}},\ }\href@noop {} {\bibfield  {journal} {\bibinfo  {journal} {Physical review letters}\ }\textbf {\bibinfo {volume} {114}},\ \bibinfo {pages} {105503} (\bibinfo {year} {2015})}\BibitemShut {NoStop}%
\bibitem [{\citenamefont {Jha}\ \emph {et~al.}(2018)\citenamefont {Jha}, \citenamefont {Ward}, \citenamefont {Paul}, \citenamefont {Liao}, \citenamefont {Choudhary}, \citenamefont {Wolverton},\ and\ \citenamefont {Agrawal}}]{ElemNet}%
  \BibitemOpen
  \bibfield  {author} {\bibinfo {author} {\bibfnamefont {D.}~\bibnamefont {Jha}}, \bibinfo {author} {\bibfnamefont {L.}~\bibnamefont {Ward}}, \bibinfo {author} {\bibfnamefont {A.}~\bibnamefont {Paul}}, \bibinfo {author} {\bibfnamefont {W.-k.}\ \bibnamefont {Liao}}, \bibinfo {author} {\bibfnamefont {A.}~\bibnamefont {Choudhary}}, \bibinfo {author} {\bibfnamefont {C.}~\bibnamefont {Wolverton}}, \ and\ \bibinfo {author} {\bibfnamefont {A.}~\bibnamefont {Agrawal}},\ }\href@noop {} {\bibfield  {journal} {\bibinfo  {journal} {Scientific reports}\ }\textbf {\bibinfo {volume} {8}},\ \bibinfo {pages} {17593} (\bibinfo {year} {2018})}\BibitemShut {NoStop}%
\bibitem [{\citenamefont {Goodall}\ and\ \citenamefont {Lee}(2020)}]{ROOST}%
  \BibitemOpen
  \bibfield  {author} {\bibinfo {author} {\bibfnamefont {R.~E.}\ \bibnamefont {Goodall}}\ and\ \bibinfo {author} {\bibfnamefont {A.~A.}\ \bibnamefont {Lee}},\ }\href@noop {} {\bibfield  {journal} {\bibinfo  {journal} {Nature communications}\ }\textbf {\bibinfo {volume} {11}},\ \bibinfo {pages} {6280} (\bibinfo {year} {2020})}\BibitemShut {NoStop}%
\bibitem [{\citenamefont {Wang}\ \emph {et~al.}(2021)\citenamefont {Wang}, \citenamefont {Kauwe}, \citenamefont {Murdock},\ and\ \citenamefont {Sparks}}]{CrabNet}%
  \BibitemOpen
  \bibfield  {author} {\bibinfo {author} {\bibfnamefont {A.~Y.-T.}\ \bibnamefont {Wang}}, \bibinfo {author} {\bibfnamefont {S.~K.}\ \bibnamefont {Kauwe}}, \bibinfo {author} {\bibfnamefont {R.~J.}\ \bibnamefont {Murdock}}, \ and\ \bibinfo {author} {\bibfnamefont {T.~D.}\ \bibnamefont {Sparks}},\ }\href@noop {} {\bibfield  {journal} {\bibinfo  {journal} {Npj Computational Materials}\ }\textbf {\bibinfo {volume} {7}},\ \bibinfo {pages} {77} (\bibinfo {year} {2021})}\BibitemShut {NoStop}%
\bibitem [{\citenamefont {Damewood}\ \emph {et~al.}(2023)\citenamefont {Damewood}, \citenamefont {Karaguesian}, \citenamefont {Lunger}, \citenamefont {Tan}, \citenamefont {Xie}, \citenamefont {Peng},\ and\ \citenamefont {G{\'o}mez-Bombarelli}}]{materials_ML_representations}%
  \BibitemOpen
  \bibfield  {author} {\bibinfo {author} {\bibfnamefont {J.}~\bibnamefont {Damewood}}, \bibinfo {author} {\bibfnamefont {J.}~\bibnamefont {Karaguesian}}, \bibinfo {author} {\bibfnamefont {J.~R.}\ \bibnamefont {Lunger}}, \bibinfo {author} {\bibfnamefont {A.~R.}\ \bibnamefont {Tan}}, \bibinfo {author} {\bibfnamefont {M.}~\bibnamefont {Xie}}, \bibinfo {author} {\bibfnamefont {J.}~\bibnamefont {Peng}}, \ and\ \bibinfo {author} {\bibfnamefont {R.}~\bibnamefont {G{\'o}mez-Bombarelli}},\ }\href@noop {} {\bibfield  {journal} {\bibinfo  {journal} {Annual Review of Materials Research}\ }\textbf {\bibinfo {volume} {53}},\ \bibinfo {pages} {399} (\bibinfo {year} {2023})}\BibitemShut {NoStop}%
\bibitem [{\citenamefont {Ward}\ \emph {et~al.}(2018)\citenamefont {Ward}, \citenamefont {Dunn}, \citenamefont {Faghaninia}, \citenamefont {Zimmermann}, \citenamefont {Bajaj}, \citenamefont {Wang}, \citenamefont {Montoya}, \citenamefont {Chen}, \citenamefont {Bystrom}, \citenamefont {Dylla} \emph {et~al.}}]{Matminer}%
  \BibitemOpen
  \bibfield  {author} {\bibinfo {author} {\bibfnamefont {L.}~\bibnamefont {Ward}}, \bibinfo {author} {\bibfnamefont {A.}~\bibnamefont {Dunn}}, \bibinfo {author} {\bibfnamefont {A.}~\bibnamefont {Faghaninia}}, \bibinfo {author} {\bibfnamefont {N.~E.}\ \bibnamefont {Zimmermann}}, \bibinfo {author} {\bibfnamefont {S.}~\bibnamefont {Bajaj}}, \bibinfo {author} {\bibfnamefont {Q.}~\bibnamefont {Wang}}, \bibinfo {author} {\bibfnamefont {J.}~\bibnamefont {Montoya}}, \bibinfo {author} {\bibfnamefont {J.}~\bibnamefont {Chen}}, \bibinfo {author} {\bibfnamefont {K.}~\bibnamefont {Bystrom}}, \bibinfo {author} {\bibfnamefont {M.}~\bibnamefont {Dylla}},  \emph {et~al.},\ }\href@noop {} {\bibfield  {journal} {\bibinfo  {journal} {Computational Materials Science}\ }\textbf {\bibinfo {volume} {152}},\ \bibinfo {pages} {60} (\bibinfo {year} {2018})}\BibitemShut {NoStop}%
\bibitem [{\citenamefont {Cortes}\ \emph {et~al.}(1993)\citenamefont {Cortes}, \citenamefont {Jackel}, \citenamefont {Solla}, \citenamefont {Vapnik},\ and\ \citenamefont {Denker}}]{learning_curves}%
  \BibitemOpen
  \bibfield  {author} {\bibinfo {author} {\bibfnamefont {C.}~\bibnamefont {Cortes}}, \bibinfo {author} {\bibfnamefont {L.~D.}\ \bibnamefont {Jackel}}, \bibinfo {author} {\bibfnamefont {S.}~\bibnamefont {Solla}}, \bibinfo {author} {\bibfnamefont {V.}~\bibnamefont {Vapnik}}, \ and\ \bibinfo {author} {\bibfnamefont {J.}~\bibnamefont {Denker}},\ }\href@noop {} {\bibfield  {journal} {\bibinfo  {journal} {Advances in neural information processing systems}\ }\textbf {\bibinfo {volume} {6}} (\bibinfo {year} {1993})}\BibitemShut {NoStop}%
\bibitem [{\citenamefont {Pernot}\ \emph {et~al.}(2020)\citenamefont {Pernot}, \citenamefont {Huang},\ and\ \citenamefont {Savin}}]{pernot2020impact}%
  \BibitemOpen
  \bibfield  {author} {\bibinfo {author} {\bibfnamefont {P.}~\bibnamefont {Pernot}}, \bibinfo {author} {\bibfnamefont {B.}~\bibnamefont {Huang}}, \ and\ \bibinfo {author} {\bibfnamefont {A.}~\bibnamefont {Savin}},\ }\href@noop {} {\bibfield  {journal} {\bibinfo  {journal} {Machine Learning: Science and Technology}\ }\textbf {\bibinfo {volume} {1}},\ \bibinfo {pages} {035011} (\bibinfo {year} {2020})}\BibitemShut {NoStop}%
\bibitem [{\citenamefont {Maxwell}(1950)}]{isotope_effect_1}%
  \BibitemOpen
  \bibfield  {author} {\bibinfo {author} {\bibfnamefont {E.}~\bibnamefont {Maxwell}},\ }\href@noop {} {\bibfield  {journal} {\bibinfo  {journal} {Physical Review}\ }\textbf {\bibinfo {volume} {78}},\ \bibinfo {pages} {477} (\bibinfo {year} {1950})}\BibitemShut {NoStop}%
\bibitem [{\citenamefont {Reynolds}\ \emph {et~al.}(1950)\citenamefont {Reynolds}, \citenamefont {Serin}, \citenamefont {Wright},\ and\ \citenamefont {Nesbitt}}]{isotope_effect_2}%
  \BibitemOpen
  \bibfield  {author} {\bibinfo {author} {\bibfnamefont {C.}~\bibnamefont {Reynolds}}, \bibinfo {author} {\bibfnamefont {B.}~\bibnamefont {Serin}}, \bibinfo {author} {\bibfnamefont {W.}~\bibnamefont {Wright}}, \ and\ \bibinfo {author} {\bibfnamefont {L.}~\bibnamefont {Nesbitt}},\ }\href@noop {} {\bibfield  {journal} {\bibinfo  {journal} {Physical Review}\ }\textbf {\bibinfo {volume} {78}},\ \bibinfo {pages} {487} (\bibinfo {year} {1950})}\BibitemShut {NoStop}%
\bibitem [{\citenamefont {Kresin}\ \emph {et~al.}(1997)\citenamefont {Kresin}, \citenamefont {Bill}, \citenamefont {Wolf},\ and\ \citenamefont {Ovchinnikov}}]{unconventional_isotope_effects_1}%
  \BibitemOpen
  \bibfield  {author} {\bibinfo {author} {\bibfnamefont {V.~Z.}\ \bibnamefont {Kresin}}, \bibinfo {author} {\bibfnamefont {A.}~\bibnamefont {Bill}}, \bibinfo {author} {\bibfnamefont {S.~A.}\ \bibnamefont {Wolf}}, \ and\ \bibinfo {author} {\bibfnamefont {Y.~N.}\ \bibnamefont {Ovchinnikov}},\ }\href@noop {} {\bibfield  {journal} {\bibinfo  {journal} {Physical Review B}\ }\textbf {\bibinfo {volume} {56}},\ \bibinfo {pages} {107} (\bibinfo {year} {1997})}\BibitemShut {NoStop}%
\bibitem [{\citenamefont {Zhao}\ \emph {et~al.}(2001)\citenamefont {Zhao}, \citenamefont {Keller},\ and\ \citenamefont {Conder}}]{unconventional_isotope_effects_2}%
  \BibitemOpen
  \bibfield  {author} {\bibinfo {author} {\bibfnamefont {G.-m.}\ \bibnamefont {Zhao}}, \bibinfo {author} {\bibfnamefont {H.}~\bibnamefont {Keller}}, \ and\ \bibinfo {author} {\bibfnamefont {K.}~\bibnamefont {Conder}},\ }\href@noop {} {\bibfield  {journal} {\bibinfo  {journal} {Journal of Physics: Condensed Matter}\ }\textbf {\bibinfo {volume} {13}},\ \bibinfo {pages} {R569} (\bibinfo {year} {2001})}\BibitemShut {NoStop}%
\bibitem [{\citenamefont {Hutcheon}\ \emph {et~al.}(2020)\citenamefont {Hutcheon}, \citenamefont {Shipley},\ and\ \citenamefont {Needs}}]{hutcheon_2020}%
  \BibitemOpen
  \bibfield  {author} {\bibinfo {author} {\bibfnamefont {M.~J.}\ \bibnamefont {Hutcheon}}, \bibinfo {author} {\bibfnamefont {A.~M.}\ \bibnamefont {Shipley}}, \ and\ \bibinfo {author} {\bibfnamefont {R.~J.}\ \bibnamefont {Needs}},\ }\href@noop {} {\bibfield  {journal} {\bibinfo  {journal} {Physical Review B}\ }\textbf {\bibinfo {volume} {101}},\ \bibinfo {pages} {144505} (\bibinfo {year} {2020})}\BibitemShut {NoStop}%
\bibitem [{\citenamefont {Roter}\ and\ \citenamefont {Dordevic}(2020)}]{roter_2020}%
  \BibitemOpen
  \bibfield  {author} {\bibinfo {author} {\bibfnamefont {B.}~\bibnamefont {Roter}}\ and\ \bibinfo {author} {\bibfnamefont {S.~V.}\ \bibnamefont {Dordevic}},\ }\href@noop {} {\bibfield  {journal} {\bibinfo  {journal} {Physica C: Superconductivity and its applications}\ }\textbf {\bibinfo {volume} {575}},\ \bibinfo {pages} {1353689} (\bibinfo {year} {2020})}\BibitemShut {NoStop}%
\bibitem [{\citenamefont {Flacke}\ and\ \citenamefont {Jacobs}(1995)}]{MP_765341_1}%
  \BibitemOpen
  \bibfield  {author} {\bibinfo {author} {\bibfnamefont {F.}~\bibnamefont {Flacke}}\ and\ \bibinfo {author} {\bibfnamefont {H.}~\bibnamefont {Jacobs}},\ }\href@noop {} {\bibfield  {journal} {\bibinfo  {journal} {Journal of alloys and compounds}\ }\textbf {\bibinfo {volume} {227}},\ \bibinfo {pages} {109} (\bibinfo {year} {1995})}\BibitemShut {NoStop}%
\bibitem [{\citenamefont {Project}({\natexlab{a}})}]{MP_765341_2}%
  \BibitemOpen
  \bibfield  {author} {\bibinfo {author} {\bibfnamefont {M.}~\bibnamefont {Project}},\ }\href@noop {} {\enquote {\bibinfo {title} {mp-765341},}\ }\bibinfo {howpublished} {\url{https://next-gen.materialsproject.org/materials/mp-765341?material_ids=mp-765341}} ({\natexlab{a}})\BibitemShut {NoStop}%
\bibitem [{\citenamefont {Project}({\natexlab{b}})}]{MP_765341_3}%
  \BibitemOpen
  \bibfield  {author} {\bibinfo {author} {\bibfnamefont {M.}~\bibnamefont {Project}},\ }\href@noop {} {\enquote {\bibinfo {title} {mp-765341},}\ }\bibinfo {howpublished} {\url{https://next-gen.materialsproject.org/materials/mp-753273?material_ids=mp-753273}} ({\natexlab{b}})\BibitemShut {NoStop}%
\bibitem [{\citenamefont {Khodel}\ and\ \citenamefont {Shaginyan}(1990)}]{flat_bands_1}%
  \BibitemOpen
  \bibfield  {author} {\bibinfo {author} {\bibfnamefont {V.}~\bibnamefont {Khodel}}\ and\ \bibinfo {author} {\bibfnamefont {V.}~\bibnamefont {Shaginyan}},\ }\href@noop {} {\bibfield  {journal} {\bibinfo  {journal} {Jetp Lett}\ }\textbf {\bibinfo {volume} {51}},\ \bibinfo {pages} {553} (\bibinfo {year} {1990})}\BibitemShut {NoStop}%
\bibitem [{\citenamefont {Heikkil{\"a}}\ and\ \citenamefont {Volovik}(2016)}]{flat_bands_2}%
  \BibitemOpen
  \bibfield  {author} {\bibinfo {author} {\bibfnamefont {T.~T.}\ \bibnamefont {Heikkil{\"a}}}\ and\ \bibinfo {author} {\bibfnamefont {G.~E.}\ \bibnamefont {Volovik}},\ }\href@noop {} {\bibfield  {journal} {\bibinfo  {journal} {Basic Physics of Functionalized Graphite}\ ,\ \bibinfo {pages} {123}} (\bibinfo {year} {2016})}\BibitemShut {NoStop}%
\bibitem [{\citenamefont {Godby}\ \emph {et~al.}(1988)\citenamefont {Godby}, \citenamefont {Schl{\"u}ter},\ and\ \citenamefont {Sham}}]{band_structure_problem_1}%
  \BibitemOpen
  \bibfield  {author} {\bibinfo {author} {\bibfnamefont {R.~W.}\ \bibnamefont {Godby}}, \bibinfo {author} {\bibfnamefont {M.}~\bibnamefont {Schl{\"u}ter}}, \ and\ \bibinfo {author} {\bibfnamefont {L.}~\bibnamefont {Sham}},\ }\href@noop {} {\bibfield  {journal} {\bibinfo  {journal} {Physical Review B}\ }\textbf {\bibinfo {volume} {37}},\ \bibinfo {pages} {10159} (\bibinfo {year} {1988})}\BibitemShut {NoStop}%
\bibitem [{\citenamefont {Project}({\natexlab{c}})}]{band_structure_problem_2}%
  \BibitemOpen
  \bibfield  {author} {\bibinfo {author} {\bibfnamefont {M.}~\bibnamefont {Project}},\ }\href@noop {} {\enquote {\bibinfo {title} {{E}lectronic {S}tructure | {M}aterials {P}roject {D}ocumentation --- docs.materialsproject.org},}\ }\bibinfo {howpublished} {\url{https://docs.materialsproject.org/methodology/materials-methodology/electronic-structure}} ({\natexlab{c}})\BibitemShut {NoStop}%
\bibitem [{\citenamefont {Vovk}(2013)}]{kernel_ridge_regression_notes_1}%
  \BibitemOpen
  \bibfield  {author} {\bibinfo {author} {\bibfnamefont {V.}~\bibnamefont {Vovk}},\ }in\ \href@noop {} {\emph {\bibinfo {booktitle} {Empirical Inference: Festschrift in Honor of Vladimir N. Vapnik}}}\ (\bibinfo  {publisher} {Springer},\ \bibinfo {year} {2013})\ pp.\ \bibinfo {pages} {105--116}\BibitemShut {NoStop}%
\bibitem [{\citenamefont {Chen}\ and\ \citenamefont {Guestrin}(2016)}]{XGBoost}%
  \BibitemOpen
  \bibfield  {author} {\bibinfo {author} {\bibfnamefont {T.}~\bibnamefont {Chen}}\ and\ \bibinfo {author} {\bibfnamefont {C.}~\bibnamefont {Guestrin}},\ }in\ \href@noop {} {\emph {\bibinfo {booktitle} {Proceedings of the 22nd acm sigkdd international conference on knowledge discovery and data mining}}}\ (\bibinfo {year} {2016})\ pp.\ \bibinfo {pages} {785--794}\BibitemShut {NoStop}%
\bibitem [{\citenamefont {Nadaraya}(1964)}]{kernel_ridge_regression_notes_2}%
  \BibitemOpen
  \bibfield  {author} {\bibinfo {author} {\bibfnamefont {E.~A.}\ \bibnamefont {Nadaraya}},\ }\href@noop {} {\bibfield  {journal} {\bibinfo  {journal} {Theory of Probability \& Its Applications}\ }\textbf {\bibinfo {volume} {9}},\ \bibinfo {pages} {141} (\bibinfo {year} {1964})}\BibitemShut {NoStop}%
\bibitem [{\citenamefont {Watson}(1964)}]{kernel_ridge_regression_notes_3}%
  \BibitemOpen
  \bibfield  {author} {\bibinfo {author} {\bibfnamefont {G.~S.}\ \bibnamefont {Watson}},\ }\href@noop {} {\bibfield  {journal} {\bibinfo  {journal} {Sankhy{\=a}: The Indian Journal of Statistics, Series A}\ ,\ \bibinfo {pages} {359}} (\bibinfo {year} {1964})}\BibitemShut {NoStop}%
\bibitem [{\citenamefont {Bierens}(1988)}]{kernel_ridge_regression_notes_4}%
  \BibitemOpen
  \bibfield  {author} {\bibinfo {author} {\bibfnamefont {H.~J.}\ \bibnamefont {Bierens}},\ }\href@noop {} {\  (\bibinfo {year} {1988})}\BibitemShut {NoStop}%
\bibitem [{\citenamefont {Malistov}\ and\ \citenamefont {Trushin}(2019)}]{trees_cant_extrapolate}%
  \BibitemOpen
  \bibfield  {author} {\bibinfo {author} {\bibfnamefont {A.}~\bibnamefont {Malistov}}\ and\ \bibinfo {author} {\bibfnamefont {A.}~\bibnamefont {Trushin}},\ }in\ \href@noop {} {\emph {\bibinfo {booktitle} {2019 18th IEEE International Conference On Machine Learning And Applications (ICMLA)}}}\ (\bibinfo {organization} {IEEE},\ \bibinfo {year} {2019})\ pp.\ \bibinfo {pages} {783--789}\BibitemShut {NoStop}%
\bibitem [{\citenamefont {Ho}(1995)}]{random_forest_notes_1}%
  \BibitemOpen
  \bibfield  {author} {\bibinfo {author} {\bibfnamefont {T.~K.}\ \bibnamefont {Ho}},\ }in\ \href@noop {} {\emph {\bibinfo {booktitle} {Proceedings of 3rd international conference on document analysis and recognition}}},\ Vol.~\bibinfo {volume} {1}\ (\bibinfo {organization} {IEEE},\ \bibinfo {year} {1995})\ pp.\ \bibinfo {pages} {278--282}\BibitemShut {NoStop}%
\bibitem [{\citenamefont {Ho}(1998)}]{random_forest_notes_2}%
  \BibitemOpen
  \bibfield  {author} {\bibinfo {author} {\bibfnamefont {T.~K.}\ \bibnamefont {Ho}},\ }\href@noop {} {\bibfield  {journal} {\bibinfo  {journal} {IEEE transactions on pattern analysis and machine intelligence}\ }\textbf {\bibinfo {volume} {20}},\ \bibinfo {pages} {832} (\bibinfo {year} {1998})}\BibitemShut {NoStop}%
\end{thebibliography}%

\newpage 
\onecolumngrid

\section{\label{sec: Supplementary Information}Supplementary Information}

\subsection{\label{sec: Toy Problem}Toy Problem}

\begin{figure*}[h]
\centering
\subfloat[\label{fig: toy problem linear}]{\includegraphics[width=0.218\textwidth]{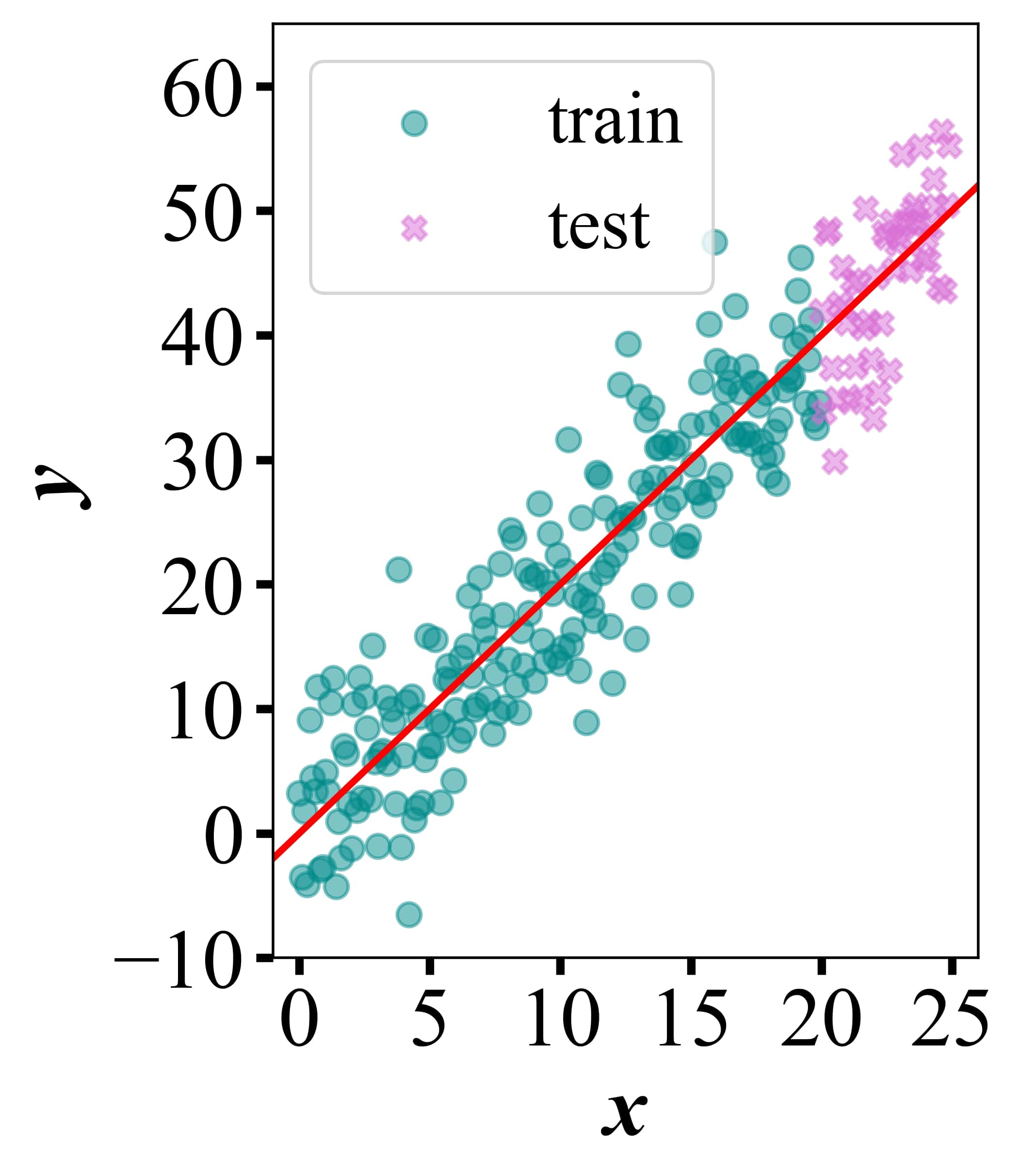}}
\centering
\subfloat[\label{fig: toy problem linear ridge regression}]{\includegraphics[width=0.20\textwidth]{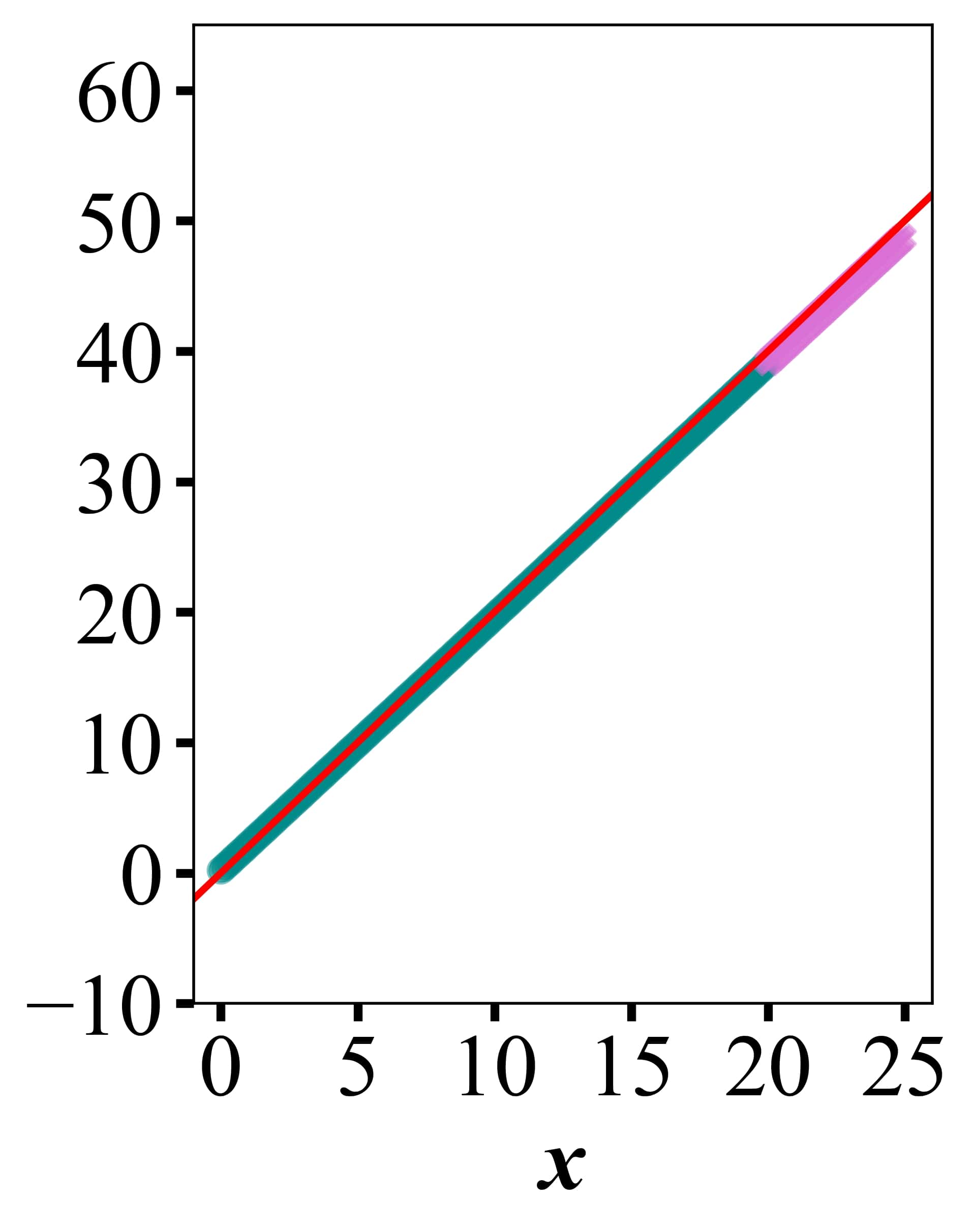}}
\centering
\subfloat[\label{fig: toy problem linear kernel ridge regression}]{\includegraphics[width=0.20\textwidth]{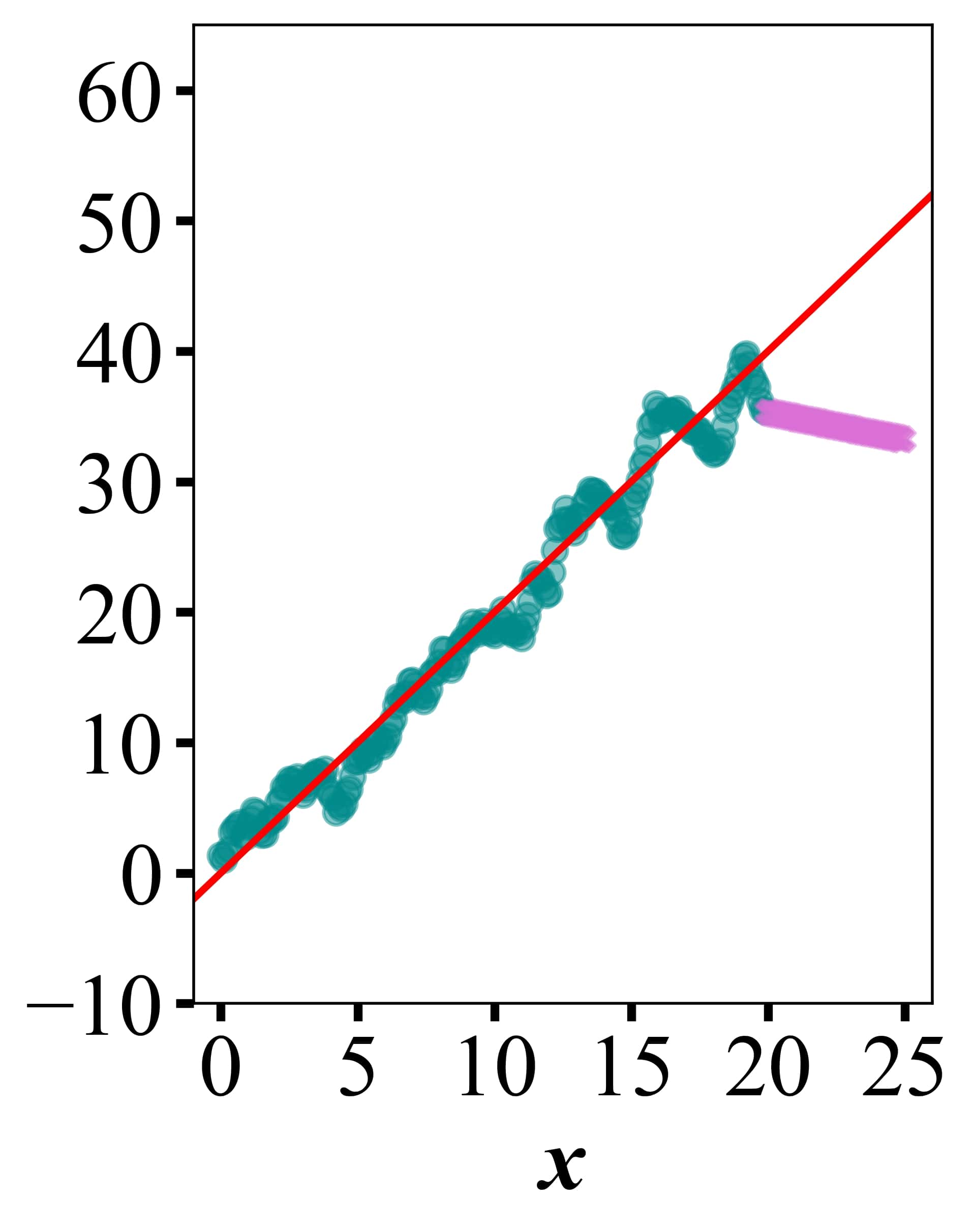}}
\centering
\subfloat[\label{fig: toy problem linear xgbr}]{\includegraphics[width=0.20\textwidth]{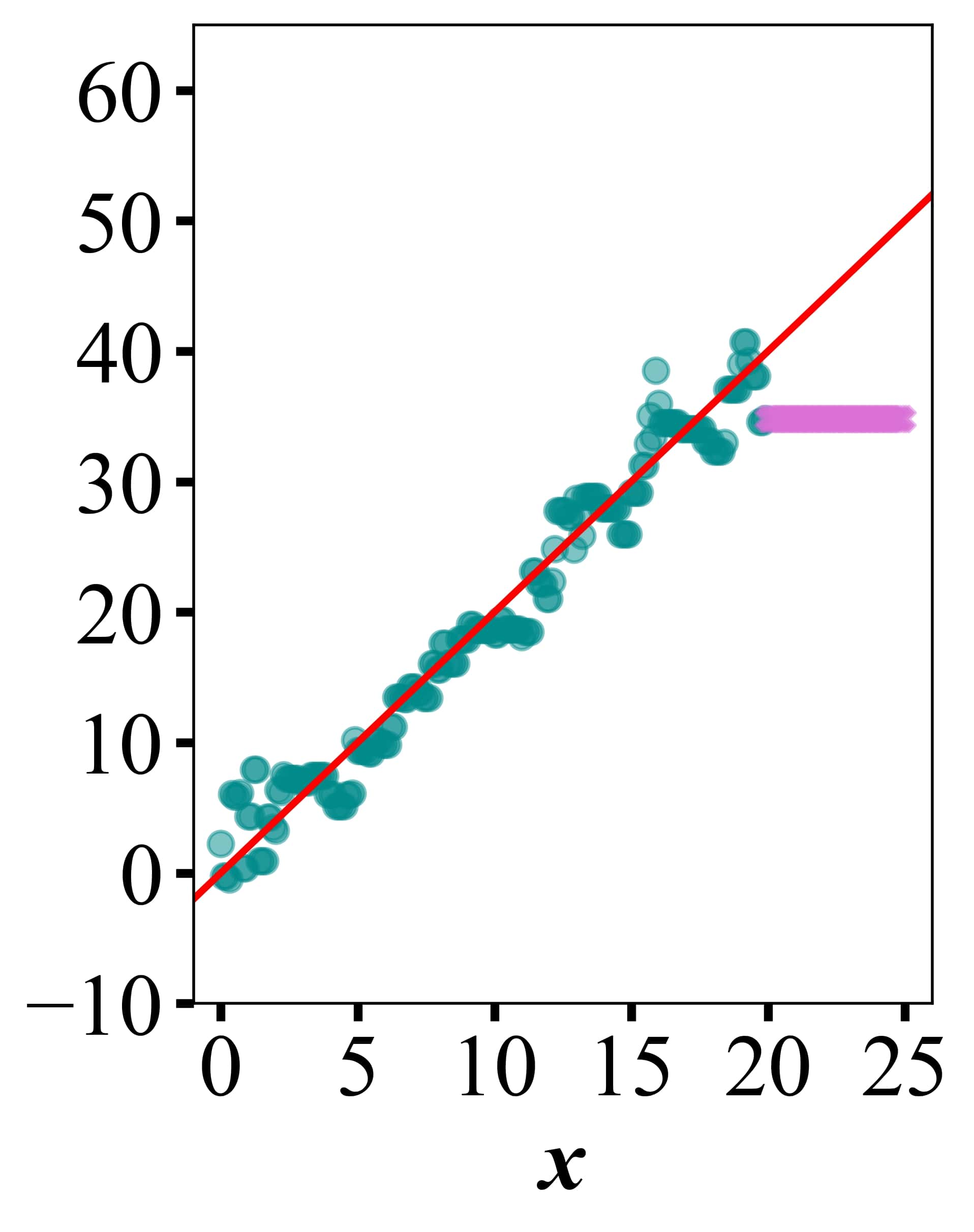}}
\hfill
\centering 
\subfloat[\label{fig: toy problem quadratic}]
{\includegraphics[width=0.209\textwidth]{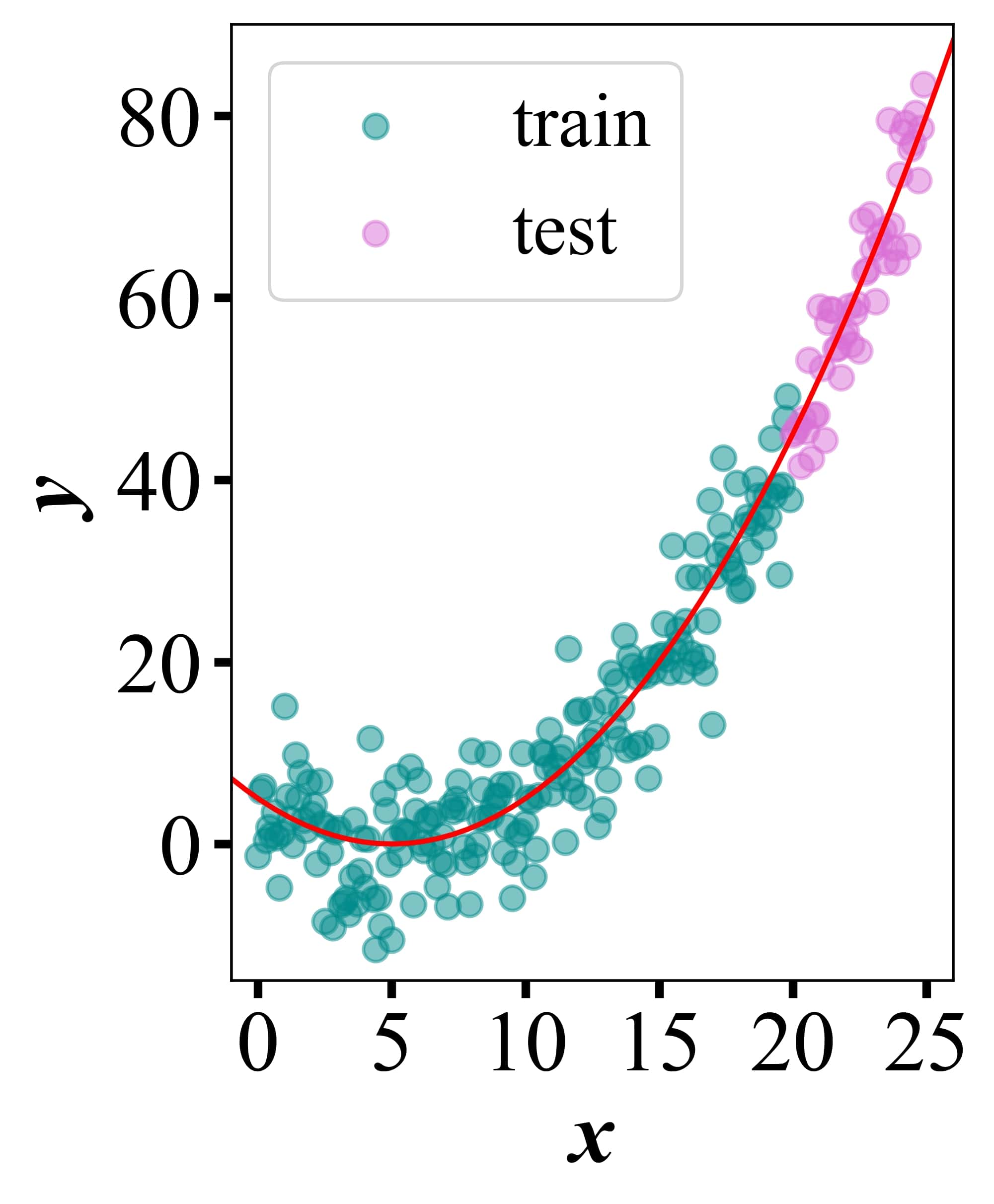}}
\centering 
\subfloat[\label{fig: toy problem quadratic ridge regression}]{\includegraphics[width=0.19\textwidth]{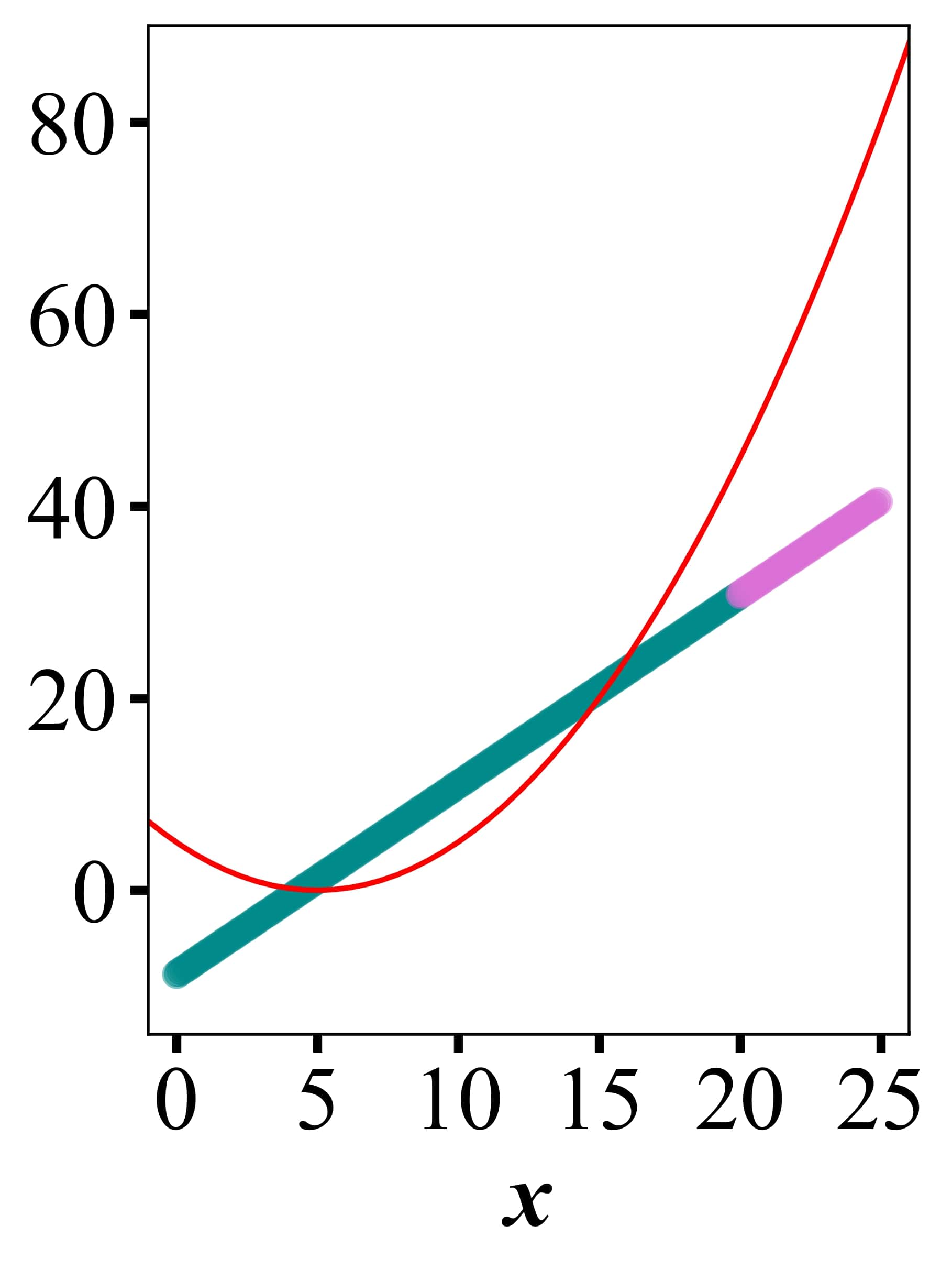}}
\centering 
\subfloat[\label{fig: toy problem quadratic kernel ridge regression}]{\includegraphics[width=0.19\textwidth]{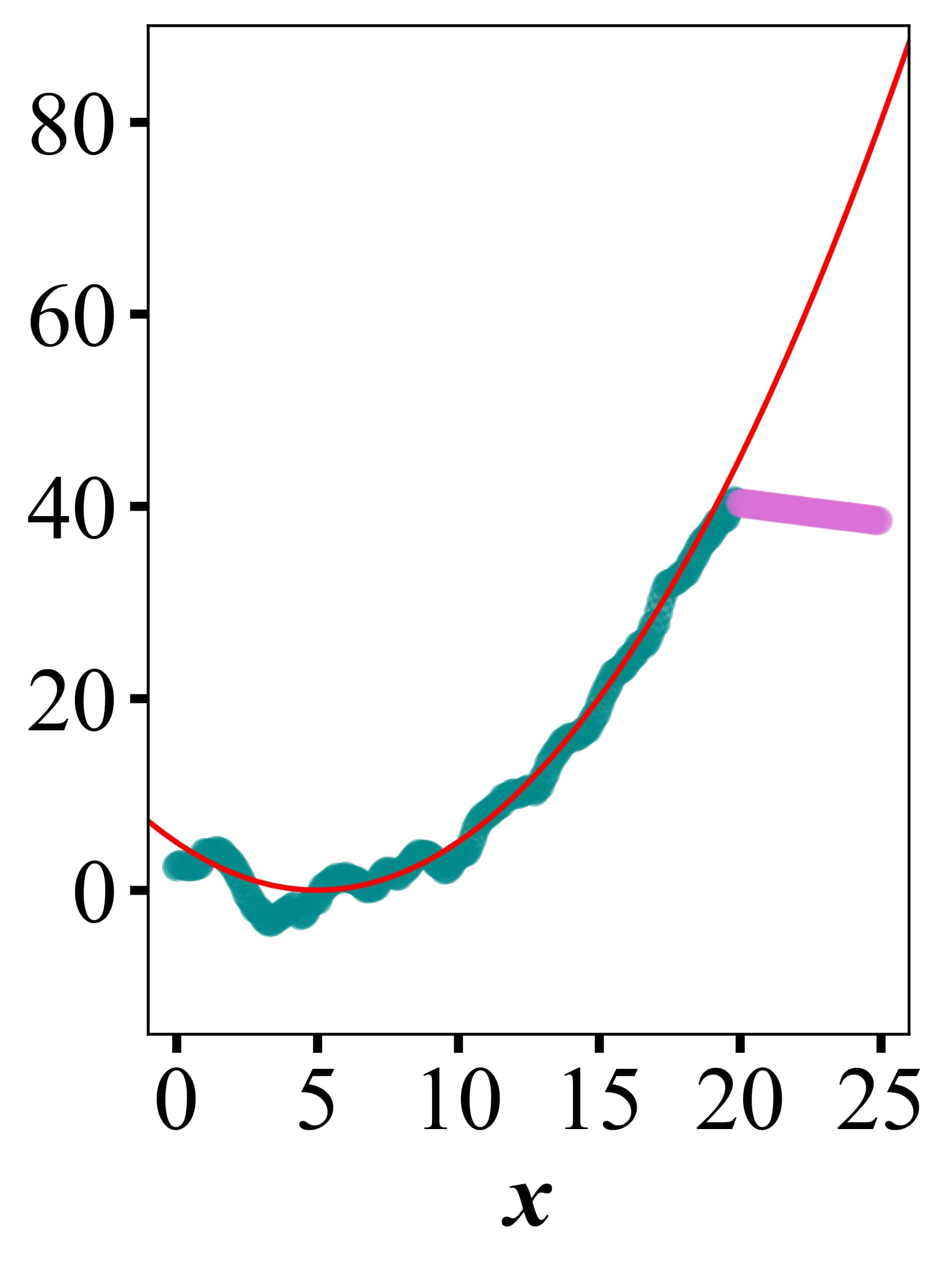}}
\centering 
\subfloat[\label{fig: toy problem quadratic xgbr}]{\includegraphics[width=0.19\textwidth]{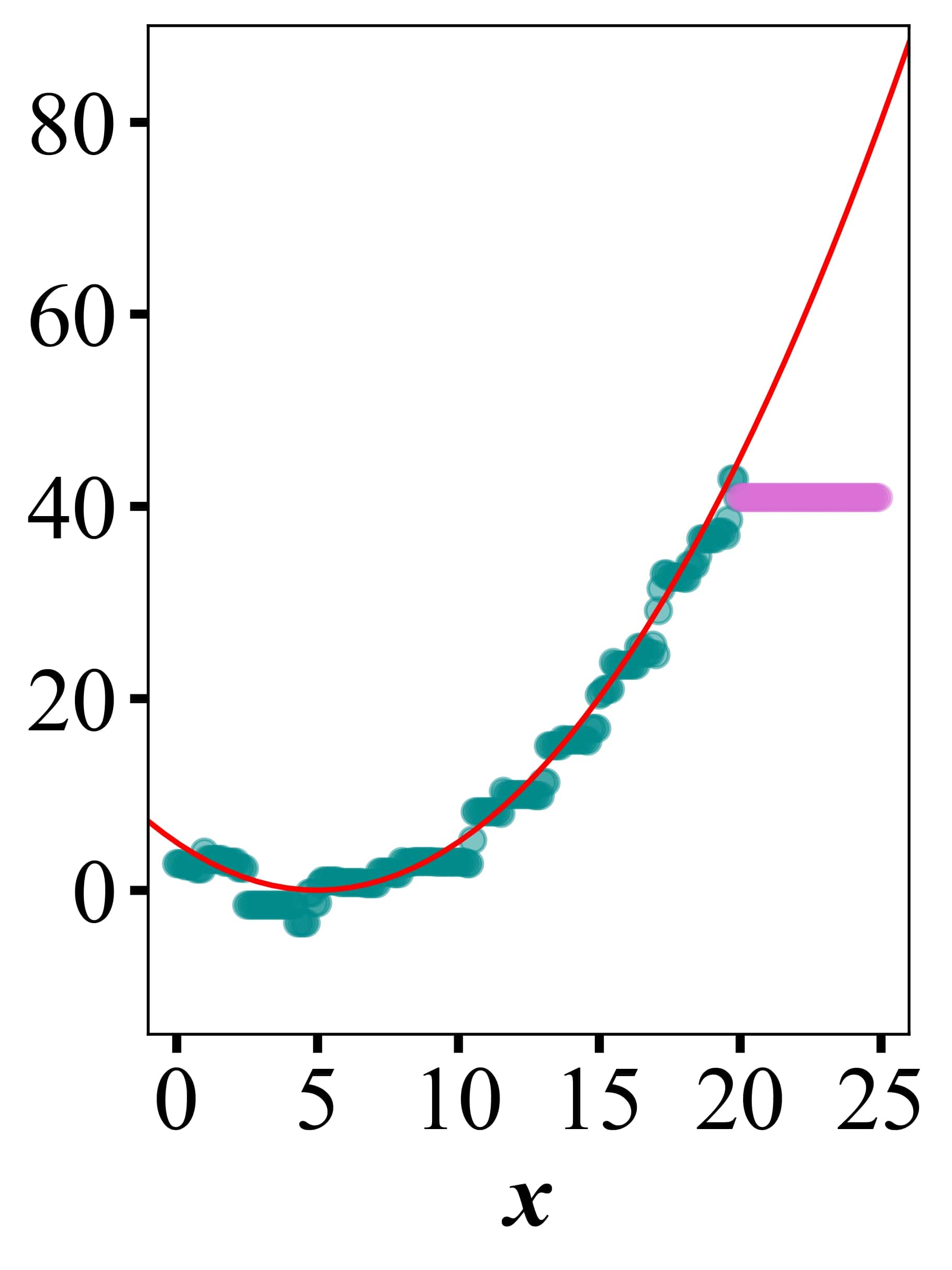}}
 \caption{\label{fig: toy problem solutions}
 A simple toy problem of \textbf{(a)} $y = 2x$ with the addition of Gaussian noise, where the label values of the training set are less than those of the test set.
 Shown are the predictions of three models on the training and test sets, developed with the following three popular machine learning algorithms: \textbf{(b)} ridge regression \textbf{(c)} kernel ridge regression and \textbf{(d)} eXtreme Gradient Boosting regression.
 Similarly, the exercise is repeated for \textbf{(e)} $y = 0.2(x - 5)^2$, with model predictions obtained using \textbf{(f)} ridge regression \textbf{(g)} kernel ridge regression \textbf{(h)} and eXtreme Gradient Boosting regression.
 }
\end{figure*}

The selection of ridge regression over other learning algorithms in our ML models is motivated by its ability to potentially make accurate predictions for samples with labels found beyond the distribution of the training set.
Consider the toy example of $y = 2x$ with the addition of Gaussian noise in \autoref{fig: toy problem linear}.
Here, the training set contains feature and label values that are both less than those found in the test set. 
Thus, with this OOD transfer learning problem, the performances of ML models trained with different learning algorithms (hyperparameters selected as those returning the lowest mean absolute error, MAE, upon five-fold cross-validation on the training set) are evaluated by their abilities to effectively extrapolate beyond their training set.\\

Ridge regression (\autoref{fig: toy problem linear ridge regression}) learns the linearity of the problem since for greater values of $x$, it correctly predicts correspondingly greater values of $y$ in both the training and test sets, although its accuracy could be improved. 
Models developed using two other popular and powerful ML algorithms of kernel ridge regression\cite{kernel_ridge_regression_notes_1} with the Laplacian kernel (\autoref{fig: toy problem linear kernel ridge regression}) and eXtreme Gradient Boosting\cite{XGBoost} (XGBoost) regression with the tree-based \texttt{gbtree} booster (\autoref{fig: toy problem linear xgbr}) exhibit, relative to the ridge regression model, improved performances on the training set. 
However, regarding the test set, both fail to make even a single prediction of $y$ that is greater than the maximum value found in the training set. 
This is, perhaps, an unsurprising result. 
After all, the objective of a kernel-based regression algorithm is to find a non-parametric mapping between the domain of feature vectors, $\textbf{X}$, and their labels, $\textbf{Y}$, where the estimated function outputs, $\hat{\textbf{Y}}$, are obtained as kernel-weighted local averages\cite{kernel_ridge_regression_notes_1, kernel_ridge_regression_notes_2, kernel_ridge_regression_notes_3, kernel_ridge_regression_notes_4}. 
Thus, this fundamentally restricts the value that a given $\hat{\textnormal{Y}}$ can take to within the range of values found in $\textbf{Y}_{\textnormal{train}}$.
The same conclusion results from a tree-based regression algorithm, which returns the weighted-average of the predictions obtained from an ensemble of decision trees\cite{trees_cant_extrapolate, random_forest_notes_1, random_forest_notes_2}.\\

In contrast, ridge regression weights a sample's feature values, which thus does not limit the model's output to within the distribution of $\textbf{Y}_{train}$. 
Therefore, ridge regression appears to be suitable for the objective of predicting label values beyond those found in the training set. 
Of course, before applying it to the prediction of out-of-distribution test samples, care must be taken to ensure that its estimates are robust and accurate since it requires the assumption that there exists some linear relationship between the features and labels space. 
If this condition is not fulfilled, then it can be seen from a quadratic toy problem of $y = 0.2(x - 5)^2$ (\autoref{fig: toy problem quadratic}) that ridge regression exhibits poor predictive power in both the train and test regimes due to the non-linearity of the problem. 
The fundamental limits on the labels space imposed by the different learning algorithms are observed again, since ridge regression can predict $y_{test}$ values greater than those found in training, albeit poorly in this case; kernel ridge regression (\autoref{fig: toy problem quadratic kernel ridge regression}) and XGBoost regression (\autoref{fig: toy problem quadratic xgbr}) exhibit good performances on the training set but they fail once more to predict a single value of $y_{test}$ beyond $\textbf{Y}_{train}$.

\clearpage

\subsection{\label{sec: SuperCon Pressure Samples}SuperCon Pressure Samples}

\begin{table}[h]
\begin{center}
\caption{\label{table: SuperCon Pressure Samples}
Compositions of materials in SuperCon that we have identified as having $T_c$'s measured under applied pressure.
}
\begin{tabular}{|p{5cm}|p{2cm}|} \textbf{Composition} & \textbf{$T_c$ (K)}\\
\hline 
\ce{IrU} & 0\\ 
\ce{SnTe} & 0\\ 
\ce{KC8} & 0\\
\ce{Zn2Zr} & 0\\ 
\ce{Ge2U} & 0\\
\ce{LaC2} & 1\\ 
\ce{Te2U} & 1\\ 
\ce{PdTe2} & 1\\ 
\ce{TiSe2} & 1\\ 
\ce{As2Cd3} & 3\\ 
\ce{OSn} & 3\\ 
\ce{S2Ta} & 3\\ 
\ce{BaBi3} & 6\\ 
\ce{B12Zr} & 6\\ 
\ce{C6Yb} & 6\\ 
\ce{Au2Pb} & 6\\ 
\ce{Te2W} & 7\\ 
\ce{Sb2Te3} & 7\\ 
\ce{NaAlSi} & 7\\ 
\ce{Mo3Sb7} & 7\\ 
\ce{MoTe2} & 8\\ 
\ce{B2Nb} & 9\\ 
\ce{C6Ca} & 11\\ 
\ce{H4Si} & 17\\ 
\ce{NNb} & 17\\ 
\ce{C3Y2} & 18\\ 
\ce{LiFeAs} & 18\\ 
\ce{Nb3Sn} & 18\\
\ce{GeNb3} & 22\\ 
\ce{NaFeAs} & 26\\ 
\ce{B2Mg} & 40\\ 
\ce{FeSe} & 80\\ 
\ce{Hg_{0.75}Ba_{2.07}Ca_{2.07}Cu_{3.11}O_{8.208}} & 135\\ 
\ce{Hg_{0.66}Pb_{0.34}Ba2Ca_{1.98}Cu_{2.9}O_{8.4}} & 143\\ 
\ce{H3S} & 147\\ 
\ce{H2S} & 203\\ 
\ce{LaH10} & 250\\

\end{tabular}
\end{center}
\end{table}
\clearpage

\subsection{\label{sec: Features Rankings}Features Rankings}

\begin{longtable}{
|>{\centering\arraybackslash\fontsize{10}{10}\selectfont}p{2.5cm}
|>{\centering\arraybackslash\fontsize{10}{10}\selectfont}p{7cm}
|>{\centering\arraybackslash\fontsize{10}{10}\selectfont}p{7cm}
|}
\caption{
Ranking of the relative average contribution of each ML feature to the $T_c$ predictions in the SuperCon implicit/ambient pressure models, from greatest to least.
}\\
\hline
\toprule
\thead{\fontsize{10}{10}\selectfont\textbf{Ranking}} &
\thead{\fontsize{10}{10}\selectfont\textbf{Implicit pressure}} &
\thead{\fontsize{10}{10}\selectfont\textbf{Ambient pressure}}\\
\midrule
\endfirsthead
\caption{(\textbf{Continued})}\\
\hline
\toprule
\thead{\fontsize{10}{10}\selectfont\textbf{Ranking}} &
\thead{\fontsize{10}{10}\selectfont\textbf{Implicit pressure}} &
\thead{\fontsize{10}{10}\selectfont\textbf{Ambient pressure}}\\
\midrule
\endhead
\hline
\multicolumn{3}{|r|}{{\textbf{Continued on next page}}} \\
\hline
\endfoot
\hline
\endlastfoot

       1 & Melting temperature, mean & Melting temperature, mean \\
       2 & Melting temperature, avg. dev. & Melting temperature, avg. dev. \\
       3 & Atomic weight, mean & Atomic weight, mean \\
       4 & Covalent radius, mean & Covalent radius, mean \\
       5 & Space group number, mean & Space group number, mean \\
       6 & Mendeleev number, mean & Mendeleev number, mean \\
       7 & Atomic weight, avg. dev. & Atomic weight, avg. dev. \\
       8 & Atomic number, mean & Atomic number, mean \\
       9 & Covalent radius, avg. dev. & Covalent radius, avg. dev. \\
      10 & Elemental solid volume, mean & Elemental solid volume, mean \\
      11 & Space group number, avg. dev. & Space group number, avg. dev. \\
      12 & Mendeleev number, avg. dev. & Mendeleev number, avg. dev. \\
      13 & Atomic number, avg. dev. & Atomic number, avg. dev. \\
      14 & Number of valence electrons, mean & Number of valence electrons, mean \\
      15 & Elemental solid volume, avg. dev. & Elemental solid volume, avg. dev. \\
      16 & Column number, mean & Column number, mean \\
      17 & Number of unfilled valence orbitals, mean & Number of unfilled valence orbitals, mean \\
      18 & Number of valence electrons, avg. dev. & Number of valence electrons, avg. dev.\\
      19 & Number of \textit{d} valence electrons, mean & Number of \textit{d} valence electrons, mean \\
      20 & Number of filled \textit{d} valence orbitals, mean & Number of filled \textit{d} valence orbitals, mean \\
      21 & Column number, avg. dev. & Column number, avg. dev. \\
      22 & Number of unfilled valence orbitals, avg. dev. & Number of unfilled valence orbitals, avg. dev. \\
      23 & Number of filled \textit{d} valence orbitals, avg. dev. & Number of filled \textit{d} valence orbitals, avg. dev. \\
      24 & Row number, mean & Row number, mean \\
      25 & Atomic weight, range & Atomic weight, range \\
      26 & Atomic weight, max. & Atomic weight, max. \\
      27 & Electronegativity, mean & Electronegativity, mean \\
      28 & Mendeleev number, range & Mendeleev number, range \\
      29 & Melting temperature, range & Melting temperature, range \\
      30 & 2-norm & 2-norm \\
      31 & Elemental solid volume, max. & Elemental solid volume, max. \\
      32 & Number of unfilled \textit{d} valence orbitals, mean & Number of unfilled \textit{d} valence orbitals, mean\\
      33 & 3-norm & 3-norm \\
      34 & Number of unfilled \textit{d} valence orbitals, avg. dev. & Number of unfilled \textit{d} valence orbitals, avg. dev. \\
      35 & Row number, avg. dev. & Row number, avg. dev. \\
      36 & 5-norm & 5-norm \\
      37 & Covalent radius, range & Covalent radius, range \\
      38 & 7-norm & 7-norm \\
      39 & 10-norm & 10-norm \\
      40 & Atomic number, range & Atomic number, range \\
      41 & Electronegativity, avg. dev. & Electronegativity, avg. dev. \\
      42 & Frac. \textit{d} valence electrons & Frac. \textit{d} valence electrons \\
      43 & Mendeleev number, min. & Mendeleev number, min. \\
      44 & Elemental solid volume, range & Elemental solid volume, range \\
      45 & Covalent radius, max. & Covalent radius, max. \\
      46 & Number of filled \textit{p} valence orbitals, mean & Number of filled \textit{p} valence orbitals, mean \\
      47 & Number of \textit{p} valence electrons, mean & Number of \textit{p} valence electrons, mean \\
      48 & Number of unfilled \textit{p} valence orbitals, mean & Number of unfilled \textit{p} valence orbitals \\
      49 & Frac. \textit{s} valence electrons & Frac. \textit{s} valence electrons \\
      50 & Atomic number, max. & Atomic number, max. \\
      51 & Number of filled \textit{p} valence orbitals, avg. dev. & Number of filled \textit{p} valence orbitals, avg. dev. \\
      52 & Number of filled \textit{s} valence orbitals, mean & Number of filled \textit{s} valence orbitals, mean \\
      53 & Number of \textit{s} valence electrons, mean & Number of \textit{s} valence electrons, mean \\
      54 & Number of valence electrons, range & Number of valence electrons, range \\
      55 & 0-norm & 0-norm \\
      56 & Number of unfilled \textit{p} valence orbitals, avg. dev. & Number of unfilled \textit{p} valence orbitals, avg. dev. \\
      57 & Number of unfilled valence orbitals, range & Number of unfilled valence orbitals, range \\
      58 & Atomic weight, min. & Atomic weight min. \\
      59 & Frac. \textit{p} valence electrons & Frac. \textit{p} valence electrons \\
      60 & Melting temperature, min. & Melting temperature, min.\\
      61 & Number of valence electrons, max. & Number of valence electrons, max. \\
      62 & Number of filled \textit{f} valence orbitals, avg. dev. & Number of filled \textit{f} valence orbitals, avg. dev. \\
      63 & Number of unfilled valence orbitals, max. & Number of unfilled valence orbitals, max. \\
      64 & Number of \textit{f} valence electrons, mean & Number of \textit{f} valence electrons, mean \\
      65 & Number of filled \textit{f} valence orbitals, mean & Number of filled \textit{f} valence orbitals, mean \\
      66 & Elemental solid volume, min. & Elemental solid volume, min. \\
      67 & Atomic weight, mode & Atomic weight, mode\\
      68 & Number of filled \textit{s} valence orbitals, avg. dev. & Number of filled \textit{s} valence orbitals, avg. dev. \\
      69 & Space group number, range & Space group number, rage \\
      70 & Transition metal fraction & Transition metal fraction \\
      71 & Number of unfilled \textit{s} valence orbitals, avg. dev. & Number of unfilled \textit{s} valence orbitals, avg. dev. \\
      72 & Covalent radius, min. & Covalent radius, min. \\
      73 & Melting temperature, max. & Melting temperature, max. \\
      74 & Electronegativity, range & Electronegativity, range \\
      75 & Mendeleev number, max. & Mendeleev number, max. \\
      76 & Number of unfilled \textit{s} valence orbitals, mean & Number of unfilled \textit{s} valence orbitals, mean \\
      77 & Atomic number, min. & Atomic number, min. \\
      78 & Electronegativity, min. & Electronegativity, min. \\
      79 & Space group number, max. & Space group number, max. \\
      80 & Elemental solid volume, mode & Elemental solid volume, mode \\
      81 & Column number, range & Column number, range \\
      82 & Melting temperature, mode & Melting temperature, mode \\
      83 & Mendeleev number, mode & Mendeleev number, mode \\
      84 & Number of unfilled \textit{d} valence orbitals, range & Number of unfilled \textit{d} valence orbitals, range \\
      85 & Number of valence electrons, min. & Number of valence electrons, min. \\
      86 & Space group number, min. & Number of unfilled \textit{d} valence orbitals, max. \\
      87 & Number of unfilled \textit{d} valence orbitals, max. & Space group number, min. \\
      88 & Number of filled \textit{d} valence orbitals, range & Number of filled \textit{d} valence orbitals, range \\
      89 & Frac. \textit{f} valence electrons & Frac. \textit{f} valence electrons \\
      90 & Covalent radius, mode & Covalent radius, mode \\
      91 & Space group number, mode & Space group number, mode \\
      92 & Electronegativity, max. & Electronegativity, max. \\
      93 & Atomic number, mode & Atomic number, mode \\
      94 & Number of unfilled valence orbitals, min. & Number of unfilled valence orbitals, min. \\
      95 & Column number, max. & Column number, max. \\
      96 & Column number, min. & Column number, min. \\
      97 & Number of valence electrons, mode & Number of valence electrons, mode \\
      98 & Electronegativity, mode & Electronegativity, mode \\
      99 & Column number, mode & Number of filled \textit{d} valence orbitals, max. \\
     100 & Number of filled \textit{d} valence orbitals, max. & Column number, mode \\
     101 & Number of unfilled \textit{p} valence orbitals, range & Number of unfilled \textit{p} valence orbitals, range \\
     102 & Number of unfilled \textit{p} valence orbitals, max. & Number of unfilled \textit{p} valence orbitals, max. \\
     103 & Elemental solid band gap, avg. dev. & Elemental solid band gap, avg. dev. \\
     104 & Elemental solid band gap, mean & Elemental solid band gap, mean \\
     105 & Number of unfilled \textit{f} valence orbitals, avg. dev. & Number of unfilled \textit{f} valence orbitals, avg. dev. \\
     106 & Number of unfilled \textit{f} valence orbitals, mean & Number of unfilled \textit{f} valence orbitals, mean \\
     107 & Row number, range & Row number, range \\
     108 & Number of filled \textit{f} valence orbitals, range & Number of filled \textit{f} valence orbitals, range \\
     109 & Number of unfilled valence orbitals, mode & Number of unfilled valence orbitals, mode\\
     110 & Number of filled \textit{f} valence orbitals, max. & Number of filled \textit{f} valence orbitals, max. \\
     111 & Row number, max. & Row number, max. \\
     112 & Elemental solid magnetic moment, avg. dev. & Elemental solid magnetic moment, avg. dev. \\
     113 & Row number, min. & Elemental solid magnetic moment, mean \\
     114 & Elemental solid magnetic moment, mean & Row number, min. \\
     115 & Elemental solid band gap, range & Elemental solid band gap, range \\
     116 & Elemental solid band gap, max. & Elemental solid band gap, max. \\
     117 & Number of filled \textit{s} valence orbitals, min. & Number of filled \textit{s} valence orbitals, min. \\
     118 & Number of filled \textit{p} valence orbitals, range & Number of filled \textit{p} valence orbitals, range \\
     119 & Number of filled \textit{p} valence orbitals, max. & Number of filled \textit{p} valence orbitals, max. \\
     120 & Number of unfilled \textit{f} valence orbitals, range & Number of unfilled \textit{f} valence orbitals, range \\
     121 & Number of unfilled \textit{f} valence orbitals, max. & Number of unfilled \textit{f} valence orbitals max. \\
     122 & Number of filled \textit{d} valence orbitals, min. & Number of filled \textit{d} valence orbitals, min. \\
     123 & Number of filled \textit{d} valence orbitals, mode & Number of filled \textit{d} valence orbitals, mode \\
     124 & Row number, mode & Row number, mode \\
     125 & Elemental solid magnetic moment max. & Elemental solid magnetic moment, max. \\
     126 & Elemental solid magnetic moment, range & Elemental solid magnetic moment, range \\
     127 & Number of filled \textit{s} valence orbitals, range & Number of filled \textit{s} valence orbitals, range \\
     128 & Number of unfilled \textit{d} valence orbitals, mode & Number of unfilled \textit{d} valence orbitals, mode \\
     129 & Number of unfilled \textit{p} valence orbitals, mode & Number of unfilled \textit{p} valence orbitals, mode \\
     130 & Elemental solid band gap, mode & Elemental solid band gap, mode \\
     131 & Number of filled \textit{s} valence orbitals, mode & Number of filled \textit{s} valence orbitals, mode \\
     132 & Number of filled \textit{p} valence orbitals, mode & Number of unfilled \textit{s} valence orbitals, range \\
     133 & Number of unfilled \textit{s} valence orbitals, range & Number of filled \textit{p} valence orbitals, mode \\
     134 & Number of unfilled \textit{d} valence orbitals, min. & Number of unfilled \textit{d} valence orbitals, min. \\
     135 & Number of unfilled \textit{s} valence orbitals, max. & Number of unfilled \textit{s} valence orbitals, max. \\
     136 & Number of filled \textit{f} valence orbitals, mode & Number of filled \textit{f} valence orbitals, mode \\
     137 & Number of filled \textit{s} valence orbitals, max. & Number of filled \textit{s} valence orbitals, max. \\
     138 & Number of unfilled \textit{s} valence orbitals, mode & Number of unfilled \textit{s} valence orbitals, mode \\
     139 & Number of unfilled \textit{p} valence orbitals, min. & Number of unfilled \textit{p} valence orbitals, min. \\
     140 & Number of filled \textit{p} valence orbitals min. & Number of filled \textit{p} valence orbitals, min. \\
     141 & Elemental solid magnetic moment, mode & Elemental solid magnetic moment, mode \\
     142 & Number of filled \textit{f} valence orbitals, min. & Number of filled \textit{f} valence orbitals, min. \\
     143 & Number of unfilled \textit{s} valence orbitals, min. & Number of unfilled \textit{s} valence orbitals, min. \\
     144 & Elemental solid band gap, min. & Elemental solid band gap min.\\
     145 & Number of unfilled \textit{f} valence orbitals, mode & Number of unfilled \textit{f} valence orbitals, mode \\
     146 & Elemental solid magnetic moment, min. & Elemental solid magnetic moment, min. \\
     147 & Number of unfilled \textit{f} valence orbitals, min. & Number of unfilled \textit{f} valence orbitals, min. \\

\end{longtable}

\clearpage

\subsection{\label{sec: Materials Project, Ambient Pressure}Materials Project, Ambient Pressure}

\begin{longtable}{
|>{\centering\arraybackslash\fontsize{10}{11}\selectfont}p{1cm}
|>{\arraybackslash\fontsize{10}{11}\selectfont}p{3cm}
|>{\centering\arraybackslash\fontsize{10}{11}\selectfont}p{3.5cm}
|>{\centering\arraybackslash\fontsize{10}{11}\selectfont}p{2cm}
|>{\centering\arraybackslash\fontsize{10}{11}\selectfont}p{2.5cm}
|>{\centering\arraybackslash\fontsize{10}{11}\selectfont}p{2.3cm}
|}
\caption{
Top 100 materials in the Materials Project with the highest $T_c$ predictions using ML model trained on ambient pressure data only, dot product spreads of less than 50 K, and energies above their convex hulls of less than 0.030 eV/atom.
}\\
\hline
\toprule
\thead{\fontsize{10}{10}\selectfont\textbf{}} &
\thead{\fontsize{10}{10}\selectfont\textbf{ID}} &
\thead{\fontsize{10}{10}\selectfont\textbf{Composition}} &
\thead{\fontsize{10}{10}\selectfont\textbf{Predicted}\\ \fontsize{10}{10}\selectfont\textbf{$T_c$ (K)}} & 
\thead{\fontsize{10}{10}\selectfont\textbf{Dot product}\\ \fontsize{10}{10}\selectfont\textbf{spread (K)}} &
\thead{\fontsize{10}{10}\selectfont\textbf{Hull energy}\\ \fontsize{10}{10}\selectfont\textbf{(eV/atom)}}\\
\midrule
\endfirsthead
\caption{(\textbf{Continued})}\\
\hline
\toprule
\thead{\fontsize{10}{10}\selectfont\textbf{}} &
\thead{\fontsize{10}{10}\selectfont\textbf{ID}} &
\thead{\fontsize{10}{10}\selectfont\textbf{Composition}} &
\thead{\fontsize{10}{10}\selectfont\textbf{Predicted}\\ \fontsize{10}{10}\selectfont\textbf{$T_c$ (K)}} & 
\thead{\fontsize{10}{10}\selectfont\textbf{Dot product}\\ \fontsize{10}{10}\selectfont\textbf{spread (K)}} &
\thead{\fontsize{10}{10}\selectfont\textbf{Hull energy}\\ \fontsize{10}{10}\selectfont\textbf{(eV/atom)}}\\
\midrule
\endhead
\hline
\multicolumn{6}{|r|}{{\textbf{Continued on next page}}} \\
\hline
\endfoot
\hline
\endlastfoot

       1 &  mp-765341, mp-753273, mp-753257 &           \ce{LiCuF4} & 316 &           46 &        0.027 \\
       2 &  mp-723002 &     \ce{Ag2H12S(NO)4} & 315 &           49 &        0.027 \\
       3 &  mp-632760 &        \ce{Na2H6PtO6} & 315 &           36 &        0.000 \\
       4 &  mp-720197 &       \ce{K4GeH4N2O3} & 314 &           49 &        0.000 \\
       5 &    mp-8090, mp-1225821, mp-3638 &          \ce{Cu2P2O7} & 310 &           45 &        0.002 \\
       6 &  mp-567820 &        \ce{K6NbTlAs4} & 307 &           39 &        0.000 \\
       7 & mp-1204687 &        \ce{Cu3P2H2O9} & 306 &           44 &        0.016 \\
       8 &   mp-26304 &        \ce{Cu9(PO4)8} & 305 &           46 &        0.018 \\
       9 & mp-1190512 &         \ce{Na2UTeO7} & 302 &           48 &        0.000 \\
      10 &  mp-721293 & \ce{Na4GeH28(Se2O7)2} & 302 &           47 &        0.021 \\
      11 &  mp-722266 &    \ce{Na2GeH16Se3O8} & 301 &           48 &        0.020 \\
      12 & mp-1023132 &       \ce{K2Cu(ClF)2} & 301 &           38 &        0.000 \\
      13 & mp-1214168 &         \ce{Cs2CoBr4} & 300 &           38 &        0.000 \\
      14 & mp-1201159 &   \ce{Na2MnH4(SeO5)2} & 297 &           31 &        0.000 \\
      15 &  mp-744084 &      \ce{MnAgH8(OF)4} & 296 &           34 &        0.011 \\
      16 &  mp-765413 &          \ce{Li2AgF6} & 295 &           32 &        0.000 \\
      17 & mp-1218805 &     \ce{Sr2LaGa11O20} & 294 &           49 &        0.015 \\
      18 &  mp-756711, mp-772434 &          \ce{Li4CrO4} & 293 &           32 &        0.018 \\
      19 & mp-1113451 &       \ce{Cs2ScAgBr6} & 293 &           37 &        0.023 \\
      20 & mp-1147629 &          \ce{Cu3PF12} & 292 &           47 &        0.005 \\
      21 & mp-1369804 &           \ce{ZnCuF5} & 291 &           48 &        0.000 \\
      22 &  mp-560555 &      \ce{Li8Bi2PdO10} & 290 &           48 &        0.000 \\
      23 & mp-1112081 &        \ce{K2TlAgCl6} & 289 &           43 &        0.014 \\
      24 & mp-1223641, mp-1223639 &         \ce{K2MgCuF6} & 288 &           37 &        0.022 \\
      25 & mp-1221355, mp-10340 &        \ce{Na2GeTeO6} & 285 &           34 &        0.001 \\
      26 &  mp-556946 &    \ce{Rb2V2Cu(PO6)2} & 282 &           47 &        0.004 \\
      27 & mp-1220230 &  \ce{Rb2MnP2H3(O4F)2} & 281 &           29 &        0.000 \\
      28 &  mp-697075 &          \ce{Ti3Zn3N} & 281 &           16 &        0.000 \\
      29 & mp-1213351 &          \ce{CsFeBr4} & 280 &           37 &        0.000 \\
      30 &   mp-20982 &          \ce{K3Cu2F7} & 279 &           38 &        0.019 \\
      31 &   mp-18782, mp-757614, mp-756497, mp-755709 &           \ce{LiFeO2} & 278 &           37 &        0.020 \\
      32 &  mp-640926, mp-662543 &         \ce{Cu(NO3)2} & 276 &           44 &        0.002 \\
      33 &  mp-551101 &         \ce{HoBi2IO4} & 274 &           39 &        0.000 \\
      34 & mp-1178131, mp-760786 &        \ce{Li2Cu3F11} & 272 &           49 &        0.028 \\
      35 & mp-1195234 &        \ce{KP2H4PtO9} & 271 &           43 &        0.000 \\
      36 &  mp-757347, mp-766878 &      \ce{Li13Ni15O28} & 267 &           37 &        0.010 \\
      37 & mp-1222038 &      \ce{MgCu2(PO4)2} & 267 &           48 &        0.009 \\
      38 &  mp-546625 &        \ce{HoBi2BrO4} & 264 &           39 &        0.000 \\
      39 &   mp-35799 &         \ce{Hf2Bi2O7} & 264 &           40 &        0.000 \\
      40 & mp-1111629 &       \ce{Rb2TlAgCl6} & 263 &           44 &        0.000 \\
      41 &  mp-549127 &        \ce{HoBi2ClO4} & 259 &           39 &        0.000 \\
      42 &  mp-760678 &       \ce{Sb2H4AuF16} & 258 &           39 &        0.000 \\
      43 & mp-1192000 &          \ce{CsH3AuN} & 257 &           49 &        0.026 \\
      44 &  mp-636950, mp-504196, mp-25266, mp-1101589 &         \ce{Mo2P2O11} & 256 &           40 &        0.016 \\
      45 & mp-1224553 &  \ce{K8Cu9S8(ClO18)2} & 255 &           36 &        0.010 \\
      46 & mp-1223920 &        \ce{K3MgCu2F9} & 255 &           42 &        0.006 \\
      47 & mp-1209342 &       \ce{Rb5Tm3Br12} & 254 &           35 &        0.024 \\
      48 & mp-1198418 &        \ce{TaTe4Cl4O} & 250 &           46 &        0.000 \\
      49 & mp-1232302, mp-568857, mp-571305, mp-23055 &          \ce{RbCuCl3} & 248 &           46 &        0.001 \\
      50 & mp-1209376 &         \ce{RbAlCuF6} & 248 &           37 &        0.000 \\
      51 &  mp-604273 &   \ce{Cu2GeP2(H5O7)2} & 246 &           36 &        0.011 \\
      52 &  mp-556787 &       \ce{K2Cu3S3O13} & 245 &           35 &        0.000 \\
      53 &  mp-556333 &    \ce{K2ZnCu3P3O12F} & 244 &           36 &        0.000 \\
      54 &  mp-559539 &       \ce{Li2U3P2O15} & 243 &           46 &        0.000 \\
      55 &  mp-568928 &          \ce{Rb2FeI4} & 243 &           37 &        0.018 \\
      56 & mp-1205840 &         \ce{Cs2TcCl6} & 243 &           21 &        0.000 \\
      57 &  mp-554726 &      \ce{KNaCu3S3O13} & 240 &           35 &        0.000 \\
      58 &    mp-3237 &          \ce{Na2CuF4} & 239 &           36 &        0.000 \\
      59 &  mp-759552 &   \ce{Li2CrP4(H4O5)4} & 237 &           30 &        0.015 \\
      60 &   mp-18124 &          \ce{KAlCuF6} & 237 &           37 &        0.003 \\
      61 & mp-1190440 &   \ce{Ho2CuTe2(SO7)2} & 237 &           43 &        0.015 \\
      62 &  mp-583853, mp-26317, mp-756865, mp-25447, mp-26579, mp-1666120 &          \ce{LiCrPO4} & 236 &           26 &        0.023 \\
      63 & mp-1223957, mp-1147665, mp-1224040 &       \ce{K4MgCu3F12} & 236 &           43 &        0.007 \\
      64 & mp-1110598 &        \ce{Rb2TlAgF6} & 235 &           40 &        0.000 \\
      65 &  mp-558998 &     \ce{K2Cu(Si2O5)2} & 234 &           35 &        0.002 \\
      66 &  mp-695830 &       \ce{Na2Cu(HO)4} & 234 &           34 &        0.025 \\
      67 & mp-1113442 &       \ce{Cs2TlAgCl6} & 233 &           45 &        0.000 \\
      68 & mp-1198285 &  \ce{Cu3P4H8N2(O3F)4} & 232 &           36 &        0.000 \\
      69 &   mp-20076 &            \ce{KCrF6} & 230 &           47 &        0.000 \\
      70 & mp-1210812 &  \ce{Mn4Zn3H10(CO8)2} & 230 &           41 &        0.009 \\
      71 &  mp-722937 &          \ce{Zr3Zn3N} & 229 &           19 &        0.000 \\
      72 & mp-1110608 &        \ce{Rb2InAgF6} & 229 &           40 &        0.000 \\
      73 & mp-1211966 &          \ce{K3GdCl6} & 228 &           47 &        0.016 \\
      74 & mp-1202248 &  \ce{ReH30Ru2(NCl)10} & 228 &           31 &        0.029 \\
      75 &  mp-574909 &           \ce{K2NbS7} & 227 &           27 &        0.003 \\
      76 & mp-1223772 &      \ce{Na4U3Te5O21} & 226 &           47 &        0.021 \\
      77 & mp-1113205 &       \ce{Cs2TlCuCl6} & 224 &           39 &        0.011 \\
      78 & mp-1539446, mp-29216 &         \ce{Cs2PtCl4} & 224 &           23 &        0.007 \\
      79 & mp-1112940 &       \ce{Cs2AgSbCl6} & 223 &           38 &        0.000 \\
      80 & mp-1211993 &           \ce{K2FeI4} & 222 &           37 &        0.000 \\
      81 & mp-1113150 &        \ce{Cs2TlAgF6} & 222 &           42 &        0.000 \\
      82 & mp-1640489, mp-510562, mp-762493 &        \ce{KMnP3HO10} & 221 &           26 &        0.001 \\
      83 &  mp-555369 &         \ce{SrCuH6O5} & 220 &           23 &        0.015 \\
      84 & mp-1113588 &        \ce{Cs2InAgF6} & 219 &           39 &        0.000 \\
      85 & mp-1197826 &      \ce{K3VMo3H4O15} & 219 &           34 &        0.024 \\
      86 &   mp-16801 &           \ce{GdSO4F} & 219 &           43 &        0.000 \\
      87 &   mp-29051 &           \ce{WS9Cl4} & 218 &           32 &        0.000 \\
      88 & mp-1191076, mp-28020 &          \ce{AlCuCl4} & 218 &           29 &        0.002 \\
      89 & mp-1147709 &          \ce{Sr2CuH6} & 216 &           47 &        0.000 \\
      90 & mp-1194841 &       \ce{SrCuH3ClO3} & 215 &           41 &        0.024 \\
      91 &  mp-758903 &    \ce{Li2VP4(H4O5)4} & 215 &           31 &        0.015 \\
      92 & mp-1113979 &       \ce{Rb2LiGdCl6} & 213 &           48 &        0.009 \\
      93 &  mp-755821 &      \ce{Li5Cu(PO4)2} & 213 &           27 &        0.029 \\
      94 &  mp-557826 &      \ce{Li3U7(PO7)5} & 213 &           45 &        0.004 \\
      95 &  mp-558875 &        \ce{HoTe2ClO5} & 213 &           44 &        0.000 \\
      96 &  mp-570831 &          \ce{U3NbSb5} & 213 &           42 &        0.000 \\
      97 &    mp-6118 &     \ce{Ca3Cu3(PO4)4} & 210 &           49 &        0.000 \\
      98 & mp-1104879 &        \ce{Li2H6PtO6} & 209 &           24 &        0.028 \\
      99 & mp-2646913 &    \ce{Cs4Mn(SbCl6)2} & 209 &           22 &        0.000 \\
     100 & mp-1110690 &       \ce{Rb2CuBiCl6} & 205 &           39 &        0.000 \\

\end{longtable}

\clearpage

\subsection{\label{sec: Materials Project, Ambient Pressure, Small Band Gaps}Materials Project, Ambient Pressure, Small Band Gaps}

\begin{longtable}{
|>{\centering\arraybackslash\fontsize{10}{11}\selectfont}p{1cm}
|>{\arraybackslash\fontsize{10}{11}\selectfont}p{3cm}
|>{\centering\arraybackslash\fontsize{10}{11}\selectfont}p{3.5cm}
|>{\centering\arraybackslash\fontsize{10}{11}\selectfont}p{2cm}
|>{\centering\arraybackslash\fontsize{10}{11}\selectfont}p{2.5cm}
|>{\centering\arraybackslash\fontsize{10}{11}\selectfont}p{2.3cm}
|>{\centering\arraybackslash\fontsize{10}{11}\selectfont}p{2.0cm}
|}
\caption{
Top 100 materials in the Materials Project with the highest $T_c$ predictions using ML model trained on ambient pressure data only, dot product spreads of less than 50 K, energies above their convex hulls of less than 0.030 eV/atom, and band gaps of less than 1.000 eV.
}\\
\hline
\toprule
\thead{\fontsize{10}{10}\selectfont\textbf{}} &
\thead{\fontsize{10}{10}\selectfont\textbf{ID}} &
\thead{\fontsize{10}{10}\selectfont\textbf{Composition}} &
\thead{\fontsize{10}{10}\selectfont\textbf{Predicted}\\ \fontsize{10}{10}\selectfont\textbf{$T_c$ (K)}} & 
\thead{\fontsize{10}{10}\selectfont\textbf{Dot product}\\ \fontsize{10}{10}\selectfont\textbf{spread (K)}} &
\thead{\fontsize{10}{10}\selectfont\textbf{Hull energy}\\ \fontsize{10}{10}\selectfont\textbf{(eV/atom)}} & 
\thead{\fontsize{10}{10}\selectfont\textbf{Band gap}\\ \fontsize{10}{10}\selectfont\textbf{(eV)}}\\
\midrule
\endfirsthead
\caption{(\textbf{Continued})}\\
\hline
\toprule
\thead{\fontsize{10}{10}\selectfont\textbf{}} &
\thead{\fontsize{10}{10}\selectfont\textbf{ID}} &
\thead{\fontsize{10}{10}\selectfont\textbf{Composition}} &
\thead{\fontsize{10}{10}\selectfont\textbf{Predicted}\\ \fontsize{10}{10}\selectfont\textbf{$T_c$ (K)}} & 
\thead{\fontsize{10}{10}\selectfont\textbf{Dot product}\\ \fontsize{10}{10}\selectfont\textbf{spread (K)}} &
\thead{\fontsize{10}{10}\selectfont\textbf{Hull energy}\\ \fontsize{10}{10}\selectfont\textbf{(eV/atom)}} & 
\thead{\fontsize{10}{10}\selectfont\textbf{Band gap}\\ \fontsize{10}{10}\selectfont\textbf{(eV)}}\\
\midrule
\endhead
\hline
\multicolumn{7}{|r|}{{\textbf{Continued on next page}}} \\
\hline
\endfoot
\hline
\endlastfoot
       1 &  mp-765341, mp-753257, mp-753273 &           \ce{LiCuF4} & 316 &           46 &        0.027 &     0.000 \\
       2 &    mp-3638, mp-1225821, mp-8090 &          \ce{Cu2P2O7} & 310 &           45 &        0.000 &     0.000 \\
       3 & mp-1204687 &        \ce{Cu3P2H2O9} & 306 &           44 &        0.016 &     0.000 \\
       4 &   mp-26304 &        \ce{Cu9(PO4)8} & 305 &           46 &        0.018 &     0.000 \\
       5 & mp-1023132 &       \ce{K2Cu(ClF)2} & 301 &           38 &        0.000 &     0.300 \\
      6 & mp-1214168 &         \ce{Cs2CoBr4} & 300 &           38 &        0.000 &     0.698 \\
      7 &  mp-744084 &      \ce{MnAgH8(OF)4} & 296 &           34 &        0.011 &     0.302 \\
      8 &  mp-765413 &          \ce{Li2AgF6} & 295 &           32 &        0.000 &     0.000 \\
      8 & mp-1147629 &          \ce{Cu3PF12} & 292 &           47 &        0.005 &     0.000 \\
      10 & mp-1369804 &           \ce{ZnCuF5} & 291 &           48 &        0.000 &     0.000 \\
      11 &  mp-560555 &      \ce{Li8Bi2PdO10} & 290 &           48 &        0.000 &     0.743 \\
      12 & mp-1112081 &        \ce{K2TlAgCl6} & 289 &           43 &        0.014 &     0.000 \\
      13 & mp-1223639, mp-1223641 &         \ce{K2MgCuF6} & 288 &           37 &        0.009 &     0.000 \\
      14 &  mp-697075 &          \ce{Ti3Zn3N} & 281 &           16 &        0.000 &     0.000 \\
      15 &   mp-20982 &          \ce{K3Cu2F7} & 279 &           38 &        0.019 &     0.000 \\
      16 &   mp-18782 &           \ce{LiFeO2} & 278 &           37 &        0.020 &     0.000 \\
      17 &  mp-662543, mp-640926 &         \ce{Cu(NO3)2} & 276 &           44 &        0.000 &     0.565 \\
      18 &  mp-760786, mp-1178131 &        \ce{Li2Cu3F11} & 272 &           49 &        0.028 &     0.000 \\
      19 &  mp-766878, mp-757347 &      \ce{Li13Ni15O28} & 267 &           37 &        0.010 &     0.000 \\
      20 & mp-1222038 &      \ce{MgCu2(PO4)2} & 267 &           48 &        0.009 &     0.309 \\
      21 & mp-1111629 &       \ce{Rb2TlAgCl6} & 263 &           44 &        0.000 &     0.000 \\
      22 &  mp-760678 &       \ce{Sb2H4AuF16} & 258 &           39 &        0.000 &     0.894 \\
      23 & mp-1192000 &          \ce{CsH3AuN} & 257 &           49 &        0.026 &     0.886 \\
      24 & mp-1224553 &  \ce{K8Cu9S8(ClO18)2} & 255 &           36 &        0.010 &     0.047 \\
      25 & mp-1223920 &        \ce{K3MgCu2F9} & 255 &           42 &        0.006 &     0.000 \\
      26 & mp-1209342 &       \ce{Rb5Tm3Br12} & 254 &           35 &        0.024 &     0.000 \\
      27 &   mp-23055, mp-571305, mp-568857, mp-1232302 &          \ce{RbCuCl3} & 248 &           46 &        0.000 &     0.186 \\
      28 & mp-1209376 &         \ce{RbAlCuF6} & 248 &           37 &        0.000 &     0.000 \\
      29 &  mp-604273 &   \ce{Cu2GeP2(H5O7)2} & 246 &           36 &        0.011 &     0.314 \\
      30 &  mp-556787 &       \ce{K2Cu3S3O13} & 245 &           35 &        0.000 &     0.141 \\
      31 &  mp-556333 &    \ce{K2ZnCu3P3O12F} & 244 &           36 &        0.000 &     0.000 \\
      32 &  mp-568928 &          \ce{Rb2FeI4} & 243 &           37 &        0.018 &     0.000 \\
      33 &  mp-554726 &      \ce{KNaCu3S3O13} & 240 &           35 &        0.000 &     0.156 \\
      34 &    mp-3237 &          \ce{Na2CuF4} & 239 &           36 &        0.000 &     0.649 \\
      35 &  mp-759552 &   \ce{Li2CrP4(H4O5)4} & 237 &           30 &        0.015 &     0.326 \\
      36 &   mp-18124 &          \ce{KAlCuF6} & 237 &           37 &        0.003 &     0.081 \\
      37 & mp-1190440 &   \ce{Ho2CuTe2(SO7)2} & 237 &           43 &        0.015 &     0.866 \\
      38 & mp-1224040, mp-1223957, mp-1147665 &       \ce{K4MgCu3F12} & 236 &           43 &        0.014 &     0.000 \\
      39 & mp-1110598 &        \ce{Rb2TlAgF6} & 235 &           40 &        0.000 &     0.552 \\
      40 &  mp-695830 &       \ce{Na2Cu(HO)4} & 234 &           34 &        0.025 &     0.000 \\
      41 & mp-1113442 &       \ce{Cs2TlAgCl6} & 233 &           45 &        0.000 &     0.000 \\
      42 & mp-1198285 &  \ce{Cu3P4H8N2(O3F)4} & 232 &           36 &        0.000 &     0.622 \\
      43 &  mp-722937 &          \ce{Zr3Zn3N} & 229 &           19 &        0.000 &     0.000 \\
      44 & mp-1202248 &  \ce{ReH30Ru2(NCl)10} & 228 &           31 &        0.029 &     0.031 \\
      45 & mp-1113205 &       \ce{Cs2TlCuCl6} & 224 &           39 &        0.011 &     0.000 \\
      46 & mp-1211993 &           \ce{K2FeI4} & 222 &           37 &        0.000 &     0.000 \\
      47 & mp-1113150 &        \ce{Cs2TlAgF6} & 222 &           42 &        0.000 &     0.805 \\
      48 &  mp-762493, mp-510562 &        \ce{KMnP3HO10} & 221 &           26 &        0.001 &     0.716 \\
      49 &  mp-555369 &         \ce{SrCuH6O5} & 220 &           23 &        0.015 &     0.652 \\
      50 &   mp-16801 &           \ce{GdSO4F} & 219 &           43 &        0.000 &     0.223 \\
      51 & mp-1147709 &          \ce{Sr2CuH6} & 216 &           47 &        0.000 &     0.000 \\
      52 & mp-1194841 &       \ce{SrCuH3ClO3} & 215 &           41 &        0.024 &     0.545 \\
      53 &  mp-570831 &          \ce{U3NbSb5} & 213 &           42 &        0.000 &     0.000 \\
      54 &    mp-6118 &     \ce{Ca3Cu3(PO4)4} & 210 &           49 &        0.000 &     0.648 \\
      55 & mp-1220937 &           \ce{NaU2F9} & 205 &           49 &        0.000 &     0.000 \\
      56 &  mp-555139 & \ce{Zr2Cu3H32(O8F7)2} & 201 &           37 &        0.012 &     0.698 \\
      57 &  mp-606617 &     \ce{CuSb2(XeF8)2} & 200 &           40 &        0.000 &     0.069 \\
      58 &   mp-28881 &          \ce{Rb2UCl5} & 199 &           42 &        0.000 &     0.116 \\
      59 &   mp-24720 &             \ce{PuH2} & 198 &           37 &        0.000 &     0.000 \\
      60 & mp-1202916 &         \ce{U3H2OF12} & 198 &           43 &        0.007 &     0.025 \\
      61 &  mp-561012, mp-560776 &     \ce{Sr3Cu3(PO4)4} & 197 &           42 &        0.009 &     0.308 \\
      62 &  mp-504759 &             \ce{MoF5} & 194 &           39 &        0.000 &     0.991 \\
      63 &  mp-555595 &       \ce{KCu3S2ClO9} & 194 &           47 &        0.013 &     0.000 \\
      64 &  mp-697458 &      \ce{KNiH9(CO5)2} & 192 &           44 &        0.009 &     0.000 \\
      65 &   mp-26223 &      \ce{Li4Cu(PO4)2} & 192 &           27 &        0.026 &     0.040 \\
      66 & mp-1226426 &     \ce{Cu12Mo3S5O36} & 190 &           42 &        0.023 &     0.000 \\
      67 &  mp-618309 &          \ce{Bi2AuO5} & 190 &           42 &        0.000 &     0.000 \\
      68 &    mp-9640 &      \ce{SrCu(SeO3)2} & 188 &           42 &        0.000 &     0.306 \\
      69 &  mp-647814 &             \ce{U2O5} & 186 &           38 &        0.019 &     0.000 \\
      70 & mp-1222873, mp-1302787 &     \ce{Li2Al(NiO2)3} & 184 &           39 &        0.020 &     0.000 \\
      71 &  mp-669347, mp-1080828, mp-3682 &            \ce{KCuF3} & 178 &           49 &        0.000 &     0.000 \\
      72 &  mp-570955 &          \ce{Na2UCl6} & 178 &           36 &        0.000 &     0.000 \\
      73 &   mp-19308 &          \ce{Li2NiO2} & 176 &           49 &        0.003 &     0.912 \\
      74 &   mp-17631 &      \ce{Fe2Ni(PO4)2} & 174 &           38 &        0.000 &     0.000 \\
      75 &    mp-9360 &        \ce{CaU(PO4)2} & 174 &           44 &        0.000 &     0.000 \\
      76 &  mp-676749 &          \ce{Li2UCl6} & 172 &           38 &        0.012 &     0.198 \\
      77 &  mp-556074 &     \ce{K2Cu3(SeO3)4} & 171 &           38 &        0.000 &     0.091 \\
      78 & mp-2646929 &         \ce{Ce(BH4)3} & 170 &           14 &        0.000 &     0.344 \\
      79 & mp-1190515 &   \ce{Gd2CuTe2(SO7)2} & 169 &           47 &        0.000 &     0.011 \\
      80 &   mp-14629 &        \ce{Li2EuSiO4} & 168 &           46 &        0.000 &     0.043 \\
      81 &    mp-5268 &           \ce{U2SbN2} & 167 &           35 &        0.000 &     0.000 \\
      82 & mp-1096938 &          \ce{Cs2ReO4} & 166 &           25 &        0.000 &     0.000 \\

     83 &  mp-556303, mp-556635 &        \ce{CuBi2(SeO3)4} & 166 &           45 &        0.000 &     0.000 \\
     84 & mp-1068157 &            \ce{Sr2CdPt2} & 165 &           27 &        0.000 &     0.000 \\
     85 & mp-1147553 &              \ce{KCuIO6} & 163 &           21 &        0.000 &     0.000 \\
     86 &   mp-31058 &              \ce{Er3InN} & 163 &           37 &        0.000 &     0.000 \\
     87 &   mp-11964 &            \ce{Ce2CdPt2} & 162 &           30 &        0.000 &     0.000 \\
     88 &  mp-554714 &               \ce{Np2O5} & 161 &           36 &        0.000 &     0.000 \\
     89 &   mp-10950 &         \ce{Sr2Cu(PO4)2} & 161 &           40 &        0.000 &     0.000 \\
     90 &   mp-27382 &               \ce{UTlF5} & 161 &           42 &        0.000 &     0.266 \\
     91 & mp-1022726 &             \ce{Ba2CuH6} & 159 &           34 &        0.023 &     0.000 \\
     92 &  mp-755363 &              \ce{Ba3SiO} & 158 &           43 &        0.000 &     0.422 \\
     93 &   mp-23108 &           \ce{Cs2NaUCl6} & 157 &           20 &        0.000 &     0.500 \\
     94 & mp-1217343 &             \ce{ThU4O10} & 157 &           23 &        0.021 &     0.000 \\
     95 &    mp-5787, mp-37514 &              \ce{SrCuO2} & 157 &           43 &        0.000 &     0.000 \\
     96 &  mp-677724 &            \ce{U8Bi2O19} & 156 &           26 &        0.000 &     0.000 \\
     97 & mp-1215989 &            \ce{YBiRu2O7} & 156 &           27 &        0.016 &     0.000 \\
     98 &   mp-24285 &                \ce{NpH2} & 156 &           29 &        0.000 &     0.000 \\
     99 &    mp-9632 &              \ce{CsReF6} & 155 &           15 &        0.000 &     0.000 \\
     100 & mp-1223068 &     \ce{Li7Eu8Si4Cl7O16} & 154 &           46 &        0.006 &     0.291 \\

\end{longtable}

\end{document}